\documentclass{aa}  
\usepackage{txfonts}
\usepackage{rotating}
\usepackage{answers}
\usepackage{xcolor}
\usepackage{graphicx}
\usepackage{aalongtable}
\usepackage{scalefnt}
\newbox\grsign \setbox\grsign=\hbox{$>$} \newdimen\grdimen \grdimen=\ht\grsign
\newbox\simlessbox \newbox\simgreatbox
\setbox\simgreatbox=\hbox{\raise.5ex\hbox{$>$}\llap
     {\lower.5ex\hbox{$\sim$}}}\ht1=\grdimen\dp1=0pt
\setbox\simlessbox=\hbox{\raise.5ex\hbox{$<$}\llap
     {\lower.5ex\hbox{$\sim$}}}\ht2=\grdimen\dp2=0pt

\def\simless{\mathrel{\copy\simlessbox}}
\begin{document}
\title{UVES analysis of red giants in the bulge globular cluster NGC 6522 
\thanks{Observations collected at the European Southern Observatory,
Paranal, Chile (ESO), under programmes 088.D-0398A, and 097.D-0175
(PI: B. Barbuy), 071.B-0617, 73.B-0074 (PI: A. Renzini.).}}
\author{
B. Barbuy\inst{1}
\and
E. Cantelli\inst{1}
\and
L. Muniz\inst{1}
\and
S.O. Souza\inst{1}
\and
C. Chiappini\inst{2}
\and
R. Hirschi\inst{3,4}
\and
G. Cescutti\inst{5,6}
\and
M. Pignatari\inst{7,8,9,10}
\and
S. Ortolani\inst{11,12}
\and
L. Kerber\inst{13}
\and
F.F.S. Maia\inst{14}
\and
E. Bica\inst{15}
\and
E. Depagne\inst{16}
}
\offprints{B. Barbuy}
\institute{
Universidade de S\~ao Paulo, IAG, Rua do Mat\~ao 1226,
Cidade Universit\'aria, S\~ao Paulo 05508-900, Brazil
\and
Astrophysikalisches Institut Potsdam, An der Sternwarte 16, Potsdam, 14482, Germany
\and
Astrophysics Group, Keele University, Keele, Staffordshire ST5 5BG, UK
\and
Kavli Institute for the Physics and Mathematice of the Universe (WPI), University of Tokyo, 5-1-5 Kashiwanoha, Kashiwa 277-8583, Japan
\and
INAF, Osservatorio Astronomico di Trieste, Via G.B. Tiepolo 11, I-34143 Trieste, Italy
\and
IFPU, Istitute for the Fundamental Physics of the Universe, Via Beirut, 2, 34151, Grignano, Trieste, Italy
\and
E.A. Milne Centre for Astrophysics, Department of Physics \& Mathematics,
University of Hull, HU6 7RX, UK
\and
Konkoly Observatory, Research Centre for Astronomy and Earth Sciences, 
Hungarian Academy of Sciences, Konkoly Thege Miklos ut 15-17, H-1121 Budapest, Hungary
\and
NuGrid Collaboration, \url{www.nugridstars.org}
\and
Joint Institute for Nuclear Astrophysics - Center for the Evolution of the
Elements, USA
\and
Universit\`a di Padova, Dipartimento di Astronomia, Vicolo
dell'Osservatorio 2, I-35122 Padova, Italy
\and
INAF-Osservatorio Astronomico di Padova, Vicolo dell'Osservatorio 5,
I-35122 Padova, Italy
\and
Universidade Estadual de Santa Cruz, Depto. de Ciências Exatas e Tecnológicas, Rodovia Jorge Amado km 16, 45662-900, Ilh\'eus, Brazil
\and
Universidade Federal do Rio de Janeiro, Instituto de Física, Av. Athos da Silveira Ramos, 149, 21941-972, Rio de Janeiro, Brazil
\and
Universidade Federal do Rio Grande do Sul, Departamento de Astronomia,
CP 15051, Porto Alegre 91501-970, Brazil
\and
South African Astronomical Observatory (SAAO), P.O. Box 9. Observatory 7935, South Africa
}
\date{Received; accepted }
 \abstract
    { NGC 6522 is a moderately metal-poor bulge globular cluster ([Fe/H]$\sim$$-$1.0), and it is a well-studied representative among 
      a number of moderately metal-poor blue horizontal branch clusters
      located in the bulge.
       The NGC 6522 abundance pattern can give hints on the earliest
       chemical enrichment  in the central Galaxy.}
   {The aim of this study is to derive abundances
     of the light elements C and N;
     alpha elements O, Mg, Si, Ca, and Ti; odd-Z elements Na and Al;
     neutron-capture elements  
  Y, Zr, Ba, La, and Nd; and the r-process element Eu.
We verify if there are first- and second-generation
stars: we find clear evidence of Na-Al, Na-N, and Mg-Al correlations, while we cannot identify the Na-O anti-correlation from our data. 
}
   {High-resolution spectra of six red giants in 
      the bulge globular cluster
     NGC 6522 were obtained at the 8m VLT UT2-Kueyen telescope with
     both the UVES and   GIRAFFE spectrographs in FLAMES+UVES configuration.  
     In light of Gaia data, it turned out that two of them are non-members,
     but these were also analysed.
   Spectroscopic parameters were derived through the excitation and ionisation 
equilibrium of \ion{Fe}{I} and \ion{Fe}{II} lines from UVES spectra. The 
 abundances were obtained with spectrum synthesis. Comparisons
 of abundances derived from UVES and GIRAFFE spectra were carried out.}
   {The  present analysis combined with previous UVES results
     gives a mean radial velocity of
     v$^{\rm hel}_{\rm r}$ = $-$15.62$\pm$7.7 km s$^{-1}$ and a metallicity of  
     [Fe/H] = $-$1.05$\pm$0.20 for NGC 6522. Mean abundances of alpha elements
     for the present four member stars are enhanced with
 [O/Fe]=+0.38, [Mg/Fe]=$\approx$+0.28, [Si/Fe]$\approx$+0.19, and
 [Ca/Fe]$\approx$+0.13, together with the iron-peak element
 [Ti/Fe]$\approx$+0.13, and the r-process element [Eu/Fe]=+0.40.
The neutron-capture elements Y, Zr, Ba, and La show enhancements in the
+0.08 $<$ [Y/Fe] $<$ +0.90, 0.11 $<$ [Zr/Fe] $<$ +0.50, 
0.00 $<$ [Ba/Fe] $<$ +0.63, 0.00 $<$ [La/Fe] $<$ +0.45, and
-0.10 $<$ [Nd/Fe] $<$ +0.70 ranges. 
We also discuss the spread in heavy-element abundances. }
   {}
\titlerunning{Red giants in NGC 6522}
\authorrunning{B. Barbuy et al.}
\maketitle
%

\section{Introduction}

The Galactic bulge formation was probably a result of early mergers and/or
dissipative collapse combined with a buckling bar, as suggested by the
excellent modern data now available in the innermost parts of the Galaxy
(Queiroz et al. 2020a,b, Rojas-Arriagada et al. 2020,
P\'erez-Villegas et al. 2020, Kunder et al. 2020, Savino et al. 2020 -
see also review by Barbuy et al. 2018a), and chemodynamical models
(e.g. Fragkoudi et al. 2020, Debattista et al. 2020, Baba \& Kawata 2020).
Within this context, globular clusters are important tracers of the early
formation of the Galactic bulge, assuming that most of them were formed in situ.
In particular, their abundance pattern could give hints
as to the early nucleosynthesis processes in the Galaxy.

The globular cluster NGC~6522 located in the Galactic bulge was identified by
Baade (1946)
as having a type II stellar population given its colour-magnitude diagram (CMD);
hence, it falling into the Population II class defined in Baade (1944).
Despite such an early identification of this cluster, it
has not been widely studied since then. 

NGC 6522 is an old globular cluster, with a moderate metallicity of
[Fe/H]$\sim$$-$1.0, and a blue horizontal branch.
Several other such clusters are present in the
Galactic bulge, such as NGC~6558 (Rich et al. 1998, Barbuy et al.
2007, 2018b), HP~1 (Barbuy et al. 2006, 2016) , AL~3 (Ortolani et al. 2006), 
Terzan~9 (Ernandes et al. 2019), and UKS~1 (Fern\'andez-Trincado et al. 2020).
 Rossi et al. (2015) presented a study gathering these inner globular clusters,
 that might represent the earliest stellar populations in the Galaxy.
  Based on their derived proper motions, Rossi et al. (2015) concluded
   that the inner bulge globular clusters
   have clearly lower transverse motions and spatial velocities than halo
   clusters, and they appear to be trapped in the bulge bar.

P\'erez-Villegas et al. (2020) computed the orbits of
78 inner Galaxy globular clusters, following the selections given
in Bica (2016).
It was found that most of the confirmed bulge-population clusters are confined in
the bar region but are not supporting the bar structure.
 For each cluster, a set of 1000 initial conditions were generated,
following the Monte Carlo technique and taking into account the observational
uncertainties. 
NGC~6522 has a 99.8\% probability of being a bulge member,
and most of the orbits among the different initial conditions
do support the bar shape.
 A study of the origin of Galactic globular clusters by Massari et al.
  (2019) also identified NGC~6522 as having been formed in the main bulge
  and not, for example, in the Gaia-Enceladus that merged with the Galaxy
  about 10 Gyr ago.
At the same time, it was confirmed to be an old
cluster by Kerber et al. (2018): 14.1, 14.2 Gyr
for [Fe/H]= $-$1.0, $-$1.15 from BaSTI isochrones
(Pietrinferni et al. 2004, 2006)
and 12.1, 12.4 Gyr for [Fe/H]=$-$1.0, $-$1.15 from Dartmouth
isochrones (Dotter et al. 2008).
This old age indicates that NGC~6522 was formed 4 Gyr prior
relatively to the estimated age
of the bar formation of 8$\pm$2 Gyr by
Buck et al. (2018) and $\sim$8 Gyr ago by Bovy et al. (2019). Therefore,
the fact that NGC~6522 follows the bar probably indicates that it was
confined within the bar when the latter formed.

Barbuy et al. (2009) analysed eight member stars based on FLAMES-GIRAFFE
 (Pasquini et al. 2002)
spectra, included in the survey by Zoccali et al. (2008).
Even with these low-resolution spectra (as compared with the UVES spectra analysed
later)
it was possible to detect some enrichment in s-process elements, which Chiappini et al.
(2011a, hereafter C11) interpreted as a possible signature of an early generation of fast rotating
stars (the so-called spinstars). This was based on the idea that, as the age-metallicity relation
in the bulge was expected to be steeper than in the halo, it would already be possible to reach metallicities
as large as [Fe/H] $\sim -$1 on a very short timescale. It was then suggested that in the bulge,
globular clusters at [Fe/H] $\sim -$1.0 would already represent tracers of the earliest chemical
enrichment phases. Subsequently, we obtained UVES spectra for four of the stars previously
analysed in Barbuy et al. (2009)
and C11. The re-analysis of these stars based on higher resolution and higher signal-to-noise (S/N)
spectra, obtained with the UVES spectrograph at the Very Large Telescope was then presented in
Barbuy et al. (2014) where some enrichment of s-process elements in the very old NGC~6522
cluster has been confirmed, although smaller than what was suggested in the earlier low-resolution spectra.
It was then necessary to expand the sample to better constrain the nature of the stars that have
polluted this very old cluster. With this aim, during our first UVES observations we also obtained parallel
observations with the FLAMES-GIRAFFE spectrograph, and new candidate members were identified. These,
in turn, were observed with UVES in a new run.
In this work, we present results for six stars in NGC~6522
  obtained in 2016 with the FLAMES-UVES spectrograph (Dekker et al. 2000).
Our main aim is to study the abundance signatures of heavy elements in the cluster.

Furthermore,
NGC~6522 was recently shown by Kerber et al. (2018) to have at least two stellar populations 
in proportions of 86\% as second generation (2G) and 14\% as first generation
(1G). 
For this reason, we inspected possible Na-O
anti-correlations and Na-Al, Na-N, and Mg-Al correlations (Gratton
et al. 2012 and references therein)
among the present sample stars together with stars analysed
in Barbuy et al. (2014).

Finally, we compare element abundance derivation from UVES and GIRAFFE
spectra for stars observed with both instruments (in their common wavelength region) to check for further use of the lower resolution
spectra.
Additionally, in Table \ref{girmag} in the appendix we list  stars
identified to be candidate members of the cluster, 
observed with FLAMES-GIRAFFE, selected from their radial velocities
together with Gaia collaboration (2018, 2021) proper motions.

   The observations are described  in Sect. 2.
  Photometric stellar parameters' effective temperature and  gravity  
are   derived in  Sect. 3.  Spectroscopic parameters are
derived in  Section  4, and   abundance ratios are   computed in  Sect.
5. A discussion is presented in Sect. 6, and conclusions are drawn in Sect. 7.

\section{Observations} 

In Barbuy et al. (2009), eight stars of NGC~6522 observed
with the GIRAFFE spectrograph, within the list
of over 600 bulge stars analysed
by Zoccali et al. (2008) (programs 71.B-0617 and 73.B-0074,
PI: A. Renzini)  
were studied. Four of these were re-observed at higher resolution
with UVES and analysed in  Barbuy et al. (2014),
in programme 88.D-0398 (PI: B. Barbuy) in 2012.
From GIRAFFE stars observed in the same field within this same programme, we 
identified stars with radial velocities and metallicities that could be cluster members.
In program 097.D-0175 (PI: B. Barbuy), we observed five of these stars, plus star B118, which was previously studied at a moderate resolution in Barbuy et al. (2009).
The log-book of observations is given in Table \ref{logbook}.

 The UVES spectra were obtained using the FLAMES setup centred at
580 nm, with a coverage ranging from 480 nm to 680 nm.
The 2012 GIRAFFE spectra  were obtained using setups
HR11 (559.7-584.0 nm) and HR12 (582.1-614.6 nm), and the 2016 GIRAFFE observations used
setups HR11, HR13 (612.0-640.5 nm), and HR14A (630.8-670.1 nm), all with
a mean resolving power of R $\approx$ 22,000.

The UVES data were reduced with the UVES pipeline v5.7.0 within the
REFLEX ambient. The extracted spectra were treated, normalised,
rest-frame aligned, and combined using the method described in Section
2.3 of Cantelli (2019). Cosmic-ray removal was done by
sigma rejection. Radial velocities
were measured using the upper wavelength chip of the UVES red arm,
through the IRAF task fxcor, using the Arcturus spectrum
as template (Hinkle et al. 2000). The measurements for each
star and each exposure and the heliocentric radial velocities computed through the IRAF task rvcorrect are reported in Table \ref{vruvesmeu}.

\begin{figure}
  \centering
  \includegraphics[width=9cm]{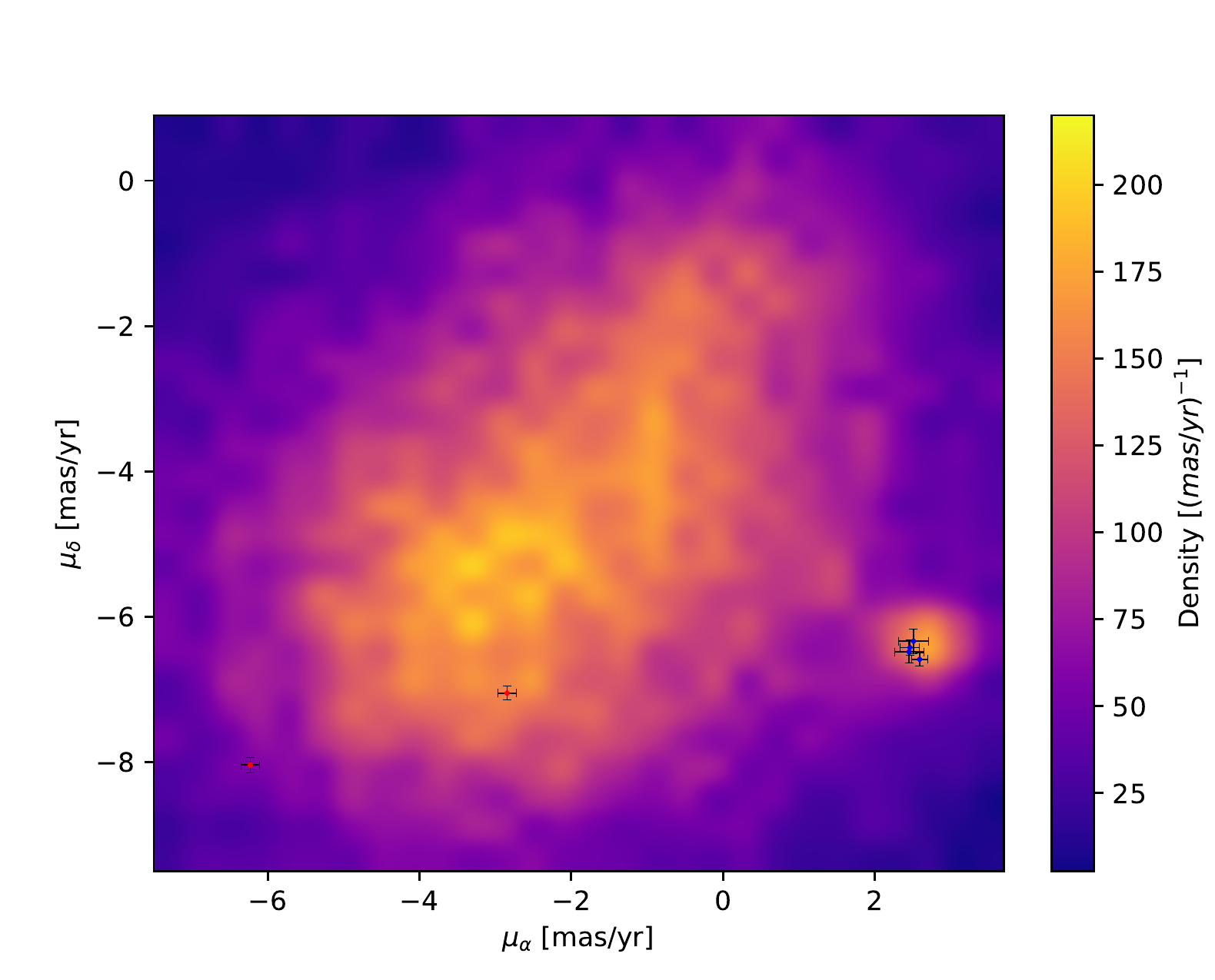}
  \caption{Gaia proper motions at the location of NGC~6522.
    The background is the proper motion density plot within 8 arcmin of the
    cluster centre. Blue crosses are the member sample stars, and red crosses
    are the non-member ones.}
\label{point} 
\end{figure}

For completeness, and thanks to the Gaia collaboration (2018, 2021)
measurements,
we were able to proceed to a more robust verification of membership.
In Table \ref{gaia},
we list the proper motions ($\mu_\alpha \cos \delta$ and $\mu_\delta$) and
G magnitude from Gaia Early Data Release 3 (EDR3, Gaia Collaboration 2020).
With the precision improvement of Gaia EDR3 on proper motions,
we recalculated the mean values for NGC~6522. We selected stars within
15 arcmin of the cluster centre and applied the Gaussian mixture
models (GMM) to separate the cluster stars from field stars. With this method,
we recalculated the cluster proper motion as
$\mu_\alpha \cos \delta = 2.55\pm0.08$
mas yr$^{-1}$ and $\mu_\delta = $-$6.45\pm0.07$ mas yr$^{-1}$.
Also, the cluster and field distributions allowed us to determine the
membership probability for each star. As shown in Table \ref{gaia},
and in the point-vector diagram of Figure \ref{point},
stars 234816 and 244523 turn out to be non-members.
This shows the power of Gaia, given that these two stars 
 have the very compatible metallicities and radial velocities, coinciding
 with the other four member stars in the CMD.
 This is even more striking given that only 0.5\% of field
 stars show a metallicity of [Fe/H]$\simless$$-$1.0 in the
 Galactic bulge (Barbuy et al. 2018a).

A mean heliocentric radial velocity of v$^{\rm hel}_{\rm r}$ = $-$16.96 km s$^{-1}$ is found for the four  UVES sample member stars. A mean value of
v$^{\rm hel}_{\rm r}$ = $-$14.3$\pm$3.3 km
s$^{-1}$ was obtained from UVES spectra of four stars analysed in Barbuy et al.
(2014).
By combining the present data with these four stars, namely, B-107, B-128, B-122 and B-130, with 
v$^{\rm hel}_{\rm r}$ = $-$7.626, $-$14.651, $-$18.043, and
$-$16.808 km s$^{-1,}$  respectively, we obtain a mean value of
v$^{\rm hel}_{\rm r}$ = -15.62 km s$^{-1}$.
A range of velocities between v$^{\rm hel}_{\rm r}$ = $-$7.63 and $-$22.57 km s$^{-1}$ gives a  dispersion of $\pm$7.7 km s$^{-1}$. A similar range of radial velocities was detected in Fern\'andez-Trincado et al. (2019), including stars with $-$21.97 $<$ v$^{\rm hel}_{\rm r}$ $<$ $-$6.61 km s$^{-1}$. 

 The GIRAFFE data were retrieved from the ESO reduced 
data\footnote{archive.eso.org/wdb/wdb/adp/phase3$_{--}$main/form} archive.
The extracted spectra belonging to the same setups were then corrected for
radial velocity, normalised, and combined by the median. 
In the appendix, we give a list of new candidate member stars observed with
GIRAFFE in 2012 and 2016, within a radial velocity range of $-$14.5$\pm$12 
km s$^{-1}$, that are confirmed members from proper motions.

\begin{table}
\caption{Log of the spectroscopic observations of programs 088.D-0398(A)
and 097.D-0175(A), carried out in 2011-2012 and 2016, respectively.
 The seeing and airmasses reported are the mean values along the exposures.}
\scalefont{0.9}
\begin{flushleft}
\begin{tabular}{llccc@{}c@{}c}
\noalign{\smallskip}
\hline
\noalign{\smallskip}
\hline
\noalign{\smallskip}
Date & UT & Julian & exp   & Air-  & Seeing& \\ 
     &    & date &  (s)  &  mass & ($''$)  & \\ 
\noalign{\smallskip}
\hline
\noalign{\smallskip}
\multicolumn{6}{c}{ Program 088.D-0398(A) } \\
\hline
\noalign{\smallskip}
2011-10-08 & 00:45:54 & 2455842.53187  & 2750 & 1.462 & 0.82$''$ & \\ 
2011-10-08 & 01:34:37 &2455842.56571 & 2750 & 1.853 & 1.29$''$ & \\ 
2012-03-06 & 07:38:32 &2455992.81843 & 2750 & 1.579 & 1.15$''$ & \\ 
2012-03-06 & 08:28:44 & 2455992.85329 & 2750 & 1.260 & 0.93$''$ & \\ 
2012-03-07 & 07:47:56 & 2455993.82495 & 2750 & 1.489 & 0.81$''$ & \\ 
2012-03-07 & 08:39:16 & 2455993.86060 & 2750 & 1.270 & 0.73$''$ & \\ 
2012-03-25 & 08:31:47 & 2456011.85541 & 2750 & 1.087 & 0.64$''$ & \\ 
\noalign{\smallskip}
\hline
\noalign{\smallskip}
\multicolumn{6}{c}{ Program 097.D-0175(A) } \\
\hline
\noalign{\smallskip}
2016-05-17 & 07:22:18 & 2457525.80716 &2400 & 1.007 & 0.40$''$ &\\ 
2016-05-17 & 08:05:08 & 2457525.83690 &2400 & 1.033 & 0.47$''$ & \\ 
2016-05-17 & 08:52:35 & 2457525.86985 & 2400 & 1.099 & 0.47$''$ & \\ 
2016-07-11 & 02:33:35 & 2457580.60666 & 2400 & 1.028 & 0.96$''$ & \\ 
2016-07-21 & 03:27:16 & 2457590.64394 & 2400 & 1.016 & 0.51$''$ & \\ 
2016-07-21 & 04:43:37 & 2457590.69696 & 2400 & 1.112 & 0.54$''$ & \\ 
2016-07-21 & 06:33:32 & 2457590.75246 & 2400 & 1.373 & 0.54$''$ & \\ 
2016-07-22 & 04:48:26 & 2457591.70031 & 2400 & 1.131 & 0.48$''$ & \\ 
2016-07-22 & 05:40:15 & 2457591.73629 & 2400 & 1.288 & 0.45$''$ & \\ 
2016-07-22 & 06:33:29 & 2457591.77326 & 2400 & 1.574 & 0.63$''$ &\\ 
\noalign{\smallskip}
\hline
\noalign{\smallskip}
\hline
\end{tabular}
\end{flushleft}
\label{logbook}
\end{table}

\begin{table*}
  \caption[2]{Observed and heliocentric radial velocities of the UVES sample stars, in each of the 10 UVES
exposures, and the mean heliocentric radial velocity.}
\scalefont{0.8}
\label{vruvesmeu}
\begin{flushleft}
\begin{tabular}{lccccccccccccccc}
\hline
\noalign{\smallskip}
{\rm OGLE~n$^{\circ}$} &             & \multispan2{ 234816} & \multispan2{244523} & \multispan2{244819} & \multispan2{256289}& \multispan2{B118}&
          \multispan2{402370} \\
\noalign{\smallskip}      
\hline
\noalign{\smallskip}
Date & UT & \multicolumn{12}{c}{ Observed and Heliocentric radial velocity ({${\rm km.s^{-1}}$}) } \\
\noalign{\smallskip}
\hline
\noalign{\smallskip}
 & &  \hbox{v$_{obs}$} & \hbox{v$_{hel}$} &
 \hbox{v$_{obs}$} & \hbox{v$_{hel}$} &
 \hbox{v$_{obs}$} & \hbox{v$_{hel}$} &
 \hbox{v$_{obs}$} & \hbox{v$_{hel}$} &
 \hbox{v$_{obs}$} & \hbox{v$_{hel}$} &
 \hbox{v$_{obs}$} & \hbox{v$_{hel}$}  \\
\noalign{\smallskip}
\hline
\noalign{\smallskip}
17-05-2016& 07:22:18.596&  $-$31.90& $-$15.21 & $-$29.59& $-$12.89 & $-$39.18& $-$22.49 &$-$31.56& $-$14.87&$-$36.48& $-$19.79&  $-$27.20& $-$10.51\\                                         
17-05-2016& 08:05:08.353&  $-$31.98& $-$15.36 & $-$29.66& $-$13.95 &  ---    &  ---   &$-$31.75& $-$15.13&$-$36.61& $-$19.99&  $-$27.53& $-$10.92\\                                           
17-05-2016& 08:52:35.292&  $-$31.67& $-$15.13 & $-$29.39& $-$12.85 & $-$39.15& $-$22.61 &$-$31.17& $-$14.63&$-$36.36& $-$19.82&  $-$27.20& $-$10.66\\                                           
11-07-2016& 02:33:35.361&  $-$06.85& $-$15.93 & $-$04.61& $-$13.68 & $-$13.99& $-$23.07 &$-$06.42& $-$15.49&$-$10.98& $-$20.05&  $-$02.02& $-$11.09\\                                           
21-07-2016& 03:27:16.562&  $-$01.53& $-$15.27 &   0.76& $-$12.99 & $-$08.47& $-$22.22 & 01.48& $-$12.27&$-$05.48& $-$19.22&   02.69& $-$11.06\\                                           
21-07-2016& 04:43:37.414&  $-$01.36& $-$15.24 &   0.77& $-$13.11 & $-$08.14& $-$22.03 &$-$0.92 & $-$14.80&$-$05.25& $-$19.13&   02.70& $-$11.19\\                                           
21-07-2016& 06:33:32.375&   $-$1.33& $-$15.33 &   0.55& $-$13.46 &  $-$8.21& $-$22.21 & $-$0.81& $-$14.82& $-$5.11& $-$19.11&    2.84& $-$11.18\\                                           
22-07-2016& 04:48:26.611&  $-$01.04& $-$15.37 &  01.11& $-$13.22 & $-$08.38& $-$22.71 & $-$0.54& $-$14.87&$-$05.36& $-$19.69&   03.78& $-$10.55\\                                           
22-07-2016& 05:40:15.273&  $-$01.39& $-$15.80 &   0.26& $-$14.15 & $-$08.26& $-$22.67 & $-$0.79& $-$15.20& $-$5.18& $-$19.59&   02.89& $-$11.52\\                                           
22-07-2016& 06:33:29.572&   $-$0.88& $-$15.36 &   1.11& $-$13.37 & $-$08.66& $-$23.14 & $-$0.44& $-$14.92&$-$05.12& $-$19.60&    3.48& $-$11.00\\
\noalign{\smallskip}
Mean v$_{hel}$&             &          & $-$15.40 &          & $-$13.37& & $-$22.57 & & $-$14.70 & & $-$19.60 & & $-$10.97 \\
\hline
\noalign{\smallskip} \hline \end{tabular}
\end{flushleft}
\end{table*}

\begin{table}  
  \caption[3]{Identifications and Gaia proper motions; Gaia G and
    Johnson B and V magnitudes
    and membership probability.
    Stars from Barbuy et al. (2009, 2014) are also included. }
\label{gaia}
\scalefont{0.80}
\begin{flushleft}
\tabcolsep 0.15cm
\begin{tabular}{l@{}cccccccccc}
\noalign{\smallskip}
\hline
\noalign{\smallskip}
\hline
\noalign{\smallskip}
\hbox{OGLE}& \hbox{Name} & pm$_{RA}$ &pm$_{DEC}$ & $Gmag$ & $V$ & $B$ &
$\mathcal{P}_{Member}$ \\
\noalign{\smallskip}
\noalign{\hrule}
&   \multicolumn{6}{c} {Present work} & \\
 234816 &        &   $-6.1170$ &   $-8.3550$ &  $15.6950$  & $16.401$ & $17.781$ &     0  \\ 
 244523 &        &   $-3.0400$ &   $-6.9400$ &  $15.3813$  & $15.988$ & $17.283$ &     0  \\ 
 244819 &        &   $2.6380$  &   $-6.6090$ &  $15.6814$  & $16.306$ & $17.614$ &   100  \\
 256289 &        &   $2.5830$  &   $-7.1310$ &  $15.3887$  & $15.918$ & $17.192$ &    98  \\
 402322 &  B118  &   $2.4800$  &   $-6.5720$ &  $15.3817$  & $16.011$ & $17.331$ &   100  \\
 402370 &        &   $2.4900$  &   $-6.6590$ &  $15.5926$  & $16.226$ & $17.551$ &   100  \\
  \multicolumn{6}{c} {Stars from Barbuy et al. (2014)} & \\
 402361 &   B107 &   $2.8460$  &   $-6.8850$ &  $15.3375$  & $15.980$ & $17.292$ &   100  \\  
 244582 &   B122 &   $2.4820$  &   $-6.0110$ &  $15.3401$  & $16.000$ & $17.354$ &    99  \\
 402607 &   B128 &   $2.4810$  &   $-6.4190$ &  $15.6001$  & $16.260$ & $17.621$ &   100  \\
 402531 &   B130 &   $3.0100$  &   $-6.1770$ &  $15.6871$  & $16.300$ & $17.597$ &   100  \\
&   \multicolumn{6}{c} {Other stars from Barbuy et al. (2009)} & \\
           
 412752 &   B008 &   $2.6170$  &   $-6.4500$ &  $15.4115$  & $15.990$ & $17.402$ &   100  \\ 
 --     &   B108 &   $3.3340$  &   $-6.2990$ &  $15.1898$  & $16.290$ & $18.313$ &    99  \\
 --     &   B134 &   $0.7350$  &   $-5.0020$ &  $15.5082$  & $16.040$ & $17.367$ &     0  \\ 
 244829 &   F121 &   $2.2580$  &   $-7.0090$ &  $15.7923$  & $16.400$ & $17.849$ &    99  \\
\noalign{\smallskip}
\hline
\end{tabular}               
\end{flushleft}                                         
\end{table}

\section{Photometric stellar parameters}

\subsection{Temperatures}

\begin{figure}
  \centering
  \includegraphics[width=9cm]{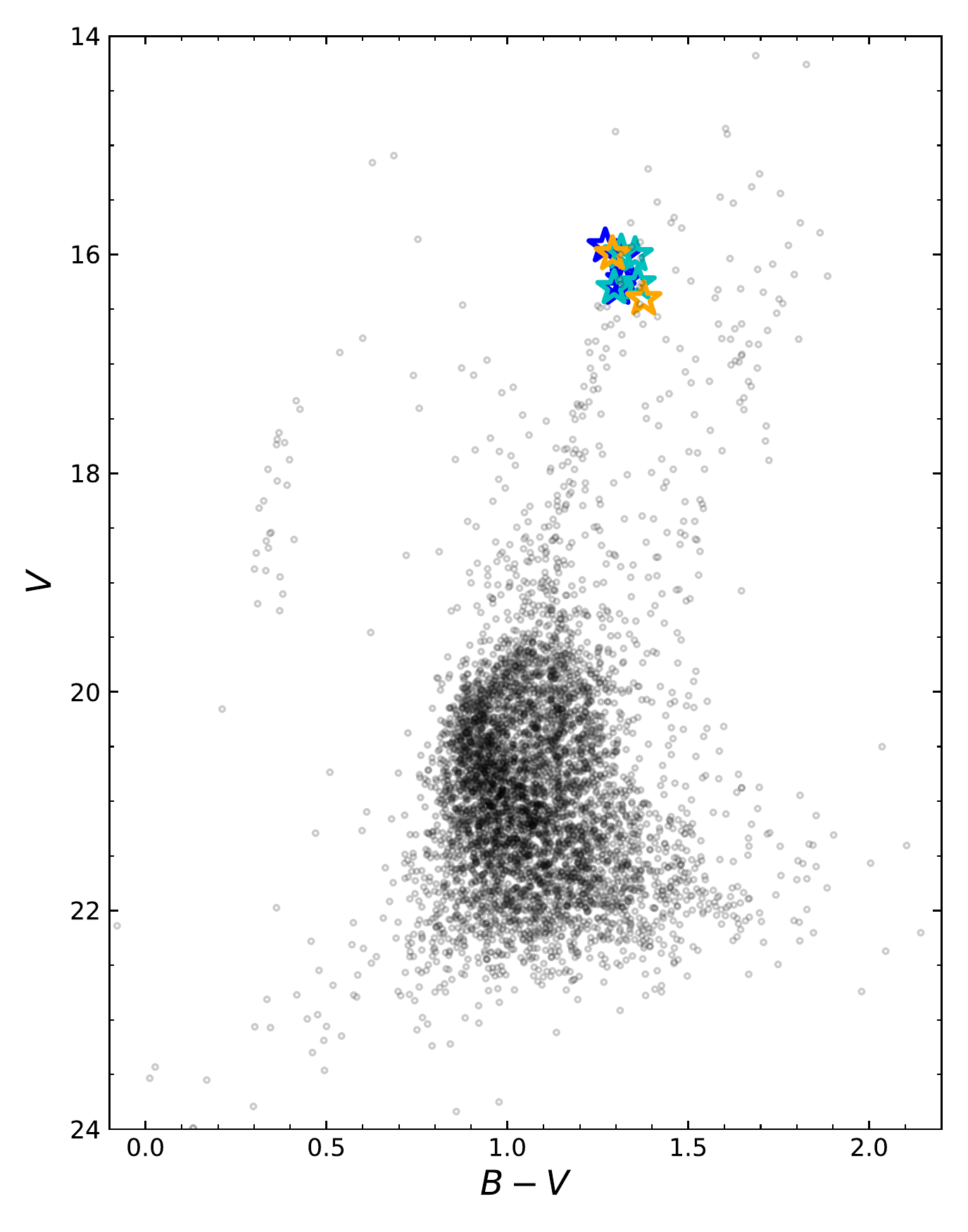}
\caption{NGC~6522 Colour-magnitude diagram $V$ versus $B-V$
  by Piotto et al. (2002) with the location of the observed stars.
  Present four stars: blue. Four stars from Barbuy et al. (2014): cyan.
  Two non-member stars: orange.}
\label{cmd} 
\end{figure}

 The selected stars, their OGLE and 2MASS designations, coordinates, and 
 $VIJHK_{\rm s}$ magnitudes are given in Table \ref{starmag}. $V$ and $I$ data
 were collected
from the Optical Gravitational Lensing Experiment (OGLE) survey,
the OGLE-II release\footnote{www.astrouw.edu.pl/$\sim$ogle/photdb},
 Field Bul-SC45 centered at 18:03:33.0, 
 $-$30:05:00 from  Udalski et al. (2002). 2MASS $J$, $H$, and $K_{\rm s}$   are
from Skrutskie et al. (2006)\footnote{
  $\mathtt{http://ipac.caltech.edu/2mass/releases/allsky/}$;
  $\mathtt{https://irsa.iapc.caltech.edu}$}, and
VVV $J$, $H$, and $K_{\rm s}$ magnitudes are from the
 Vista Variables in the Via Lactea survey
 (Saito et al. 2012)\footnote{$\mathtt{horus.roe.ac.uk/vsa}$}.
Table \ref{gaia} reports the Gaia (2018, 2020) 
  G magnitudes and deduced B magnitudes by applying
  the transformation $G_{\rm mag} - V_C = f(B-V)$ from Riello et al. (2021).
In Fig. \ref{cmd}, we show the location in $B$, $V$ of the sample stars,
in the CMD of NGC 6522 from data observed in F435W and F555W
with the Hubble Space Telescope, and converted to $B$ and $V$ by Piotto et al. (2002). 

Photometric effective temperatures and bolometric corrections
 were derived from 
 $V-I$, $V-K$, and $J-K$ using the colour-temperature calibrations 
 of Alonso et al.  (1999, hereafter AAM99) and Casagrande et al.
 (2010, hereafter C10).
 For the transformation of $V-I$ from the Cousins to Johnson system,
 we used ($V-I$)$_{C}$=0.778($V-I$)$_{J}$    (Bessell 1979).
 $J$,$H$, and $K_S$ 2MASS magnitudes  and colours were  transformed from the 2MASS
 system to  the California Institute of  Technology (CIT), and from this to
 TCS (Telescopio Carlos S\'anchez),  following Carpenter (2001)  and
 Alonso et  al.  (1998).  
The VVV $JHK_{s}$ colours were transformed to the 2MASS $JHK_{s}$ system using
relations by Soto et al. (2013) and then transformed to CIT as above.

 The derived photometric effective
 temperatures, which are adopted as initial guesses, resulting from relations
 from AAM99 and C10  are both listed in  Table~\ref{tabteff}. 
 The differences in effective temperatures, in the 
$\Delta$T$_{eff}$(C10-AAM99) sense, are of +64.7 K, +54.2 K and $\pm$140 K
for $V-I$, $V-K$ and $J-K,$ respectively. These temperatures are used only
as a guide to start fitting them from the \ion{Fe}{I} and \ion{Fe}{II} lines.

\subsection{Gravities}

For a derivation of photometric gravities, we used the classical
formula, 
adopting T$_{\rm eff,\odot}$=5770 K, M$_*$=0.85 M$_{\odot,}$ and M$_{\rm bol, \odot}$
= 4.75. For the cluster, we used a distance modulus of ($m-$M)$_0$ = 14.40
 and a reddening of  E(B-V)=0.52 and A$_V$    = 1.61, based on
  Kerber et al. (2018).
  Bolometric  corrections were derived using AAM99 and C10 assuming
  BC$_{\rm V,\odot}=$-$0.07$, M$_{\rm V,\odot}=4.81$, and M$_{\rm I,\odot}=4.10$
  from Willmer (2018).
 The computed bolometric magnitudes and gravities are
 given in Table~\ref{tabteff}.

\begin{table*}  
\caption[3]{Identifications, coordinates, and magnitudes.
  $JHK_{s}$ from both 2MASS and VVV surveys.  }
\label{starmag}
\scalefont{0.75}
\begin{flushleft}
\tabcolsep 0.15cm
\begin{tabular}{lcccccccccccccccccccc}
\noalign{\smallskip}
\hline
\noalign{\smallskip}
\hline
\noalign{\smallskip}
{\rm  OGLE}& \phantom{-}\phantom{-}\phantom{-}2MASS ID & $\alpha_{2000}$ & \phantom{-}\phantom{-}\phantom{-}$\delta_{2000}$ 
& \phantom{-}\phantom{-}\phantom{-}$V$ & \phantom{-}\phantom{-}\phantom{-}$I$ & \phantom{-}\phantom{-}\phantom{-}$J$ 
& \phantom{-}\phantom{-}\phantom{-}$H$ & \phantom{-}\phantom{-}\phantom{-}$K_{\rm s}$ &   \phantom{-}\phantom{-}\phantom{-}$J_{\rm VVV}$ 
& \phantom{-}\phantom{-}\phantom{-}\hbox{$H_{\rm VVV}$} & {$K_{\rm VVV}$} & \cr
\noalign{\smallskip}
\noalign{\hrule}
\noalign{\smallskip}
234816 &  18032652-3006385 & 18:03:26.52 & $-$30:06:38.1 & 16.401 &  14.604&  13.198&  12.488 & 12.302 & 13.1735& 12.5075& 12.3052 &\cr
244523 &  18032757-3003455 & 18:03:27.56 & $-$30:03:45.1 &  15.988 & 14.325& 12.667 & 11.872& 12.274 & 13.0323& 12.4093& 12.233 &\cr
244819 & 18033354-3002254 & 18:03:33.51& $-$30:02:25.2 & 16.306 & 14.672 & 13.020 & 12.284 & 11.421 & 13.3148& 12.6694& 12.499 &\cr
256289 & --- &   18:03:31.58& $-$30:00:51.0 &  15.918 &  14.337 &  ---  &   --- &    --- &   13.0892& 12.4721& 12.289 &\cr
402322 &18034225-3003403&  18:03:42.25& $-$30:03:40.0 & 16.011 & 14.313 & 13.056 & 12.305 & 12.142 & 12.9661& 12.324 &\cr  
402370 &18034235-3002088&  18:03:42.35& $-$30:02:08.5 & 16.226 & 14.554 & 13.391 & 12.673&  12.550 & 13.2643& 12.628 &\cr
\noalign{\smallskip}
\hline
\end{tabular}               
\end{flushleft}                                         
\end{table*}

\begin{table*}
\caption{Photometric stellar parameters derived using the calibrations
 by Alonso et al. (1999) (AAM99) and Casagrande et al. (2010) (C10) 
 for $V-I$, $V-K$, $J-K$, bolometric corrections, bolometric magnitudes
and corresponding gravity log $g$.}
\scalefont{0.75}
\label{tabteff}
\centering
\begin{tabular}{cccccccccccccccc}
\noalign{\smallskip}
\hline
\noalign{\smallskip}
\hline
\noalign{\smallskip}
{\rm star} & T($V-I$) & T($V-K$) & T($J-K$) & T($V-K$) & T($J-K$) & log(T$_{\rm eff}$) & ${\rm BC_{V}}$ & ${\rm M_{bol}}$ & log g & Calib \\
           &          &  2MASS   &  2MASS   &   VVV    &   VVV    &   (mean)           &                &                 &       & \\
           &    (K)   &   (K)    &   (K)    &   (K)    &   (K)    &                    &                &                 &       & \\
\noalign{\smallskip}
\noalign{\hrule}
\noalign{\smallskip}
  234816 &   $4517.1$  &  $4504.8$ &   $4693.7$ &   $4501.0$ &   $4579.4$ &   $3.655$ &  $-0.475$ & $0.86$ &  $2.363$ & AAM99\cr
         &   $4600.0$  &  $4568.4$ &   $4843.0$ &   $4564.5$ &   $4712.6$ &   $3.668$ &  $-0.477$ & $0.86$ &  $2.414$ & C10\cr
\noalign{\smallskip}
  244523 &   $4792.5$  &  $4848.0$ &   $7391.8$ &   $4799.3$ &   $4806.5$ &   $3.681$ &  $-0.344$ & $0.32$ &  $2.249$ & AAM99\cr
         &   $4854.4$  &  $4912.6$ &   $7372.1$ &   $4865.1$ &   $4969.3$ &   $3.732$ &  $-0.351$ & $0.32$ &  $2.453$ & C10\cr
\noalign{\smallskip}
  244819 &   $4858.8$  &  $4027.3$ &   $3340.5$ &   $4749.2$ &   $4746.3$ &   $3.687$ &  $-0.319$ & $0.62$ &  $2.391$ & AAM99\cr
         &   $4917.7$  &  $4039.6$ &   $3115.6$ &   $4815.7$ &   $4902.1$ &   $3.639$ &  $-0.325$ & $0.61$ &  $2.199$ & C10\cr
\noalign{\smallskip}
  256289 &   $4986.7$  &    ---    &     ---    &   $4928.7$ &   $4805.2$ &   $3.698$ &  $-0.277$ & $0.19$ &  $2.264$ & AAM99\cr
         &   $5042.1$  &    ---    &     ---    &   $4990.7$ &   $4967.9$ &   $3.699$ &  $-0.279$ & $0.18$ &  $2.264$ & C10\cr
\noalign{\smallskip}
  402322 &   $4715.8$  &  $4698.8$ &   $4638.5$ &   $4681.9$ &   $4686.9$ &   $3.674$ &  $-0.376$ & $0.37$ &  $2.242$ & AAM99\cr
         &   $4782.2$  &  $4765.6$ &   $4780.3$ &   $4748.8$ &   $4835.3$ &   $3.680$ &  $-0.383$ & $0.37$ &  $2.266$ & C10\cr
\noalign{\smallskip}
  402370 &   $4772.5$  &  $4887.1$ &   $4873.3$ &   $4786.4$ &   $4778.9$ &   $3.679$ &  $-0.352$ & $0.57$ &  $2.341$ & AAM99\cr
         &   $4835.4$  &  $4950.6$ &   $5043.1$ &   $4852.4$ &   $4938.6$ &   $3.692$ &  $-0.359$ & $0.56$ &  $2.393$ & C10\cr
\noalign{\smallskip} \hline \end{tabular}
\end{table*}

\begin{table}
\caption[1]{Final UVES spectroscopic parameters.  }
\label{tabparam}
\small
\begin{flushleft}
\begin{tabular}{cccc@{}c@{}cc}
\noalign{\smallskip}
\hline
\noalign{\smallskip}
\hline
\noalign{\smallskip}
{\rm star} & \phantom{-}T$_{\rm eff}$ &\phantom{-}log g & \phantom{-}[FeI/H]
 & \phantom{-}[FeII/H]  & \phantom{-}[Fe/H] & ${\rm v_t }$  \\
 & (K) & & & & & km s$^{-1}$  \\
\noalign{\smallskip}
\noalign{\hrule}
\noalign{\smallskip}
 234816  &  4650 & 2.25 & $-$1.03 & $-$1.08&  $-$1.05 & 1.65  \\
 244523  &  4800 & 2.00 & $-$1.10 & $-$1.12&  $-$1.11&2.30  \\
 244819  &  4690 & 2.30 & $-$1.23 & $-$1.19&  $-$1.21&1.51  \\
 256289  &  4770 & 2.10 & $-$1.12 & $-$1.11&  $-$1.11&1.25  \\
B118    &  4820 & 2.20 & $-$1.18 & $-$1.16&  $-$1.17 &2.10  \\
 402370  &  4700 & 2.20 & $-$1.15 & $-$1.16&  $-$1.15&1.15  \\
\noalign{\smallskip} \hline \end{tabular}
\end{flushleft}
\end{table}

\section{Spectroscopic stellar parameters}

The equivalent widths (EW) of \ion{Fe}{I} and \ion{Fe}{II} lines were measured
using IRAF.
The EWs, together with wavelength ({\rm \AA}); excitation potential (eV),
damping constant C$_6$ and
oscillator strengths from
VALD3 (Piskunov et al. 1995, Ryabchikova et al. 2015), National
Institute of Standards \& Technology (NIST, Martin et al.
2002)\footnote{http://physics.nist.gov/PhysRefData/ASD/lines$_-$form.html} and Kur\'ucz 
(1993)\footnote{http://www.cfa.harvard.edu/amp/amp\-data/ku\-rucz23/\-se\-kur.\-html},
\footnote{http://kurucz.harvard.edu/atoms.html} and adopted values are
given in Table \ref{EW}. For \ion{Fe}{I,} we chose NIST values when available,
otherwise they are from VALD3. In most cases, these coincide with the value
from Kurucz, as can be seen in Table \ref{EW}. For \ion{Fe}{II}, values from Mel\'endez
\&  Barbuy (2009) are used.
The solar abundances were adopted from Grevesse \& Sauval (1998),
including $\epsilon$(Fe)=7.50 for Fe, except for
A(O)=8.76 for oxygen from Steffen et al. (2015).

The models were interpolated in the
MARCS model atmospheres grid (Gustafsson et al. 2008).
We adopted the spherical
and mildly CN-cycled set ([C/Fe]$=-0.13$, [N/Fe]$=+0.31$).
These models consider [$\alpha$/Fe]=+0.20 for [Fe/H]=-0.50 and 
[$\alpha$/Fe]=+0.40 for [Fe/H]$\leq-$1.00.
The LTE  abundance  analysis and  the spectrum  synthesis calculations
were performed using the code described  in  Barbuy et al. (2018c).
 The code is an update of the Meudon ABON2 code by M. Spite, continously
  improved along the years, which adopts
  local thermodynamic equilibrium (LTE).  In Trevisan et al. (2011),
  the calculation of lines and in particular the
  continuum opacity calculation were cross-checked
with the code by the Uppsala group BSYN/EQWI (Edvardsson
et al. 1993, and updates until that date).
  The basic atomic line list is from VALD3 (Ryabchikova et al. 2015).
Molecular lines
of CN  (A$^2$$\Pi$-X$^2$$\Sigma$), C$_2$ Swan (A$^3$$\Pi$-X$^3$$\Pi$), TiO
(A$^3$$\Phi$-X$^3$$\Delta$) $\gamma$, TiO (B$^3$$\Pi$-X$^3$$\Delta$)
$\gamma$', TiO $\alpha$ C$^3$$\Delta - $X$^3$$\Delta,$
and TiO $\beta$ c$^1$$\Phi - $a$^1$$\Pi$
systems  are taken   into  account, as described in Barbuy et al. (2018c).

\begin{figure*}
\centering
\includegraphics[width=\hsize]{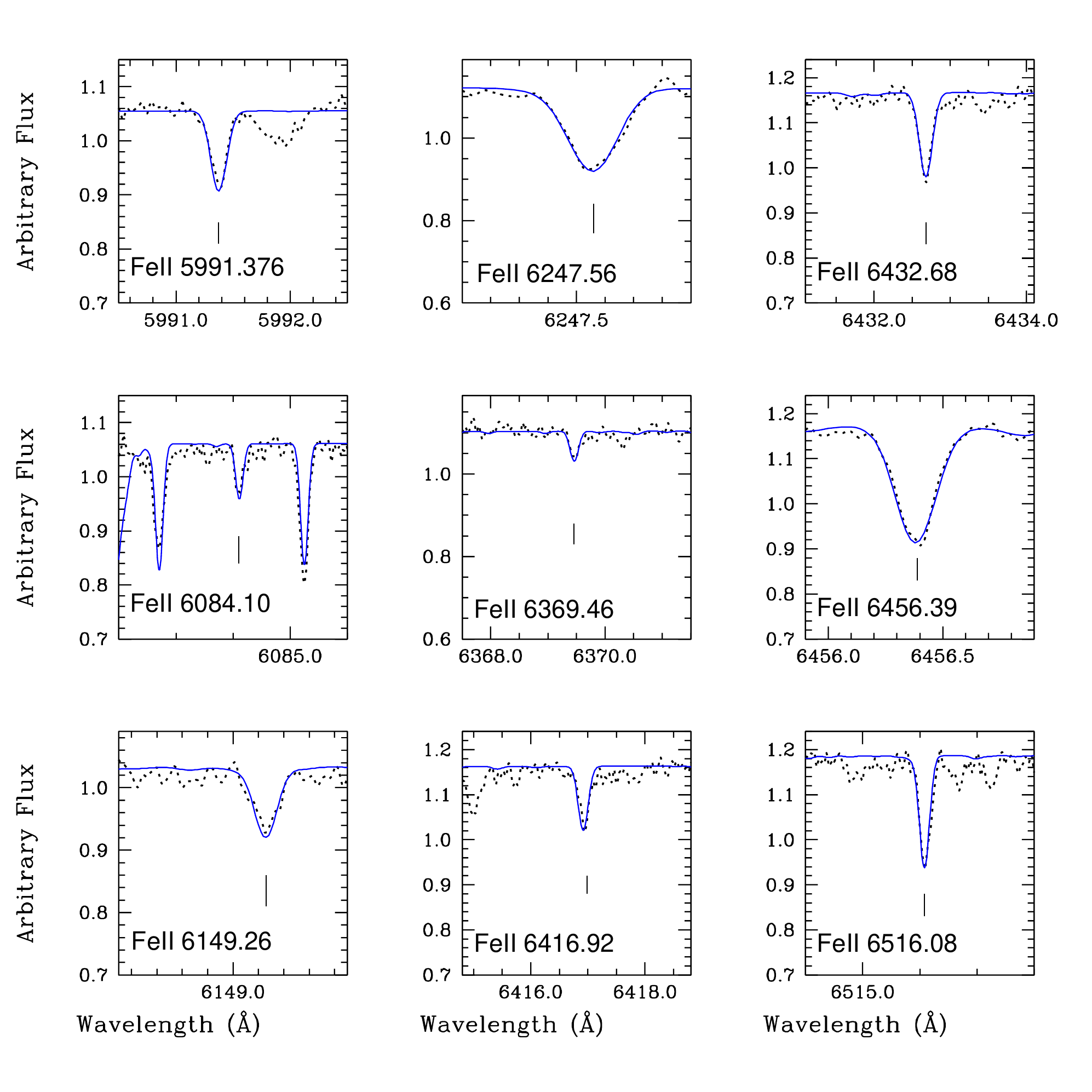}
\caption{\ion{Fe}{II} lines in star 256289.}
\label{feii} 
\end{figure*}

\begin{figure}
\centering
\includegraphics[width=9cm]{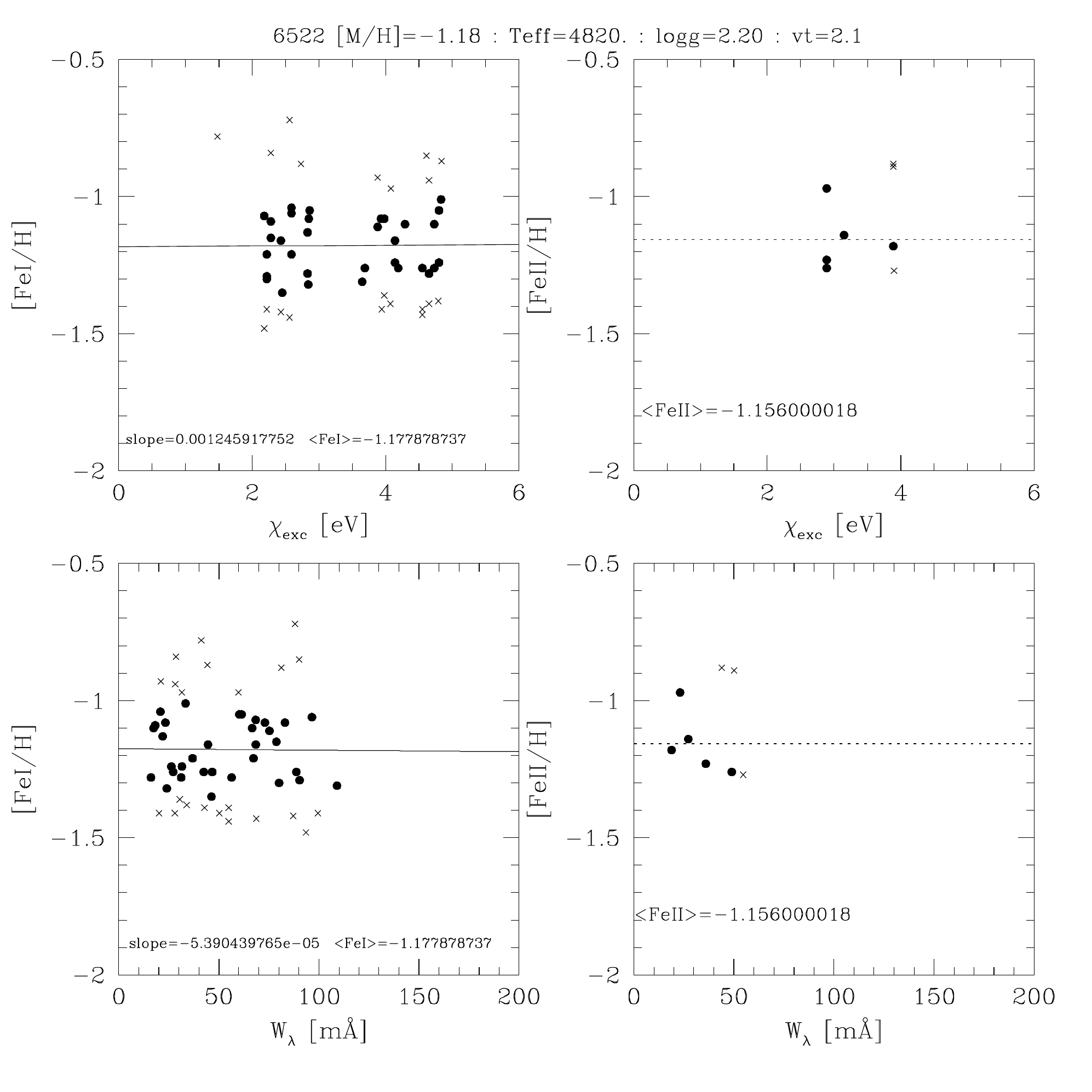}
\caption{Excitation and ionisation equilibrium in 
star B118.}
\label{abonB118final} 
\end{figure}

We initially  adopted  the
photometric effective  temperature  and  gravity,  and   then  further
constrained  the  temperature by imposing  an excitation equilibrium for
\ion{Fe}{I} lines and gravities by
imposing ionisation equilibrium from lines of \ion{Fe}{I} and \ion{Fe}{II}. 
Microturbulence velocities v$_t$  (km s$^{-1}$)
were  determined by cancelling the trend of \ion{Fe}{I} abundance versus EW.
Fits to the observed spectra in regions  containing the \ion{Fe}{II} lines
were carried out, as shown in Fig. \ref{feii}
for star 256289. The good match of the \ion{Fe}{II}
lines indicates that these lines correspond to the equivalent
widths measured, that they are not plagued by blends, and that the  stellar parameters are suitable.
The final spectroscopic parameters T$_{\rm eff}$, log g, [\ion{Fe}{I}/H],
 [\ion{Fe}{II}/H],  [Fe/H] and
 v$_t$ values  are reported in  Table~\ref{tabparam}. An example of
excitation and ionisation equilibrium using \ion{Fe}{I} and \ion{Fe}{II} lines
is shown in Fig. \ref{abonB118final} for star B118.

\section{Abundance ratios}

Abundances ratios  were   obtained  by means of line-by-line  spectrum
synthesis calculations compared to the observed spectra.
The abundance derivation details are explained below,
and the results are reported in Table \ref{c2cn} for C, N, and O; and
Table \ref{abundancias} for the $\alpha$-elements
O, Mg, Si, Ca, Ti, odd-Z elements Na, Al, 
and  heavy elements  Y, Zr, Ba, La, Nd, and Eu.

\subsection{Carbon, nitrogen, and oxygen}

\begin{table}
\begin{flushleft}
\caption{Carbon, nitrogen, and oxygen abundances derived
from C$_2$ (0,1), CN (5,1), and [OI] lines.}
\label{c2cn}      
\centering 
\small         
\begin{tabular}{l@{}r@{}r@{}r@{}r@{}r@{}r@{}rr}     
\noalign{\smallskip}
\hline\hline    
\noalign{\smallskip}
\noalign{\vskip 0.1cm} 
\hbox{line} & \phantom{-}\phantom{-}\phantom{-}\hbox{$\lambda$({\AA})} &
  \phantom{-}\phantom{-}234816 &
\phantom{-}\phantom{-}244523 &\phantom{-}\phantom{-}244819 &
\phantom{-}\phantom{-}256289 &  \phantom{-}\phantom{-}B118  &  \phantom{-}\phantom{-}402370 \\
\noalign{\smallskip}
\noalign{\hrule\vskip 0.1cm}
\hbox{C$_2$(0,1)} & \phantom{-}5635.5   & \phantom{-}$\leq$+0.0  & \phantom{-}$\leq$+0.2
 & \phantom{-}$\leq$+0.2  & \phantom{-}$\leq$+0.2 & \phantom{-}$\sim$+0.2 &\phantom{-}$\leq$+0.0   \\
\hbox{CN(5,1)} & \phantom{-}6332.2       & \phantom{-}$\leq$+0.8  & \phantom{-}$\leq$+0.8
  & \phantom{-}$\leq$+1.2  & \phantom{-}$\leq$+0.3 & \phantom{-}$\leq$+0.8 &\phantom{-}$\leq$+0.3   \\
\hbox{[OI]} & \phantom{-}6300.3        & +0.60  & +0.40  & +0.40 & +0.40 & --- &+0.40  \\
\noalign{\smallskip} \hline \end{tabular}
\end{flushleft}
\end{table}

\begin{figure}
\centering
\includegraphics[width=9cm]{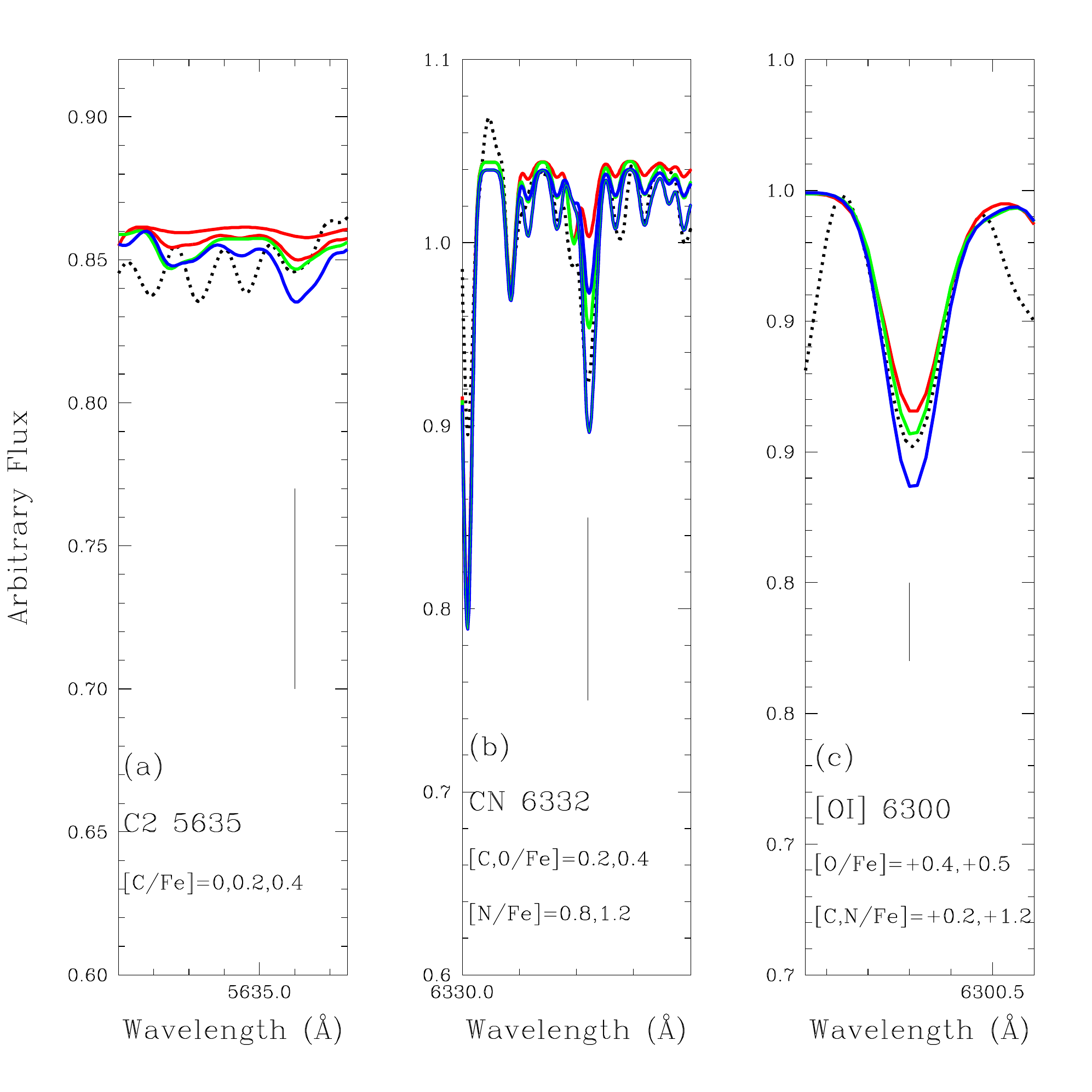}
\caption{C, N, and O lines in star 244819:
Panel (a): C$_2$ 5635 {\rm \AA} computed with ([C/Fe], [N/Fe], [O/Fe]) =
(0.0, 0.8, 0.4), (0.2, 0.8, 0.4) in red,  (0.2, 1.2, 0.4) in green,
(0.4, 1.2, 0.4) in blue. Panel (b): CN 6332 {\rm \AA} computed with:
(0.2, 0.8, 0.4) in red, (0.2, 1.2, 0.4) in green, (0.4, 0.8, 0.4) and
(0.4, 1.2, 0.4) in blue. Panel (c): [OI] 6300 {\rm \AA} computed with
(0.2, 0.8, 0.4) in red, (0.2, 1.2, 0.4) in green, (0.2, 1.2, 0.5) in blue.
In all cases, black dotted lines are the observed spectra.}
\label{cno} 
\end{figure}

Table \ref{c2cn} gives the results for C, N, and O abundances.
The carbon abundances were estimated from
the molecular C$_{2}$(0,1) Swan bandhead
at 5635.5 {\rm \AA}. These bandheads  are
faint, and in these stars they allow us to give an upper limit only.
The atomic \ion{C}{I} 5380.3 {\rm \AA} lines are essentially absent in
these stars and cannot be used.
The nitrogen abundance is derived from the
CN(5,1) red system bandhead at 6332.2 {\rm \AA}.
For the oxygen-forbidden line at [OI] 6300.311 {\rm \AA,}
a selection among the original spectra where telluric lines
did not contaminate the line was needed,
since most of the observations were contaminated.
A few spectra were retrieved showing
a clean [OI] 6300.311 {\rm \AA} line, and the oxygen abundance was derived.
 Figure \ref{cno} shows fits to C$_2$, CN and [OI] lines  for star 244819.

\begin{figure}
\centering
\includegraphics[width=\hsize]{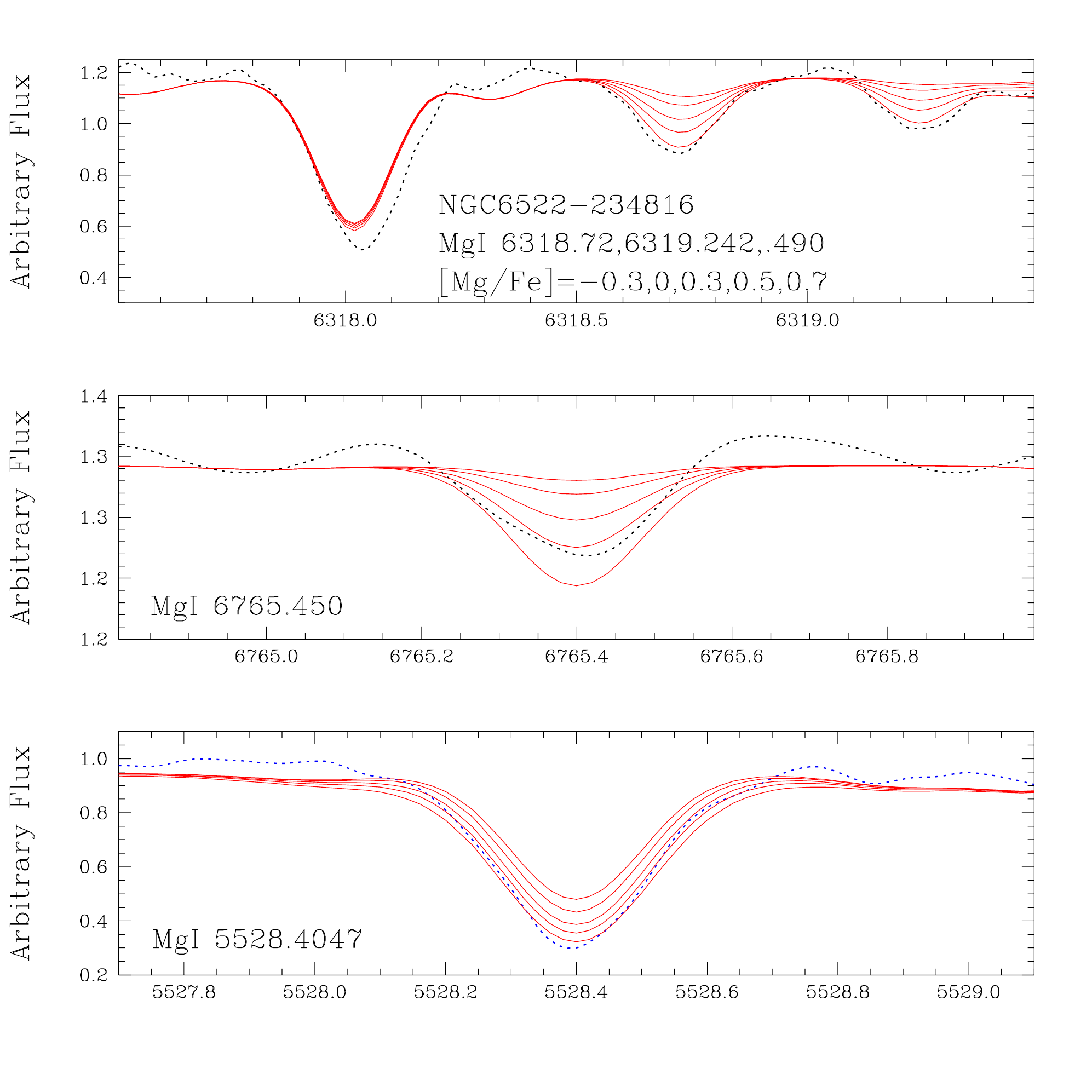}
\caption{Fits to \ion{Mg}{I} lines  in star 234816.}
\label{mg234} 
\end{figure}

\begin{figure}
\centering
\includegraphics[width=\hsize]{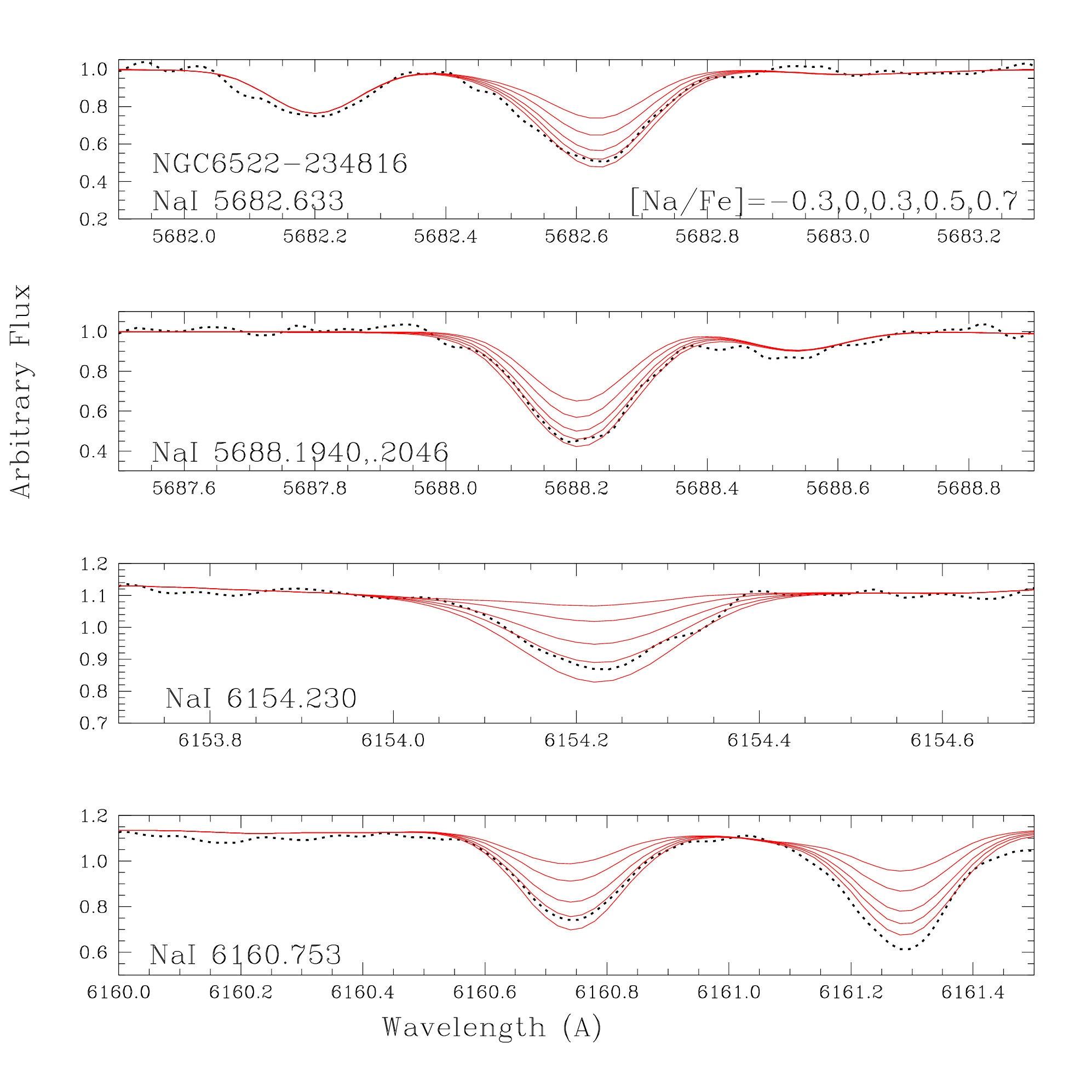}
\caption{Fits to \ion{Na}{I} lines  in star 234816.}
\label{na234} 
\end{figure}

\subsection{Odd-Z and alpha elements}

Line-by-line abundances of the odd-Z elements Na and Al
and alpha elements Mg, Si, Ca, and  Ti 
are reported in Table \ref{abundancias}. Ti is
 an iron-peak element, but given its behaviour following
the alpha-elements, it is often considered as an alpha.

For star 234816, there is a clear overenhancement of the alpha elements
as well as of the r element Eu.
Figure \ref{mg234} shows the MgI lines studied showing agreement for
a high enhancement of [Mg/Fe]=+0.7. 
The odd-Z elements Na and Al are also enhanced in this star, 
with [Na/Fe]=[Al/Fe]=+0.5, as illustrated for Na lines in Figure \ref{na234}.

We exhaustively remeasured equivalent widths using other tools
than IRAF, and we redetermined stellar parameters, and even with somewhat
different stellar parameters. Models of (T$_{eff}$, log~g, [Fe/H], v$_{t}$) =
  (4530 K, 2.2, $-$1.04, 1.2 km.s$^{-1}$) and
  (4440 K , 2.02, $-$0.78, 1.11 km.s$^{-1}$) were employed in Cantelli (2019),
and the overenhancement in alpha elements persists.

\subsection{Heavy elements}

Line-by-line abundances of Y, Zr, Ba, La, Nd, and Eu are reported
in Table \ref{abundancias}.  
In Table \ref{mean} the mean abundances are reported, including
results from Barbuy et al. (2014).
Below we describe details on the lines of the heavy elements studied.

{\it Barium, Lanthanum and Europium:}
 The hyperfine structure (HFS) for the studied lines of 
 \ion{Ba}{II} 5853.675, 6141.713 and 6496.897 {\rm \AA}
 and \ion{Eu}{II} 6645.064 {\rm \AA} were taken into account.
For \ion{Ba}{II} 5853.675, we computed the splitting of lines
by employing a code made available by Andrew McWilliam
(McWilliam et al. 2013). 
For \ion{Ba}{II} 6141.713 {\rm \AA} and 
\ion{Ba}{II} 6496.897 {\rm \AA} lines, as well as for \ion{La}{II}
lines, the  HFS structure was reported in
Barbuy et al. (2014),
and for \ion{Eu}{II} 6645.064 {\rm \AA} the HFS was adopted
from Hill et al. (2002). For \ion{Ba}{II} 5853.675 {\rm \AA},
the magnetic dipole A-factor,
and the electric quadrupole B-factor
were adopted from Biehl (1976) and Rutten (1978),
as given in Table \ref{balines1}.
 The nuclear spin is I=1.5 and the 
 isotopic nuclides Ba$^{138}$ and Ba$^{137}$
Ba$^{136}$, Ba$^{135}$ and Ba$^{134}$
contribute with 71.7\% and 11.23\%, 7.85\%, 6.59\%, 2.42\%,
respectively (Asplund et al. 2009). The hyperfine splitting applies
only to the odd-Z nuclides Ba$^{135}$ and Ba$^{137}$.
The line list taking into account hyperfine
structure for \ion{Ba}{II} 5853.675 
 {\rm \AA} is given in Table \ref{hfsBa}.

\begin{figure*}
\centering
\includegraphics[width=\hsize]{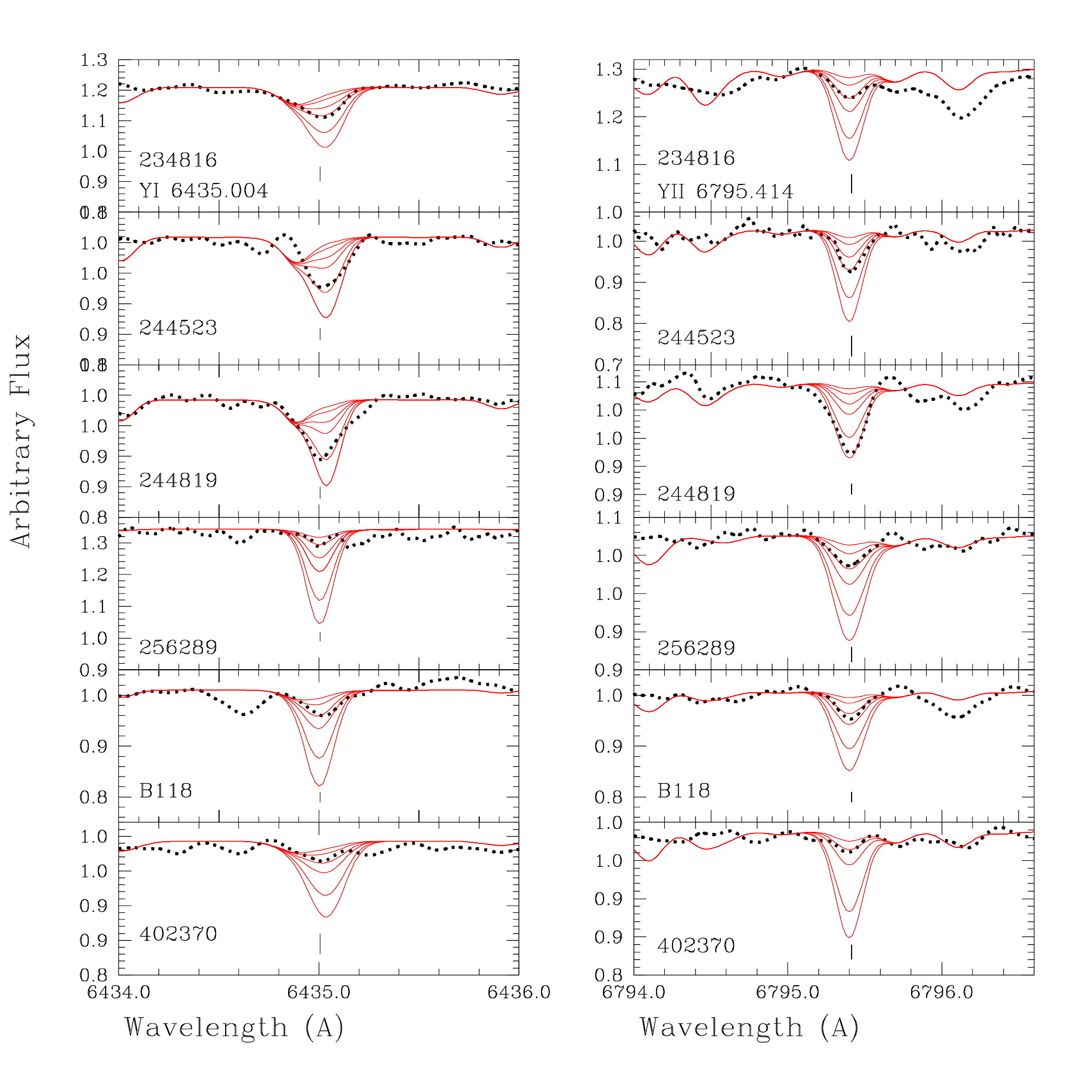}
\caption{Fits to \ion{Y}{I} 6435.004, and \ion{Y}{II}  6795.414 {\rm \AA} in the six sample stars. Observed spectra (black dotted lines) are compared with synthetic spectra (red lines) computed for [Y/Fe]=$-$0.3,0,0.3,0.5,0.8,1.0.}
\label{yall} 
\end{figure*}

\begin{figure*}
\centering
\includegraphics[width=\hsize]{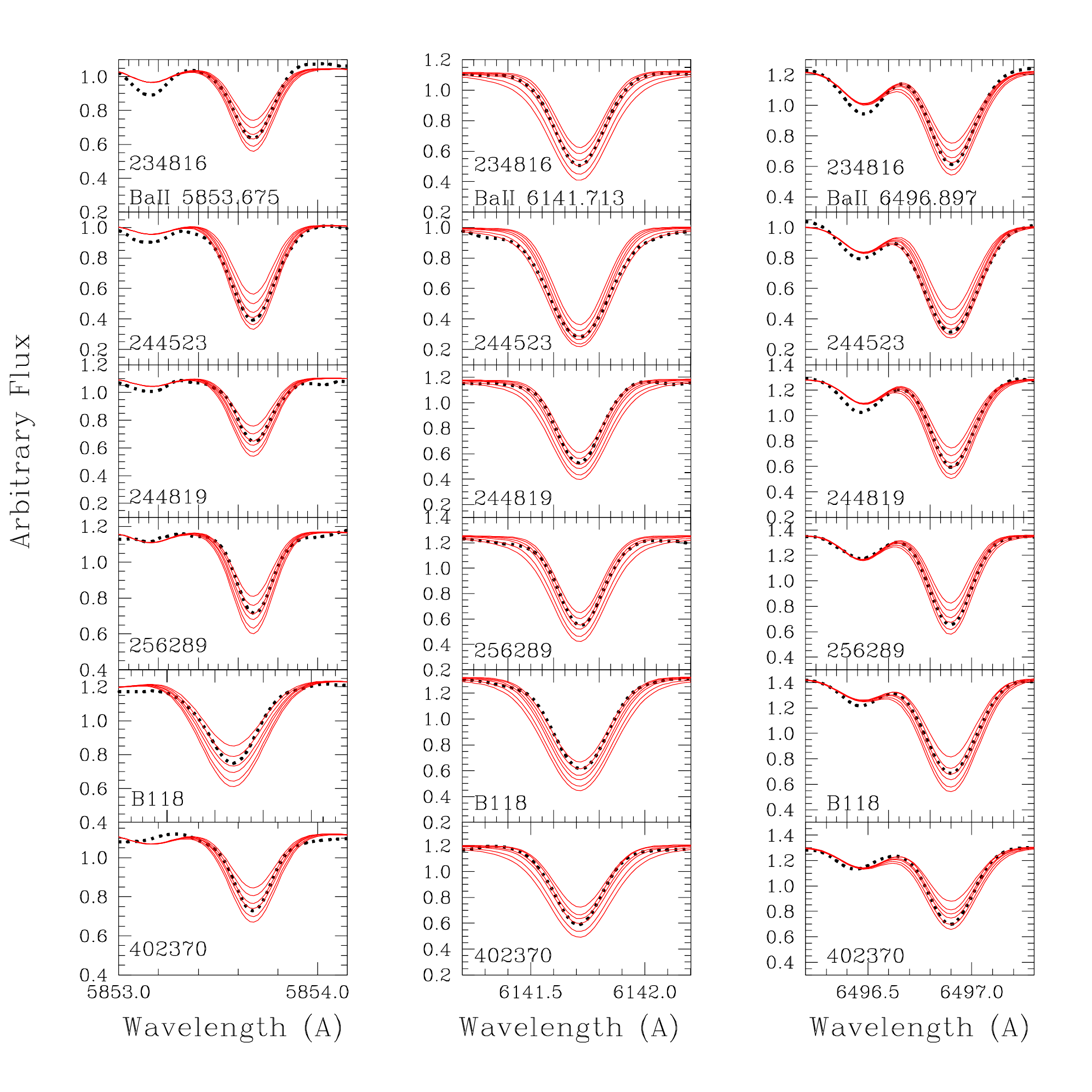}
\caption{Fits to \ion{Ba}{II} 5853.675, 6141.713 and 6496.897 {\rm \AA} in the six sample stars. Observed spectra (black dotted lines) are compared with synthetic spectra (red lines) computed for [Ba/Fe]=$-$0.3,0,0.3,0.5,0.8,1.0.}
\label{ball} 
\end{figure*}

\begin{figure*}
\centering
\includegraphics[width=\hsize]{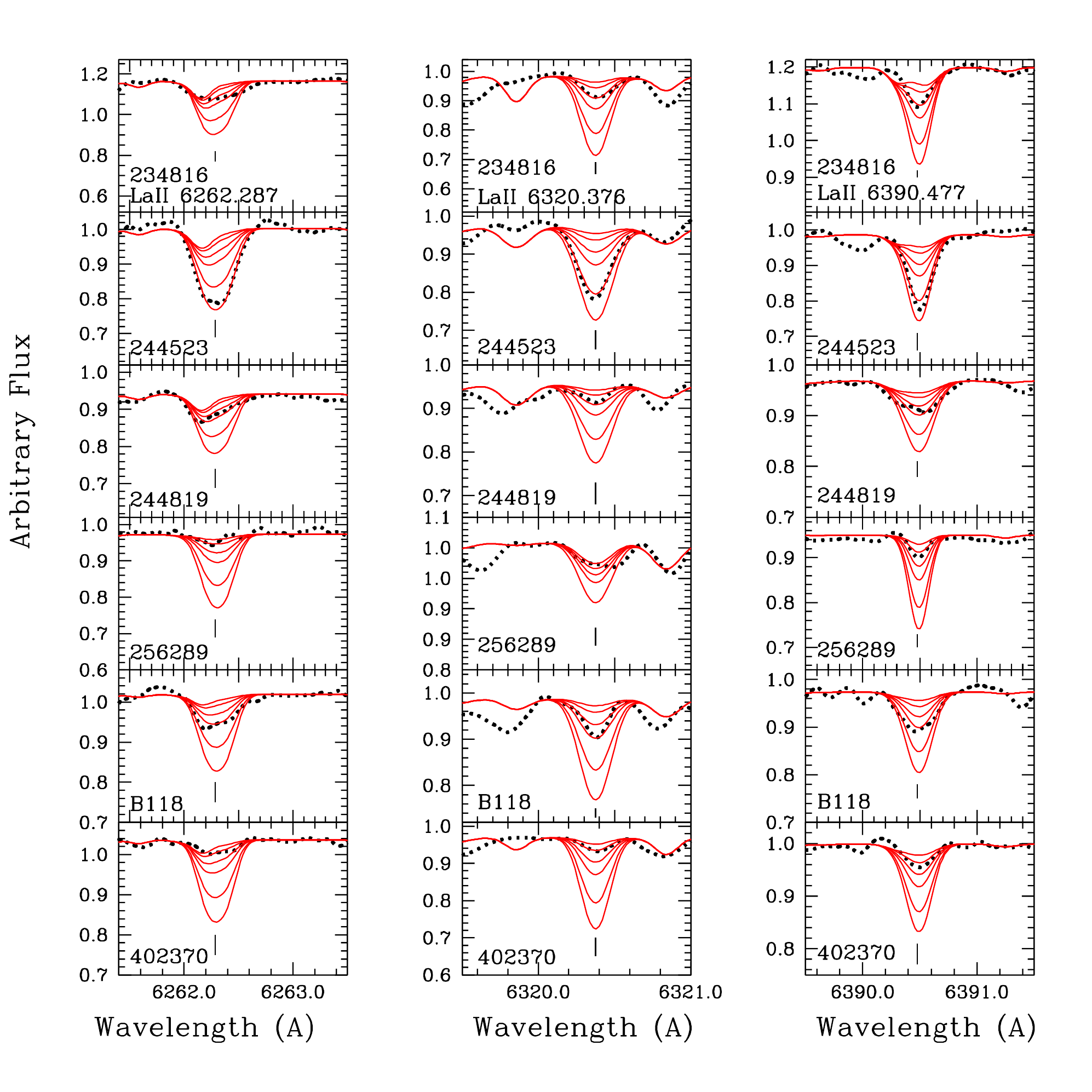}
\caption{Fits to \ion{La}{II} 6262.287, 6320.376, and 6390.477 {\rm \AA} in the six sample stars. Observed spectra (black dotted lines) are compared with synthetic spectra (red lines) computed for [La/Fe]=$-$0.3,0,0.3,0.5,0.8,1.0.}
\label{lall} 
\end{figure*}

\begin{figure*}
\centering
\includegraphics[width=\hsize]{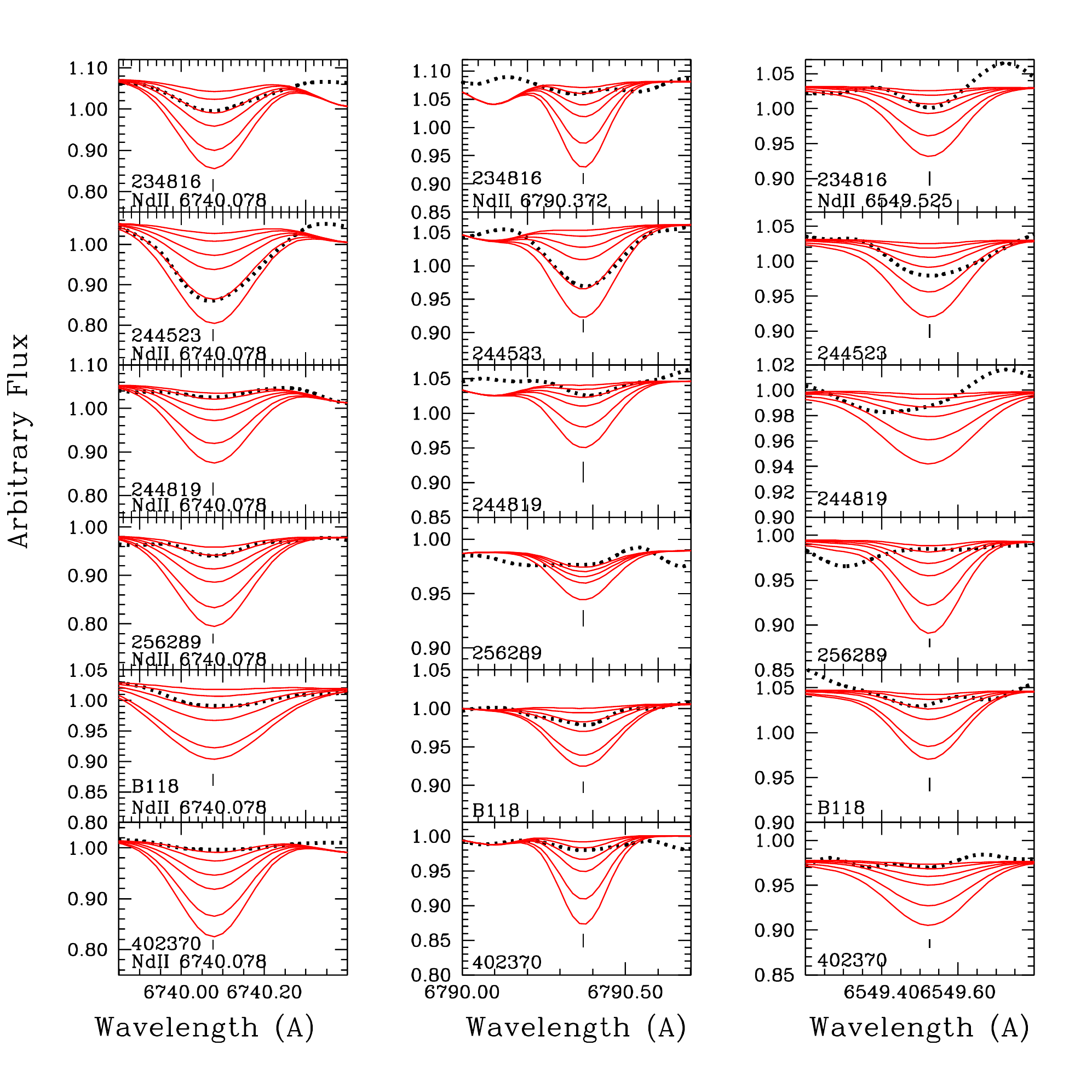}
\caption{Fits to \ion{Nd}{II} 6740.078, 6790.372 and 6549.525 {\rm \AA} in the six sample stars. Observed spectra (black dotted lines) are compared with synthetic spectra (red lines) computed for [Nd/Fe]=$-$0.3,0,0.3,0.5,0.8,1.0.}
\label{ndall} 
\end{figure*}

\begin{figure*}
\centering
\includegraphics[width=\hsize]{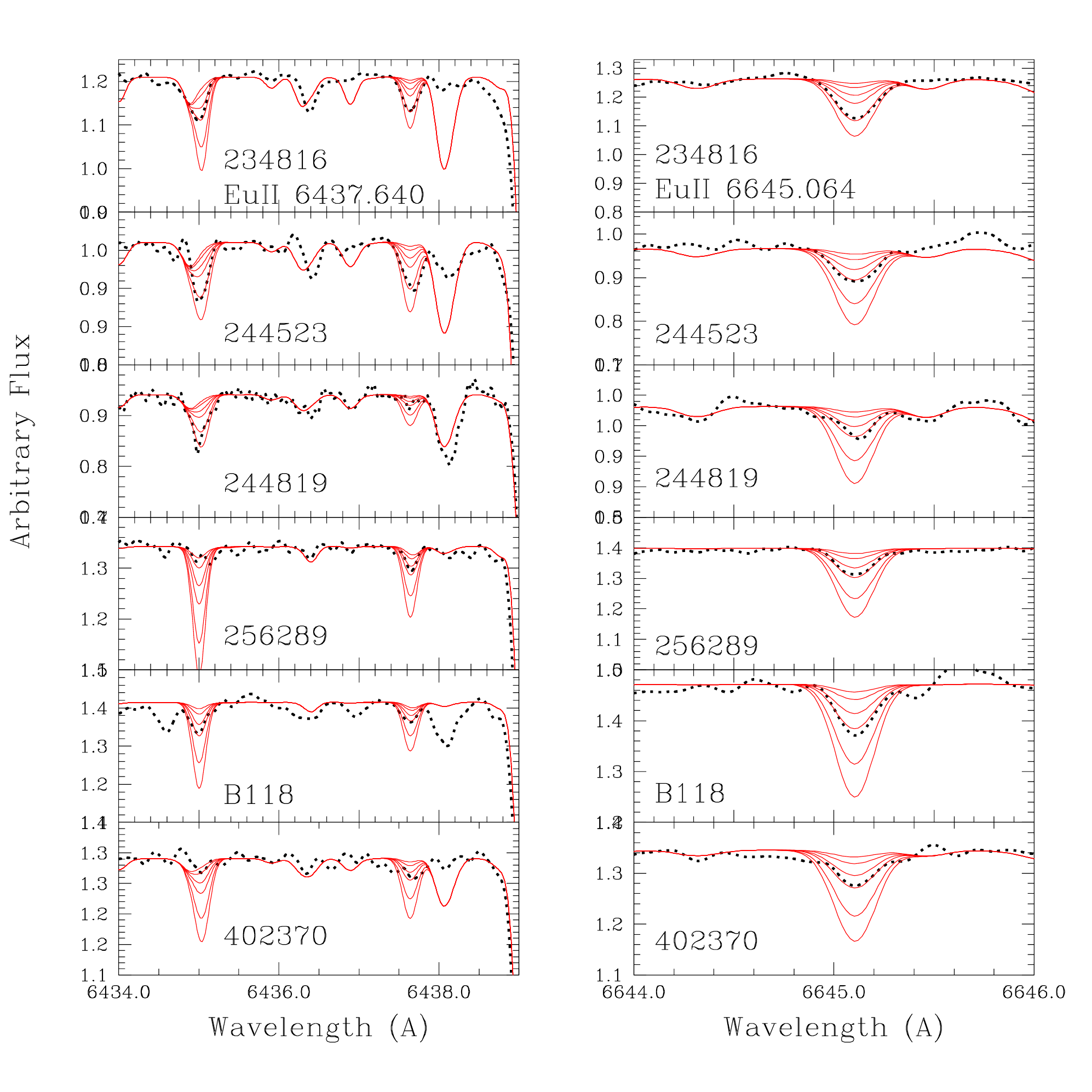}
\caption{Fits to \ion{Eu}{II} 6437.650 and 6645.064 {\rm \AA} in the six sample stars. Observed spectra (black dotted lines) are compared with synthetic spectra (red lines) computed for [Eu/Fe]=$-$0.3,0,0.3,0.5,0.8,1.0.}
\label{euall} 
\end{figure*}

 Figures \ref{yall},  \ref{ball}, \ref{lall},
 \ref{ndall}, and \ref{euall} show, respectively,
 the fits to the  \ion{Y}{I} and  \ion{Y}{II}, 
three \ion{Ba}{II}, \ion{La}{II}, \ion{Nd}{II} lines,
and \ion{Eu}{II} studied lines in the six sample stars.
 
\begin{figure}
\centering
\includegraphics[width=\hsize]{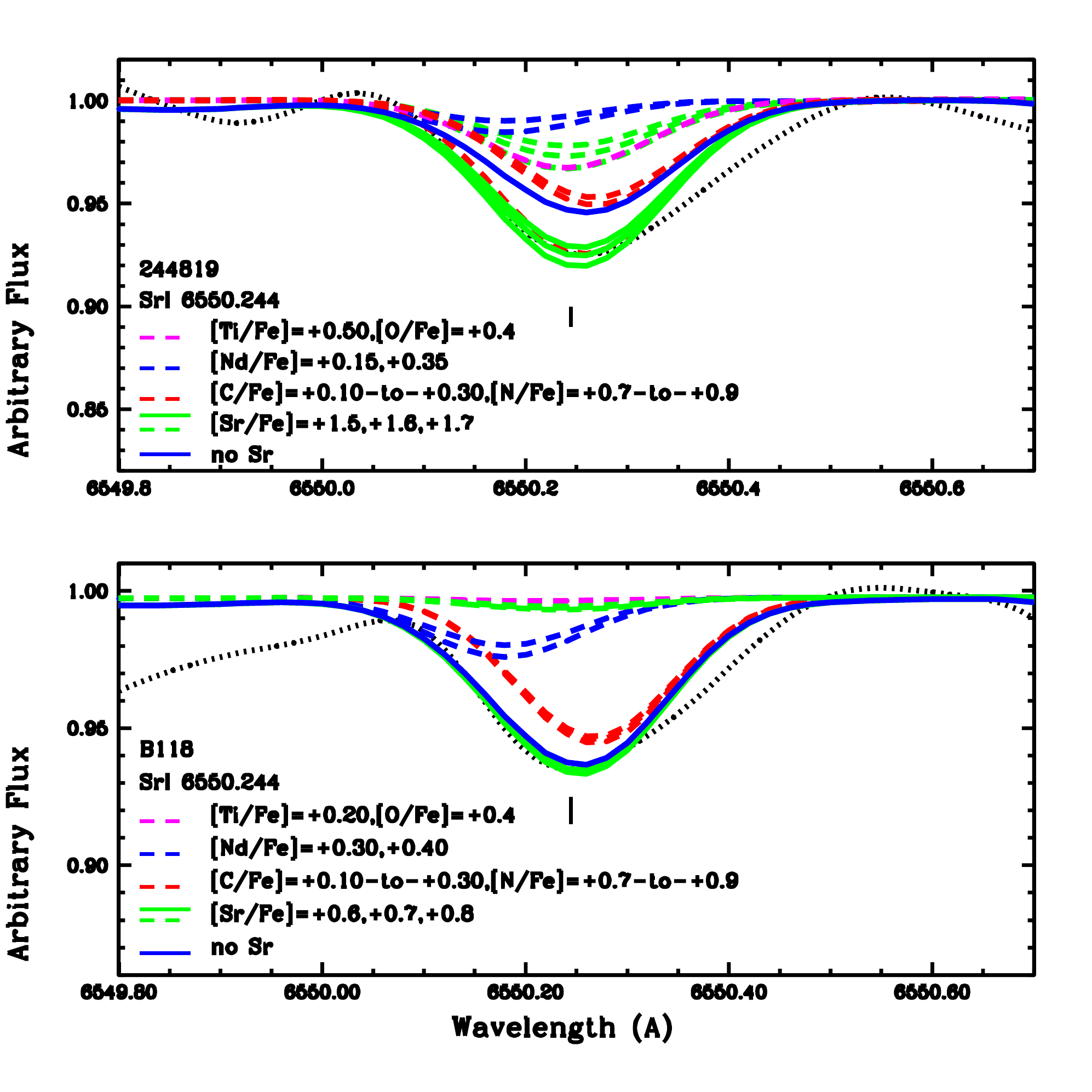}
\caption{ \ion{Sr}{I} 6550.244 {\rm \AA} line in stars 244819 and B118,
    showing the contribution of blends to the line.
    Observed spectra (black dotted lines) are compared to synthetic spectra
    for lines as indicated in the panels.
    Dashed lines correspond to the contribution of a particular element only. 
    Full lines correspond to all lines included.
    The values [Sr/Fe]=+1.6$\pm$0.1 for
    244819 and  [Sr/Fe]=+0.7$\pm$0.1 for B118 are indicated by full green lines. These results are
    not considered reliable, however, due to blends.
}
\label{sr2stars} 
\end{figure}

 {\it Strontium:} 
We carefully inspected the Sr lines and give the conclusions here.
The line \ion{Sr}{I} 6503.989 {\rm \AA} is in the wing of another line
and is very shallow; \ion{Sr}{I}
  6791.016 {\rm \AA} is also very shallow, and only at  higher S/N it
  could be used  (as we did in Barbuy et al. 2014).
We here describe the blends contained in the
  \ion{Sr}{I} 6550.244 {\rm \AA} line in detail.
  In Figure \ref{sr2stars}, we show that:
  a) despite the presence of several TiO lines from
  the $\gamma$, $\gamma$', $\alpha,$ and
  $\beta$ systems, they are faint, given that TiO only gets
  stronger is very cool stars. To take the TiO lines into account,
  we adopted [O/Fe], [Ti/Fe] from Table \ref{mean};
  b) There are lines of
   \ion{Mn}{II}, \ion{Cr}{I}, \ion{Ca}{I}, \ion{Fe}{II}, 
  \ion{Sc}{II}, \ion{Ni}{I}, \ion{Si}{I},  and \ion{Tm}{II}, but all
  of these are extremely faint and do not influence the strength of the
  blend;
 c) The  \ion{Nd}{II} 6550.178 {\rm \AA} line is not very strong,
  but it does contribute to the strength of the blend.
We used three \ion{Nd}{II} lines to derive [Nd/Fe] for the sample stars,
and the fits are presented in Fig. \ref{ndall}.
The resulting mean Nd abundance is then fixed in order to compute
the Sr abundance. In Fig. \ref{sr2stars}, we show the Nd line
for the abundance derived and also for a +0.1 or +0.2dex increase; 
 d) The main contributors to a blend are 
  C$_2$ and CN lines. Although the C and N abundances are upper
  limits, by adopting these values the blend is strong. The computations
  were done for [C/Fe]=0.2 and [N/Fe]=0.8 (that are the upper limits
  given in Table \ref{cno}). Since C and N are anticorrelated we compute
  also with [C/Fe]=0.1 and [N/Fe]=0.9 and [C/Fe]=0.3 and [N/Fe]=0.7:
  the strong variation in the C$_2$ and CN line strengths make the
  derivation of Sr abundance very uncertain.
  The identified lines are
  C$_2$ Swan system (v',v'') = (2,5):
  R3(30) 6550.660, R2(31) 6550.398 and R1(32) 6550.296 and
  CN red system  (v',v'') = (6,2): Q1(22) 6550.269 {\rm \AA};
 e) In conclusion, we estimate values of [Sr/Fe]=+1.6 for
  star 244816 and [Sr/Fe]=+0.7 for star B118 but these cannot be considered
  reliable.

\begin{table*}
\begin{flushleft}
\scalefont{0.8}
\caption{Abundances in the six UVES sample stars. }            
\label{abundancias}      
\scalefont{0.8}  
\begin{tabular}{lccccccccccccccccc}     
\noalign{\smallskip}
\hline\hline    
\noalign{\smallskip}
species & {\rm $\lambda$} ({\rm \AA}) & \phantom{-}\phantom{-}{\rm $\chi_{ex}$ (eV)} & \hbox{log~gf} &
234816 & 244523 & 244819 & 256289 &  B118 & 402370 & \\ 
\noalign{\hrule\vskip 0.1cm}
NaI     & 5682.633 & 2.10  & $-$0.71       &+0.50 &+0.00 &+0.30 &$-$0.40   &+0.30 &$-$0.30 &   \\
NaI     & 5688.194 & 2.10   & $-$1.40      &+0.50 &$-$0.30 &+0.30 &$-$0.30   &+0.30 &$-$0.30 & \\
NaI     & 5688.205 & 2.10   & $-$0.45      &+0.50 &$-$0.30 &+0.30 &$-$0.30   &+0.30 &$-$0.30 & \\
NaI & 6154.230 & 2.10  & $-$1.56               &+0.50 &+0.15 &+0.50 &$-$0.30   &+0.30 &+0.00: & \\
NaI & 6160.753 & 2.10 &  $-$1.26               &+0.50 &+0.25 &+0.30 &$-$0.30   &+0.30 &$-$0.30 & \\     
AlI     & 6696.185 & 4.02 & $-$1.58            &+0.50 &+0.30 &+0.40 &$-$0.30   &+0.60 &$-$0.30 & \\    
AlI     & 6696.788 & 4.02 & $-$1.42            &+0.50 &---   &+0.40 &---     &--- &$-$0.30 & \\     
AlI     & 6696.788 & 4.02 &$-$2.72             &+0.50 &---   &+0.40 &---     &--- &$-$0.30 & \\    
AlI     & 6698.673 & 3.14 & $-$1.65            &+0.50 &+0.30 &+0.40 &$-$0.10   &+0.60 &$-$0.30 & \\ 
MgI & 5528.405 & 4.34 &$-$0.50                 &+0.70 &+0.00 &+0.30 &+0.30   &+0.60 &+0.30 & \\     
MgI & 6318.720 & 5.11 &$-$2.10                 &+0.70 &+0.45 &+0.30 &+0.30   &+0.30 &+0.30 & \\    
MgI & 6319.242 & 5.11 &$-$2.36                 &+0.70 &+0.30 &+0.30 &+0.30   &+0.30 &---   & \\     
MgI & 6319.490 & 5.11 &$-$2.80                 &+0.70 &+0.30 &+0.30 &+0.30   &+0.30 &+0.30 & \\    
MgI & 6765.450 & 5.75  &  $-$1.94              &+0.50 &---   & ---  & ---    &---   &+0.25 & \\   
SiI     & 5665.555 & 4.92  &$-$2.04        &+0.50 &---   &--- &+0.20   &+0.20 &+0.20 & \\
SiI     & 5666.690 & 5.62   & $-$1.74      &+0.50 &---   &--- &---     &+0.30 &+0.30 & \\
SiI     & 5690.425 & 4.93   & $-$1.87      &+0.50 &---   &--- &+0.25   &+0.30 &+0.20 & \\
SiI     & 5948.545 & 5.08   & $-$1.30      &+0.50 &+0.30 &+0.30 &+0.20   &+0.30 &+0.10 & \\ 
SiI & 6142.494 & 5.62   & $-$1.50              &+0.50 &+0.50 &+0.30 &+0.30   &+0.30 &+0.30 & \\  
SiI & 6145.020 & 5.61  & $-$1.45               &+0.50 &+0.35 &+0.30 &+0.25   &+0.40 &+0.30 & \\ 
SiI & 6155.142 & 5.62  & $-$0.85               &+0.50 &+0.35 &+0.30 &+0.00   &+0.30 &+0.15 & \\  
SiI & 6237.328 & 5.61   &  $-$1.01             &+0.50 &+0.40 &--- &+0.15   &+0.30 &+0.05 & \\ 
SiI & 6243.823 & 5.61   & $-$1.30              &+0.50 &+0.40 &+0.20 &+0.15   &+0.25 &+0.30 & \\  
SiI & 6414.987 & 5.87    & $-$1.13             &+0.50 &+0.45 &--- &+0.25   &+0.30 &+0.30 & \\ 
SiI & 6721.844 & 5.86  & $-$1.17               &+0.50 &+0.45 &--- &+0.15   &+0.55 &+0.30 & \\  
                                            
CaI     & 5601.277 & 2.53 & $-$0.52        &+0.50 &$-$0.30 &$-$0.30 &$-$0.15   &+0.00 &+0.12 & \\
CaI     & 5867.562 & 2.93 & $-$1.55        &+0.50 &+0.20 &+0.15 &+0.10   &$-$0.05 &$-$0.10 & \\
CaI     & 6102.723 & 1.88 & $-$0.79        &+0.30 &$-$0.30 &+0.00 &+0.30   &+0.00 &+0.00 & \\
CaI     & 6122.217 & 1.89 & $-$0.20        &+0.30 &$-$0.30 &+0.00 &+0.00   &+0.00 &+0.00 & \\
CaI & 6156.030 & 2.52   &$-$2.39               &+0.40 &+0.30 &---   &+0.05   &+0.00 &+0.25 &   \\    
CaI & 6161.295 & 2.51   & $-$1.02              &+0.50 &+0.30 &+0.30 &+0.30   &+0.30 &+0.30 & \\     
CaI & 6162.167 & 1.89   & $-$0.09              &+0.40 &$-$0.30 &+0.00 &+0.40   &+0.15 &+0.30 & \\     
CaI & 6166.440 & 2.52   & $-$0.90              &+0.50 &+0.15 &+0.15 &+0.00   &+0.10 &+0.00 & \\     
CaI & 6169.044 & 2.52   & $-$0.54              &+0.50 &+0.00 &+0.30 &+0.30   &+0.25 &+0.00 & \\    
CaI & 6169.564 & 2.52   & $-$0.27              &+0.50 &$-$0.20 &+0.15 &+0.30   &+0.00 &+0.00 & \\     
CaI & 6439.080 & 2.52   & +0.3                 &+0.50 &$-$0.30 &+0.00 &+0.40   &$-$0.20 &+0.30 & \\    
CaI & 6455.605 & 2.52    & $-$1.35             &+0.60 &+0.30 &---  &+0.30   &+0.40 &+0.15 &   \\    
CaI & 6464.679 & 2.52     &$-$2.10             &+0.00 & ---  &---   &+0.00   &---   &--- & \\     
CaI & 6493.788 & 2.52     & -2.44             &+0.30 &+0.00 &+0.30 &+0.30   &$-$0.20 &+0.10 & \\     
CaI & 6499.654 & 2.52     & $-$0.85           &+0.50 &+0.10 &+0.30 &+0.25   &+0.00 &+0.10 & \\    
CaI & 6572.779 & 0.00     & $-$4.32           &--- &+0.30 &+0.30 &+0.00   &+0.30 &+0.00 & \\     
CaI & 6717.687 & 2.71     & $-$0.61           &--- &+0.30 &+0.30 &...     &+0.50 &+0.50 & \\    
TiI     & 5689.459 & 2.29  & $-$0.44      &+0.50      &+0.00 &+0.30 &---     &+0.30 &+0.00 & \\
TiI     & 5866.449 & 1.07  & $-$0.84      &+0.50 &+0.15 &+0.30 &+0.00   &+0.10 &+0.00 & \\
TiI     & 5922.108 & 1.05  & $-$1.46      &+0.50 &+0.20 &+0.15 &+0.00   &+0.30 &+0.00 & \\
TiI     & 5941.750 & 1.05  & $-$1.5       &+0.60 &+0.30 &+0.30 &+0.00   &+0.30 &+0.00 & \\
TiI     & 5965.825 & 1.88  & $-$0.42      &+0.50 &+0.30 &+0.10 &+0.00   &+0.30 &+0.10 & \\
TiI     & 5978.539 & 1.87  & $-$0.53      &+0.50 &+0.35 &+0.25 &+0.00   &+0.30 &+0.00 & \\
TiI     & 6064.623 & 1.05  & $-$1.94      &+0.60 &+0.30 &+0.10 &---     &---   &+0.00 & \\
TiI     & 6091.169 & 2.27  & $-$0.42      &+0.60 &+0.30 &+0.10 &+0.00   &+0.20 &+0.00 & \\  
TiI & 6126.214 & 1.07   & $-$1.43             &+0.55 &+0.30 &+0.15 &+0.10   &+0.30 &+0.00 & \\  
TiI & 6258.110 & 1.44  & $-$0.36              &+0.30 &+0.05 &$-$0.30 &$-$0.15   &$-$0.30 &+0.15 & \\      
TiI & 6261.106 & 1.43  & $-$0.48              &+0.60 &+0.20 &+0.05 &$-$0.10   &+0.00 &+0.00 & \\       
TiI & 6266.010 & 1.75  & $-$2.98              & ---  &---   &--- & ---    &---   &--- & \\       
TiI & 6303.767 & 1.44  & $-$1.57              &+0.50 &+0.30 &+0.25 &+0.00   &+0.30 &+0.00 & \\      
TiI & 6312.240 & 1.46  & $-$1.60              &+0.60 &+0.30 &+0.30 &+0.10   &+0.30 &+0.00 & \\      
TiI & 6336.113 & 1.44  & $-$1.74              &+0.55 &---   &+0.30 &---     &+0.20 &$-$0.15 & \\      
TiI & 6508.150 & 1.43   & $-$2.05             & ---  &---   &--- &---     &---   &--- & \\       
TiI & 6554.238 & 1.44   & $-$1.22             &+0.35 &+0.30 &+0.15 &---     &+0.40 &$-$0.30 & \\       
TiI & 6556.077 & 1.46   & $-$1.07             &+0.50  &+0.30 &+0.30 &+0.00   &+0.30 &+0.00 & \\ 
TiI & 6599.113 & 0.90   &$-$2.09              &+0.60 &+0.40 &+0.30 &+0.00   &+0.35 &+0.00 & \\ 
TiI & 6743.127 & 0.90   & $-$1.73             &+0.50 &+0.30 &+0.30 &+0.00   &+0.30 &$-$0.10 & \\ 
TiII & 5336.771 & 1.58 & $-$1.70          &+0.50 &+0.10 &+0.30 &+0.00   &---   &+0.30 & \\  
TiII & 5381.0212 & 1.57 &$-$2.08          &+0.50 &+0.10 &+0.30 &+0.00   &$-$0.30 &+0.00 & \\ 
TiII & 5418.751 & 1.58  &$-$2.13          &+0.50 &+0.30 &+0.50 &+0.20   &$-$0.20 &+0.10 & \\ 
TiII& 6491.580 & 2.06     &$-$2.10            &+0.35 &+0.30 &+0.30 &+0.30   &+0.40 &+0.15 & \\ 
TiII& 6559.576 & 2.05    &$-$2.35             &+0.35 &+0.30 &+0.30 &+0.30   &+0.30 &+0.30 & \\  
TiII& 6606.970 & 2.06     &$-$2.85            &+0.50 &+0.30 &--- &+0.20   &+0.25 &+0.15 & \\  
\ion{Y}{I} & 6435.004     & 0.07  & $-$0.82       &+0.50 &+0.80 &+0.80 &$-$0.15   &+0.30 &+0.30 & \\    
\ion{Y}{II} & 6795.414    & 1.74 & $-$1.19        &+0.30 &+0.50 &+1.00:&+0.30   &+0.40 &+0.00 & \\
\ion{Zr}{I}  & 6127.475 & 0.15  & $-$1.06 &+0.50 &---   &--- &+0.15   &+0.50:&--- & \\
\ion{Zr}{I}  & 6134.585 & 0.00      &  $-$1.426 &+0.65 &---   &--- &+0.10   &+0.50:&--- & \\ 
\ion{Zr}{I} & 6140.535 & 0.52      &  $-$1.6  &---  &---   &--- &+0.10   &---   &--- & \\ 
\ion{Zr}{I} & 6143.252 & 0.07   & $-$1.1& +0.65 &---   &--- &+0.10   &+0.50 &--- & \\
\ion{Ba}{II} & 5853.675 & 0.60 & $-$1.1       &+0.50 &+0.65 &+0.30 &+0.30   &+0.15 &+0.60 & \\
\ion{Ba}{II} & 6141.713 & 0.70 & $-$0.08      &+0.50 &+0.60 &+0.30 &+0.30   &+0.00 &+0.60 & \\
\ion{Ba}{II} & 6496.897 & 0.60 & $-$0.32      &+0.60 &+0.65 &+0.50 &+0.50   &+0.30 &+0.80 & \\    
\ion{La}{II} & 6262.287  & 0.40 & $-$1.60        &+0.40 & +1.00  &+0.40 &+0.00   &+0.60 &+0.00 & \\ 
\ion{La}{II} & 6320.376  & 0.17 & $-$1.56        &+0.30 &+0.80 &+0.30 &+0.00   &+0.50  &+0.00 & \\ 
\ion{La}{II} & 6390.477  & 0.32 & $-$1.41        &+0.40 &+0.90 &+0.40 &+0.00   &+0.60&+0.10 & \\ 
\ion{Nd}{II} & 6549.525  & 0.06 & $-$2.01        &+0.30 &+0.60 &---   &+0.00   &+0.30 &+0.00 & \\ 
\ion{Nd}{II} & 6740.078  & 0.06 & $-$1.53        &+0.00 &+0.80 &+0.30 &+0.00   &+0.30 &+0.00 & \\ 
\ion{Nd}{II} & 6790.372  & 0.18 & $-$1.77        &+0.30 &+0.80 &+0.00 &+0.00   &+0.30 &-0.30 & \\ 
\ion{Eu}{II}& 6173.029 & 1.32 & $-$0.86          &+0.80: &--- &--- &---   &--- &--- & \\
\ion{Eu}{II}& 6437.640 & 1.32 & $-$0.32           &+0.80 &+0.80 &+0.50   &+0.40   &+0.50 &+0.50: & \\
\ion{Eu}{II}& 6645.064 & 1.38 & +0.12             &+0.80 &+0.50 &+0.50 &+0.45   &+0.60 &+0.50 & \\ 
\noalign{\hrule\vskip 0.1cm} 
\hline                  
\end{tabular}
\end{flushleft}
\end{table*}

\begin{table*}
\caption[1]{Mean abundances of C, N, odd-Z elements Na, Al,
  $\alpha$-elements O, Mg, Si, Ca, Ti, and heavy elements Y, Zr, Ba, La, and Eu.
  The four stars analysed in Barbuy et al. (2014) are also included. For the mean values,
only the eight member stars were considered.}
\label{mean}
\scalefont{0.7}
\begin{flushleft}
\tabcolsep 0.15cm
\begin{tabular}{cccccccccccccccccc}
\noalign{\smallskip}
\hline
\noalign{\smallskip}
\hline
\noalign{\smallskip}
        {\rm star} & [C/Fe] & [N/Fe] & [Na/Fe] & [Al/Fe] & [O/Fe] & [Mg/Fe] & [Si/Fe] & [Ca/Fe] & [TiI/Fe] &[TiII/Fe] &
        [Y/Fe] & [Zr/Fe] & [Ba/Fe] & [La/Fe] & [Nd/Fe] & [Eu/Fe] \\
\noalign{\vskip 0.2cm}
\noalign{\hrule\vskip 0.2cm}
\noalign{\smallskip}
\hline
\noalign{\smallskip}
\multicolumn{15}{c}{The two non-member stars} \\
\noalign{\smallskip}
\hline
\noalign{\smallskip}
234816 &$<$0.2 &+0.80 &+0.50&+0.50&+0.60 &+0.66&+0.50&+0.42&+0.52&+0.45      &+0.40&+0.60&+0.53 &+0.36 &+0.20 &+0.80 \\
244523 &$<$0.2      &$<$0.8 &-0.04&+0.30&+0.40&+0.26&+0.40&+0.02&+0.26&+0.18 &+0.65&---  &+0.63 &+0.90 &+0.70 &+0.65 \\
\noalign{\smallskip}
\hline
\noalign{\smallskip}
\multicolumn{15}{c}{Four stars from the present work} \\
\noalign{\smallskip}
\hline
\noalign{\smallskip}
244819 &$<$0.2      &$<$1.2 &+0.34&+0.40&+0.40&+0.30&+0.28&+0.14&+0.19&+0.34 &+0.90&---   &+0.36 &+0.36 & +0.15 &+0.50 \\
256289 &$<$0.2      &$<$0.3 &$-$0.32&$-$0.20&+0.40&+0.30&+0.19&+0.18&+0.00&+0.17 &+0.08&+0.11 &+0.36 &+0.00 &+0.00 &+0.43 \\
B118& $\simless$0.2 &$<$0.8 &+0.30&+0.60&---  &+0.40&+0.32&+0.10&+0.23&+0.09 &+0.35&+0.50 &+0.15 &+0.57& +0.30 &+0.55 \\
402370 &$<$0.0      &$<$0.3 &$-$0.24&$-$0.30&+0.40&+0.29&+0.23&+0.13&$-$0.02&+0.16 &+0.15&---   &+0.66 &+0.03& -0.10 &+0.50 \\
\noalign{\smallskip}
\hline
\noalign{\smallskip}
\multicolumn{15}{c}{Four stars from Barbuy et al. (2014)} \\
\noalign{\smallskip}
\hline
\noalign{\smallskip}
B-107 & +0.00   & ---   & +0.03 & +0.28 & +0.50 & +0.33 & +0.17 & +0.16 & +0.03 & +0.17  & +0.32 & +0.20 & +0.45 & +0.20&--- & +0.40 \\
B-122 & $-$0.20 & +0.70 & +0.09 & +0.18 & +0.20 & +0.10 & +0.06 & +0.00 & +0.03  & +0.15 & +0.20 & +0.10 & +0.05 & +0.35&--- & +0.30 \\
B-128 & +0.10   & +0.60 & +0.01 & +0.08 & +0.23 & +0.23 & +0.14 & +0.20 & +0.05  & +0.17 & +0.43 & +0.40 & +0.55 & +0.35&--- & +0.30 \\
B-130 & +0.00   & +0.70 & +0.05 & +0.26 & +0.50 & +0.27 & +0.13 & +0.15 & +0.03  & +0.18 & +0.23 & +0.00 & +0.22 & +0.00&--- & +0.20 \\
\noalign{\vskip 0.2cm}
\noalign{\hrule\vskip 0.2cm}
\noalign{\vskip 0.2cm}
Mean & $\sim$+0.14   & +0.66 & +0.09 &  +0.24 & +0.38 & +0.28  & +0.19 & +0.13 & +0.07  & +0.18 & +0.33 & +0.23 & +0.35 & +0.23 &+0.09 & +0.40 \\
\noalign{\smallskip} \hline \end{tabular}
\end{flushleft}
\end{table*}

\begin{table}
\begin{flushleft}
  \caption{Heavy element abundance ratios.}
\label{ratios}      
\centering 
\small         
\begin{tabular}{lrrrrrrrrr}     
\noalign{\smallskip}
\hline\hline    
\noalign{\smallskip}
\noalign{\vskip 0.1cm} 
star & [Ba/La]  &  [Ba/Eu]& [La/Eu] & [Y/Ba] & [Y/La] &  \\
\noalign{\smallskip}
\noalign{\hrule\vskip 0.1cm}
234816 & +0.17 & $-$0.27 & $-$0.44 & $-$0.13 & +0.04 &  \\
244523 & $-$0.27 & $-$0.02 & +0.25 & +0.02 & $-$0.25 &  \\
\hline
244819 & +0.00 & $-$0.14 & $-$0.14 & +0.54 & +0.54   &  \\
256289 & +0.36 & $-$0.07 & $-$0.43 & $-$0.28 & +0.08 &  \\
B118   & $-$0.42 & $-$0.40 & +0.02 & +0.20 & $-$0.22 &  \\
402370 & +0.63 & +0.16 & $-$0.47 & -0.51 & +0.12 &  \\
\hline
B-107  & +0.25 & +0.05 & $-$0.20 & $-$0.13 & +0.12 &  \\
B-122  & $-$0.30 & $-$0.25 & +0.05 & +0.15 & $-$0.15 &  \\
B-128  & +0.20 & +0.25 & +0.05 & $-$0.12 & +0.08 &  \\
B-130  & +0.22 & +0.02 & $-$0.20 & +0.01 & +0.23 &  \\
\noalign{\smallskip} \hline \end{tabular}
\end{flushleft}
\end{table}

\subsection{Errors}

 Uncertainties in spectroscopic parameters are given 
 in Table \ref{errors} for star NGC 6522: 402370.
 For each stellar parameter,
 we adopted the usual uncertainties as for similar samples
 (Barbuy et al. 2014, 2016, 2018b):
$\pm$100 K in effective temperature, $\pm$0.2 on gravity,
and $\pm$0.2 km~s$^{-1}$ on the microturbulence velocity.
Errors were computed by employing models with these modified parameters,
 with changes of  $\Delta$T${\rm
eff}$=+100 K, $\Delta$log  g  =+0.2,  $\Delta$v$_{\rm  t}$ =  0.2  km~s$^{-1}$,
and recomputing lines of different elements.
The error given is the abundance difference needed
to reach the adopted abundances.
Uncertainties due to non-LTE effects
are negligible for these stellar parameters
as discussed in Ernandes et al. (2018). 
The same error analysis and estimations can be applied to other stars in our sample.
A more careful discussion is required for Ba. The heavy element abundances for star
B118 reported in Table \ref{bab118}
show that the abundance ratios are confirmed from one work to another,
except for Ba. This is due to the use of strong lines that fall
in the saturated part of the curve of growth, where the abundance
is a function of the square root of the number of atoms; the bottom of the lines
reaches a maximum, and the increase of abundance causes an increase in the
line wings. Therefore, abundance derivations from strong lines are in general
avoided, since they are too sensitive to stellar parameters and
spectral resolution.
The La lines are, on the other hand, faint and they are at least not affected by the same problem. 

Finally, it is important to note that the main uncertainties in stellar
parameters are due to uncertainties in the effective temperature, as can be seen
in Table \ref{tabteff}. The second most important source of error
are the EWs, given the limited S/N of the spectra, which can be estimated using the formula from Cayrel (1988): $\sigma_{EW}$ = 1.5 $\sqrt{FWHM.\delta_{x}}/(S/N)$, where $\delta_{x}$ is the pixel size. The difference in
the mean metallicities between the present work and
Barbuy et al. (2014) are probably due to a difference
in the measurements of EWs, and in particular in the placement of continua.

\begin{table*}
\begin{flushleft}
  \caption{Heavy element results for star B118 from Barbuy et al. (2009, B09),
    Chiappini et al. (2011, C11), Ness et al. (2014, N14) and present work.}
\label{bab118}      
\centering 
\small         
\begin{tabular}{lrrrrrrrrrrrrr}     
\noalign{\smallskip}
\hline\hline    
\noalign{\smallskip}
\noalign{\vskip 0.1cm} 
\hbox{work} & \hbox{T$_{\rm eff}$} & \hbox{log~g} & \hbox{[Fe/H]} &\hbox{v$_{rm t}$} &
 [Y/Fe]  &  [Zr/Fe]& [Sr/Fe] & [Ba/Fe] & [La/Fe] & [Eu/Fe] & \\
\noalign{\smallskip}
\noalign{\hrule\vskip 0.1cm}
\hbox{B09} & 4700 & 2.6  & $-$0.84 & 1.30 & ---   & ---  & ---   & +1.00 & +0.50 & +0.50 & \\
\hbox{C11} & same & same & same  & same & +0.50 & ---  & +1.50 & same  & same  & same  & \\
\hbox{N14} & 5000 & 2.25 & $-$1.04 & 2.45 & +0.30 & ---  & ---   & +0.30 & +0.55 & +0.40 & \\
\hbox{this}& 4820 & 2.20 & $-$1.17 & 2.10 & +0.25 & +0.50& +0.70 & +0.15 & +0.57 & +0.55 & \\
\noalign{\smallskip} \hline \end{tabular}
\end{flushleft}
\end{table*}

{\it Comparison between results from UVES and GIRAFFE spectra:}
For the sample stars, we have the UVES spectra ranging from
4800-5800 {\rm \AA,} with a
gap at   5777-5824 {\rm \AA}, and the GIRAFFE spectra in the setups H11 
(5597-5840) and H12 (5821-6146) only. Therefore, since most lines used for the
stellar parameter analysis are in the UVES red arm, and most lines for
deriving abundances are also in the UVES red arm, we cannot
compare the stellar parameter derivation. Moreover, we only compare
the main lines in common between the two sets of spectra, which are
located at $\lambda$$<$6142 {\rm \AA}.

By comparing the abundances for a list of lines
in common between UVES and GIRAFFE, we give another indicator
of uncertainty. In Table \ref{comp}, we compare the abundances
derived from UVES spectra to those derived from GIRAFFE spectra.
The results show an excellent agreement. In order to be clear,
the stellar parameters are much better derived from UVES spectra,
in particular because of the measurement of the \ion{Fe}{II} lines;
whereas, given a set of stellar parameters,
the abundances from the same lines at the different resolutions
are both reliable.

\begin{table}
\caption{Abundance uncertainties for star N6522:402370,
 for uncertainties of $\Delta$T$_{\rm eff}$ = 100 K,
$\Delta$log g = 0.2, $\Delta$v$_{\rm t}$ = 0.2 km s$^{-1}$ and
corresponding total error. The errors are to be
added to reach the reported abundances. }
\label{errors}
\begin{flushleft}
\small
\tabcolsep 0.15cm
\begin{tabular}{lcccc@{}c@{}}
\noalign{\smallskip}
\hline
\noalign{\smallskip}
\hline
\noalign{\smallskip}
\hbox{Element} & \hbox{$\Delta$T} & \hbox{$\Delta$log $g$} & 
\phantom{-}\hbox{$\Delta$v$_{t}$} & \phantom{-}\hbox{($\sum$x$^{2}$)$^{1/2}$}&
\\
\hbox{} & \hbox{100 K} & \hbox{0.2 dex} & \hbox{0.2 kms$^{-1}$} & & \\
\hbox{(1)} & \hbox{(2)} & \hbox{(3)} & \hbox{(4)} & \hbox{(5)}  &\\
\noalign{\smallskip}
\hline
\noalign{\smallskip}
\noalign{\hrule\vskip 0.1cm}
\hbox{[FeI/H]}       &  $-$0.10        &+0.01       & +0.05 &\phantom{+}0.11 &\\
\hbox{[FeII/H]}      &  +0.10          &  $-$0.07   & +0.04 &\phantom{+}0.13 &\\
\hbox{[C/Fe]}        &  +0.02          &  +0.02   & \phantom{+}0.00        &\phantom{+}0.03  \\
\hbox{[O/Fe]}        &  +0.00          & +0.05      &+0.00  &\phantom{+}0.05 &\\ 
\hbox{[NaI/Fe]}      &  +0.05          & +0.00      &+0.00  &\phantom{+}0.05 &\\ 
\hbox{[AlI/Fe]}       & +0.06          & +0.00      &+0.00  &\phantom{+}0.06 &\\ 
\hbox{[MgI/Fe]}      &  +0.00          & +0.01      &+0.00  &\phantom{+}0.01 &\\ 
\hbox{[SiI/Fe] }     & +0.03           & +0.00      &+0.00  &\phantom{+}0.03 &\\ 
\hbox{[CaI/Fe]}      &  +0.08          & +0.00      &+0.01  &\phantom{+}0.08 &\\ 
\hbox{[TiI/Fe]}      &  +0.12          & +0.01      &+0.00  &\phantom{+}0.12 &\\ 
\hbox{[TiII/Fe]}     & $-$0.05         & +0.07      &+0.00  &\phantom{+}0.09 &\\  
\hbox{[YI/Fe]}       & +0.15           &+0.04      &+0.00  &\phantom{+}0.10  &\\
\hbox{[YII/Fe]}      &  +0.20          & +0.15           & \phantom{+}0.00  &\phantom{+}0.25 \\
\hbox{[ZrI/Fe]}      &  +0.20          & $-$0.01           & \phantom{+}0.00  &\phantom{+}0.20  \\
\hbox{[BaII/Fe]}     &  +0.10          & +0.15           & $-$0.15          &\phantom{+}0.23  \\
\hbox{[LaII/Fe]}     &  +0.05          & +0.15           &\phantom{+}0.00   &\phantom{+}0.16  \\
\hbox{[LaII/Fe]}     &  +0.12          & +0.05           &\phantom{+}0.00   &\phantom{+}0.13  \\
\hbox{[EuII/Fe]}     &  $-$0.05        & +0.05      &+0.00  &\phantom{+}0.07 &  \\
\noalign{\smallskip} 
\hline 
\end{tabular}
\end{flushleft}
 \end{table}

\begin{table*}
\begin{flushleft}
\scalefont{0.8}
\caption{Comparison of abundances from UVES and GIRAFFE. }            
\label{comp}      
\scalefont{0.8}  
\begin{tabular}{lccccccccccccccccc}     
\noalign{\smallskip}
\hline\hline    
\noalign{\smallskip}
species & {$\lambda$ (\AA)} &
234816 &234816 & 244523 &244523&    244819 &244819& 256289 &256289&  B118 &B118& 402370 &402370& \\
 &  & 
UVES &GIRAFFE & UVES &GIRAFFE&  UVES &GIRAFFE& UVES &GIRAFFE& UVES
&GIRAFFE& UVES &GIRAFFE& \\
\noalign{\hrule\vskip 0.1cm}
\ion{Na}{I}     & 5682.633      &+0.50 &+0.50&+0.00 &+0.00&+0.30 &+0.30&-0.40   &$-$0.30&+0.30 &+0.40&$-$0.30 &+0.00&   \\
\ion{Na}{I}     & 5688.194      &+0.50 &+0.50&$-$0.30 &+0.00&+0.30 &+0.30&$-$0.30   &$-$0.30&+0.30 &+0.50&$-$0.30 &+0.00& \\
\ion{Na}{I}     & 5688.205      &+0.50 &+0.50&$-$0.30 &+0.00&+0.30 &+0.30&$-$0.30   &$-$0.30&+0.30 &+0.50&$-$0.30 &+0.00& \\                                                                           
\ion{Si}{I}     & 5665.555      &+0.50 &+0.50&---   &+0.35&---   &+0.35&+0.20  &+0.00&+0.20 &+0.35&+0.20 &+0.00& \\
\ion{Si}{I}     & 5666.690      &+0.50 &---  &---   &---&---     &---&---&+0.00&+0.30 &+0.00&+0.30 &+0.30& \\
\ion{Si}{I}     & 5690.425      &+0.50 &+0.50&---   &+0.50  &---   &+0.35&+0.25   &+0.20&+0.30 &+0.30&+0.20 &---& \\
\ion{Si}{I}     & 5948.545      &+0.50 &+0.00&+0.30 &+0.15&+0.30 &+0.30&+0.20   &+0.00&+0.30 &+0.15&+0.10 &+0.00& \\ 
\ion{Si}{I}     & 6142.494      &+0.50 &+0.50&+0.50 &---&+0.30 &---&+0.30   &---&+0.30 &+0.15&+0.30 &+0.30& \\                                                                 
\ion{Ca}{I}     & 5601.277      &+0.50 &+0.50&$-$0.30 &+0.10&$-$0.30 &+0.30&$-$0.15   &+0.15&+0.00 &+0.30&+0.12 &+0.00& \\
\ion{Ca}{I}     & 5867.562      &+0.50 &+0.50&+0.20 &+0.30&+0.15 &+0.30&+0.10   &+0.10&$-$0.05 &+0.30&$-$0.10 &+0.00& \\
\ion{Ca}{I}     & 6102.723      &+0.30 &+0.30&$-$0.30 &+0.00&+0.00 &+030&+0.30   &+0.30&+0.00 &+0.15&+0.00 &+0.15& \\
\ion{Ca}{I}     & 6122.217      &+0.30 &+0.30&$-$0.30 &+0.00&+0.00 &+0.30&+0.00   &+0.15&+0.00 &+0.00&+0.00 &+0.00& \\
\ion{Ti}{I}     & 5689.459     &+0.50 &+0.50&+0.00 &+0.30&+0.30 &+0.30&---     &---&+0.30 &+0.30&+0.00 &---& \\
\ion{Ti}{I}     & 5866.449     &+0.50 &+0.50&+0.15 &+0.20&+0.30 &+0.30&+0.00   &+0.00&+0.10 &+0.15&+0.00 &+0.00& \\                                                                     
\ion{Ti}{I}     & 5922.108     &+0.50 &+0.50&+0.20 &+0.45&+0.15 &+0.30&+0.00   &+0.00&+0.30 &+0.30&+0.00 &+0.00& \\
\ion{Ti}{I}     & 5941.750     &+0.60 &+0.50&+0.30 &+0.50&+0.30 &+0.30&+0.00   &+0.20&+0.30 &---&+0.00 &+0.00& \\
\ion{Ti}{I}     & 5965.825     &+0.50 &+0.50&+0.30 &+0.30&+0.10 &+0.30&+0.00   &+0.00&+0.30 &+0.30&+0.10 &+0.00& \\
\ion{Ti}{I}     & 5978.539     &+0.50 &+0.50&+0.35 &+0.30&+0.25 &+0.30&+0.00   &+0.00&+0.30 &+0.30&+0.00 &+0.00& \\
\ion{Ti}{I}     & 6064.623     &+0.60 &+0.50&+0.30 &+0.30&+0.10 &+0.30&---     &+0.00&---   &+0.15&+0.00 &+0.00& \\
\ion{Ti}{I}     & 6091.169     &+0.60 &+0.50&+0.30 &+0.30&+0.10 &+0.30&+0.00   &---&+0.20 &+0.30&+0.00 &+0.00& \\                                                                         
\ion{Ti}{I}     & 6126.214     &+0.55 &+0.50&+0.30 &+0.30&+0.15 &+0.30&+0.10   &+0.00&+0.30 &+0.30&+0.00 &+0.00& \\                                                                                               
\ion{Zr}{I}     & 6127.475  &+0.50 &+0.50&---   &+0.30&---   &---&+0.15   &+0.30&+0.50:&+0.50:&---   &---& \\
\ion{Zr}{I}     & 6134.585  &+0.65 &+0.50 &---   &+0.50&---   &---&+0.10   &---&+0.50:&---&---   &---& \\ 
\ion{Zr}{I}     & 6140.535  &---   &---&---   &---&---   &---&+0.10   &---&---   &--- &---   &---& \\ 
\ion{Zr}{I}     & 6143.252  &+0.65 &+0.50&---   &---&---   &---&+0.10   &---&+0.50 &+0.30&---   &---& \\ 
\ion{Ba}{II}&5853.675     &+0.50 &+0.50&+0.50 &+0.60&+0.30 &+0.30&+0.30   &+0.30&+0.00 &+0.20&+0.60 &+0.30& \\
\ion{Ba}{II} & 6141.713   &+0.50 &+0.40&+0.40 &---  &+0.30 &---  &+0.30   &---  &$-$0.30 &$-$0.30&+0.60 &+0.15& \\
\noalign{\hrule\vskip 0.1cm} 
\hline                  
\end{tabular}
\end{flushleft}
\end{table*}


\section{Discussion}

The inspection on abundances of heavy elements
in NGC 6522 was triggered by the variation in
Ba abundances reported in Barbuy et al. (2009).
In C11, we tentatively tried to connect these
abundances with the  s-process  nucleosynthesis
calculations in spinstars first presented by Pignatari et al. (2008)
and computed by
Frischknecht et al. (2012). We did so for a spinstar of  40 M$_{\odot}$, a metallicity  of
[Fe/H] = $-$3.8, and a rotational velocity of V$_{rot}$ = 500 km.s$^{-1}$. 
The lower resolution data (R$\sim$22,000) from GIRAFFE spectra were compatible with the s-process yields of spinstars 
boosted by up to four orders of magnitude with respect to a non-rotating star of the same mass and metallicity  (see their Fig. 2).

A next step was presented in Barbuy et al. (2014), where the analysis
of higher resolution data from UVES (R$\sim$45,000), was studied in
terms of  an extended grid of spinstar models from
Frischknecht et al. (2016). This paper reported enhancement of the
heavy elements Sr, Y, La, and Ba measurable in stellar spectra.
The new results were shown to be
compatible with expectations from massive spinstars,
but allowing for other mechanisms to be invoked.

In a third step using the observations of 2012, besides the
four stars observed with UVES and reported in Barbuy et al. (2014),
we also identified possible new cluster member stars from
the GIRAFFE spectra.
In 2016, we obtained new UVES observations of six such newly identified
member stars, which we analyse here.

\subsection{Analysis of the present results}

We  derived  a mean metallicity of [Fe/H]=$-$1.16$\pm$0.05, somewhat
lower than the B09, B14, and Fern\'andez-Trincado et al. (2019)
values of [Fe/H]= -1.0, -0.95 and -1.04, respectively,
and closer to the value of [Fe/H] = -1.15 from Ness et al. (2014).
By gathering the metallicities of the present four member stars
and the other four stars from Barbuy et al. (2014), we obtain
[Fe/H]=$-$1.05$\pm$0.20.

The mean abundances for the six sample stars, as well
as the four stars studied in Barbuy et al. (2014), are
reported in Table \ref{mean}.
In the mean we see a normal expected enhancement of the bona fide alpha elements O and Mg, and a 
mild enhancement of Si and Ca (and Ti, noting that Ti behaves as
an alpha, but it is an iron-peak element). There has been evidence
that the alpha elements O and Mg formed during the hydrostatic
phases of massive stars nucleosynthesis are more enhanced than the
other alpha elements, Si, Ca, and Ti, which are formed predominantly during
explosive nucleosynthesis (Woosley \& Weaver 1995, McWilliam 2016).

The enhancements of N, Na, and Al vary from star to star, indicating that
some stars are probably second generation ones.
We note that the enhancement in N is also due to a large 
scatter in its early enrichment history (e.g. Cescutti \& Chiappini 2010) as well as stellar evolution
effects.
In Fig. 14, we 
do not find a clear anti-correlation between the [O/Fe] and [Na/Fe] ratios in our stellar sample, such as, for instance, in NGC 6121,
which is a well-populated cluster in this diagram (Carretta et al. 2009).
In  Fig. 14,we also show the correlated abundance signatures of  [Al/Fe] versus [Na/Fe],
 [N/Fe] versus [Na/Fe], and [Mg/Fe] versus [Al/Fe].
 These diagrams confirm the presence of at least two stellar
 populations in NGC~6522, as found by Kerber et al. (2018) from photometry.

For the heavy elements, we present 
 the plot of  abundances including the new results.
 In Fig. 15, we compare the present results on heavy element
 abundances of Sr, Y, Zr, La, Ba, and Eu relative to Fe,
 together with those from Barbuy et al. (2014) for another four stars
 in NGC~6522. Literature  abundances from
  field bulge red giants are from 
  Johnson et al. (2012), for Zr, La, and Eu in Plaut's field,
 Siqueira-Mello et al. (2016), van der Swaelmen et al. (2016),
 and metal-poor giants from Howes et al. (2016) and Lamb et al. (2017).
 Also included are the abundances
 Bensby et al. (2017 and references therein)  for 39  microlensed bulge
 dwarfs and subgiants that are older than 9.5 Gyr,
 selected among their 90 stars. Finally, abundances
 in the  bulge globular cluster
 HP~1 (Barbuy et al. 2016), NGC 6558 (Barbuy et al. 2018b),
 and M62 (Yong et al. 2014) are also shown.

 From Fig. 15, the most striking feature is the abundance variation of Sr, Y, Zr, and to lesser extent Ba and La
 at the metallicity of [Fe/H] $\sim$ $-$1.0, where the bulge globular clusters are found.
 For Sr, we report literature data only, given the unreliability of Sr derivation in the
   present sample due to blends with CN and C$_2$ lines.
 We note that the spread is clearly larger in halo metal-poor stars with
 [Fe/H]$\simless$-2.5 (Cescutti et al. (2013), Hansen et al. (2014).
 For Eu, the behaviour of [Eu/Fe] versus [Fe/H] is more well-defined, indicating a spread at low metallicities and a declining abundance ratio with increasing metallicities.

 From Table \ref{mean} and Fig. 15, we find that
a) Y tends to be enhanced, showing strong star-to-star variations; 
b) Ba tends to be enhanced, showing star-to-star variations.
and c) Eu is enhanced similarly to the alpha elements O and Mg.

 As mentioned,  Y and Ba variations are compatible with a large number of nucleosynthesis processes, with the member stars 244819 and B118 showing [Y/Ba] excesses of +0.54 and +0.35, respectively. The observation of more heavy elements would be
 necessary to differentiate between the potential astrophysical sources. The variation in [Y/Ba] data compared with a chemical evolution model for the Galactic bulge is shown in Figure 16.

 We note that in C11 there was no model, but instead
 a calculation for only one mass and
that was showing the impact of rotation already. The argument was that even
if that calculation was done for a very metal-poor metallicity,
because in the bulge the metallicity grows quickly, we would see
its effect in the very old bulge stars at [Fe/H]~$-$1.0 as well. 

In Fig. 16, we present the result of stochastic models, as presented
in Cescutti et al. (2018). This can be summarised as follows.
The nucleosynthesis adopted for the s process from rotating massive stars
comes from Frischknecht et al. (2016). 
In this set of yields, the s process for massive stars is
computed for a rotation velocity of v$_{ini}$/v$_{crit}$ = 0.4
and is composed of a grid of four stellar masses (15, 20, 25, and 40
M$_{\odot}$) and three metallicities (solar metallicity, 10$^{-3}$, 10$^{-5}$)
(Cescutti \& Chiappini 2014, Cescutti et al. 2013).
The model considers the enrichment produced by 
r-process events as originated from magneto-rotationally driven supernovae (MRD SNe; see Winteler et al 2012, Nishimura et al. 2017); 
MRD SNe are assumed to be  10\% of  all the SNe II.
The model also takes into account the s-process production from
1.5 to 3 M$_{\odot}$ stars and SNIa enrichment, as in Cescutti et al. (2006).
In summary, this model considers the fact that the enrichment in heavy elements takes place both in spinstars and in MRD supernovae.

A spread in 
abundances of these elements is observed in metal-poor halo stars (e.g. Fran\c cois et al. 2003; Cescutti \& Chiappini 2014; Rizzutti et al. 2021) and is expected from spinstar models (Frischknecht et al. 2012, 2016; Choplin et al. 2018, Limongi $\&$ Chieffi 2018)  and from the contribution of neutrino-driven winds in CCSNe (e.g. Roberts et al. 2010). The observed heavy element  abundance ratios tend to show a higher spread of abundance ratios at the metallicity of
NGC~6522 relative to the models.

 However, the models presented in Fig. 16 were optimised for old field stars of the Galactic bulge, adopting the same nucleosynthesis that worked well for
the Galactic halo (Cescutti et al. 2013).
Although this model is not specifically made for a globular cluster, it is still useful. For instance, it shows the 
extension of the dispersion that the enrichment due to rotating massive stars can produce on these abundance ratios.
The goal is to show that the predicted scatter is indeed compatible with the dispersion observed in  NGC~6522. However, this scatter in field stars seems to appear at lower metallicities (see Barbuy et al. 2018 where our model is compared with field bulge stars). It is then plausible that other physical mechanisms are at play in the cluster evolution (involving dynamical effects, and mass loss through winds). The model for the field bulge stars would just give an idea of the mean cluster abundances but not its scatter.
 A detailed description of the models presented in Fig. 16,
with a focus on the expected differences in the abundance ratio scatter in the bulge and halo, will be presented in Cescutti et al. (2021,  in preparation).

In order to better interpret the heavy-element abundances of star B118,
in Fig. 17 we show the abundance pattern of  stars 244819 and B118
in terms of A(X)-A(Eu) (where A(X)=log(N$_{X}$/N$_{H}$+12),
in a diagram idealised by Honda et al. (2007) and Roederer et al. (2010, their Fig. 11).
In this figure, A(X)-A(Eu) versus Z of  244819 and B118 are
compared with data from the typical r-element star CS 31082-001 (Hill et al. 2002),
and the  typical LEPP star HD 122563 (Honda et al. 2007, Montes et al. 2007),
the identified spinstar-enriched star ROA 276 in $\omega$ Centauri 
(Yong et al. 2017), and the reference dwarf stars HD 94028 and
HD 140283 (Peterson et al. 2020; Siqueira-Mello et al. 2015).
First of all, Figure 17 indicates  
that  244819 and B118 are
weakly enriched in r elements. 
Figure 18 shows
A(X)-A(Eu) versus Z for the four member stars from this paper plus the four
stars from Barbuy et al. (2014) for Y, Zr, Ba, and La. It shows that
the sample stars essentially follow the same pattern, whereas 244819 and
B118 show a larger abundance difference between first peak and second peak elements.

Spite et al. (2018) suggested that the heavy element enrichment should take place first due to a pure r process, followed by an enrichment of first-peak elements only, and that this second mechanism would be 
detectable only in weak-r-process stars. 
On the other hand,  spinstars could be progenitors of magneto-rotational supernovae, but in case the conditions do not allow r-process elements to form in the final explosion, it is also possible that we can only observe the signature of the s-process production in spinstars today. 

There are a number of points that it is important to consider in our analysis. First of all, the s-process efficiency in spinstars varies greatly if we consider different theoretical stellar yields. While for instance the s-process production in models by Frischknecht et al. (2016) and Choplin et al. (2018) would stop in the Ba mass region, in models by Limongi $\&$ Chieffi (2018) heavier elements up to lead could be produced.
This uncertainty of course affects Galactic chemical evolution (GCE) predictions
(e.g. Cescutti et al. 2013, Rizzuti et al. 2019, Prantzos et al. 2020, Rizzuti et al. 2021). Additionally, CCSNe generated from slowly rotating progenitors or spinstars can also eject other nucleosynthesis components made before the SN explosion (similarly to the intermediate neutron capture process or i process - see e.g. Roederer et al. 2016 and Banerjee et al 2018) or by explosion (similarly to the zoo of neutrino-driven wind components - see e.g. Qian $\&$ Wasserburg 2008, Farouqi et al. 2009, Roberts et al. 2010, Arcones $\&$ Montes 2011).
As we mentioned earlier, all of these processes may contribute to the production of Sr, Y, and Zr, while at low metallicities the i process can potentially produce elements across the whole mass region below and beyond Fe, including Sr, Y, Zr, and Ba, in different types of stars (e.g. 
Abate et al. 2016, Roederer et al. 2016, Clarkson et al. 2018, Banerjee et al 2018). For instance, the high [Ba/La] in stars 256289, 402370, B-107, B-128, and B-130 would be compatible with the i process (see e.g. Hampel et al. 2016).

 Alternative possibilities of neutron-capture element enrichment are the 
magneto-rotationally driven explosions of core-collapse supernovae (Winteler et al. 2012), or s process taking place in asymptotic giant branch (AGB) stars and subsequent mass transfer within a binary system (e.g. Cristallo et al. 2015).
A study on possible nucleosynthesis processes is given in Hansen et al. (2014).

This makes the observation of more elements per stellar target and at high resolution of different stars in globular clusters such as NGC~6522 crucial. Within this scenario the large star-to-star variations of heavy-element enrichment could be a natural outcome of an intrinsic scatter of s-process efficiencies in spinstars or the varying contribution of different processes active before and during SN explosions in massive stars. On the other hand, when abundances of several heavy elements are available it becomes possible to disentangle the dominant nucleosynthesis component(s) that made the whole observed abundance pattern (see e.g.,
Roederer et al. 2016, Peterson et al. 2020). 
Stars in NGC~6522 are carriers of the same signatures of the nucleosynthesis processes active in the early galaxy and observed in halo stars,
even if they are more metal-rich as a result of a steeper age-metallicity
relation in the Galactic bulge (as suggested in C11).

\begin{figure}
  \includegraphics[width=9cm]{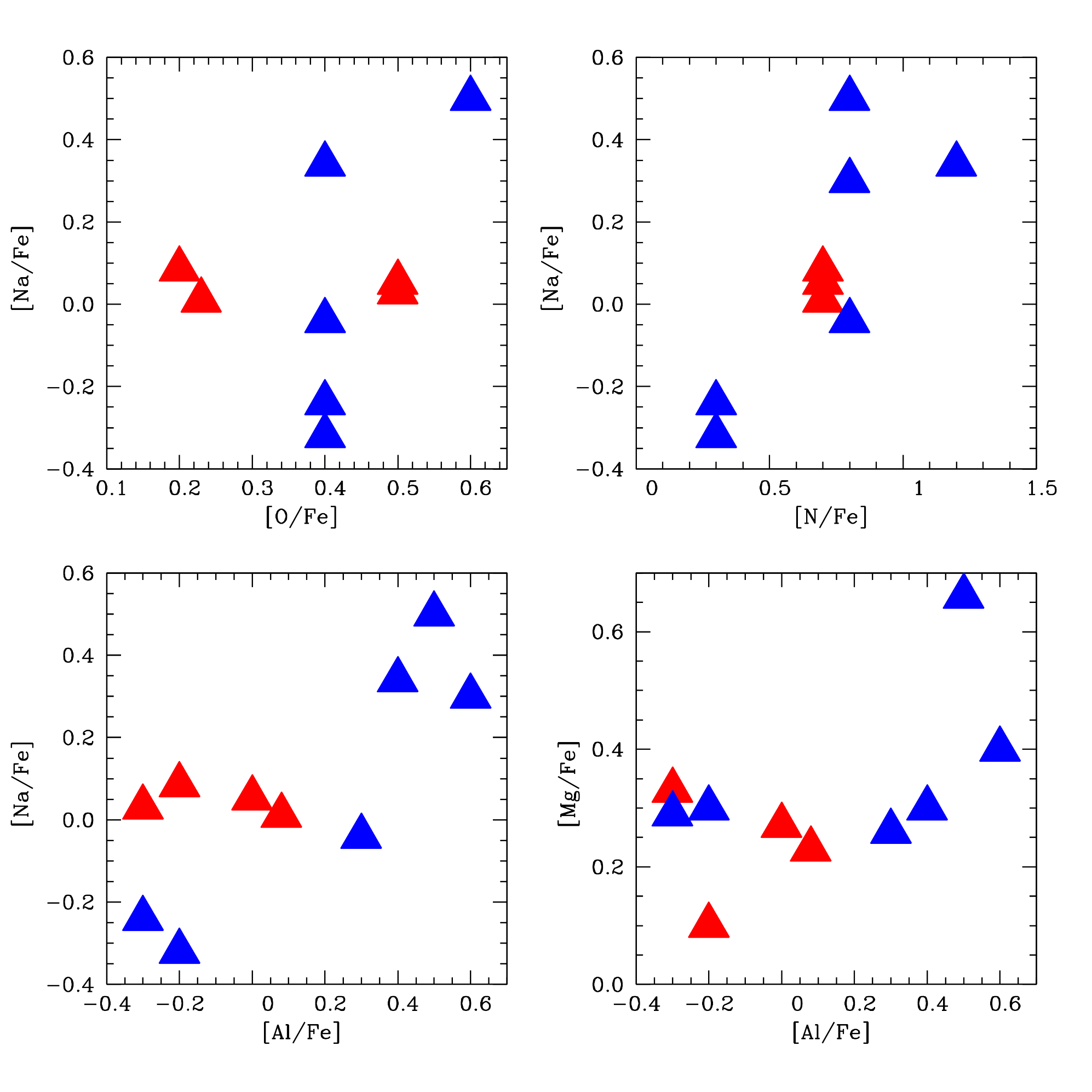}
  \label{plotnao}
  \caption{ [Na/Fe] versus [O/Fe] anti-correlation in NGC 6522 stars compared with
    stars in NGC 6121; and [Na/Fe] versus [N/Fe], [Al/Fe] and [Mg/Fe] versus [Al/Fe]
    correlations.
    Symbols: blue-filled triangles represent the present results for NGC 6522;
    red-filled triangles denote
    results for NGC 6522 from Barbuy et al. (2014).} 
\end{figure}

\begin{figure*}
  \includegraphics[width=17cm]{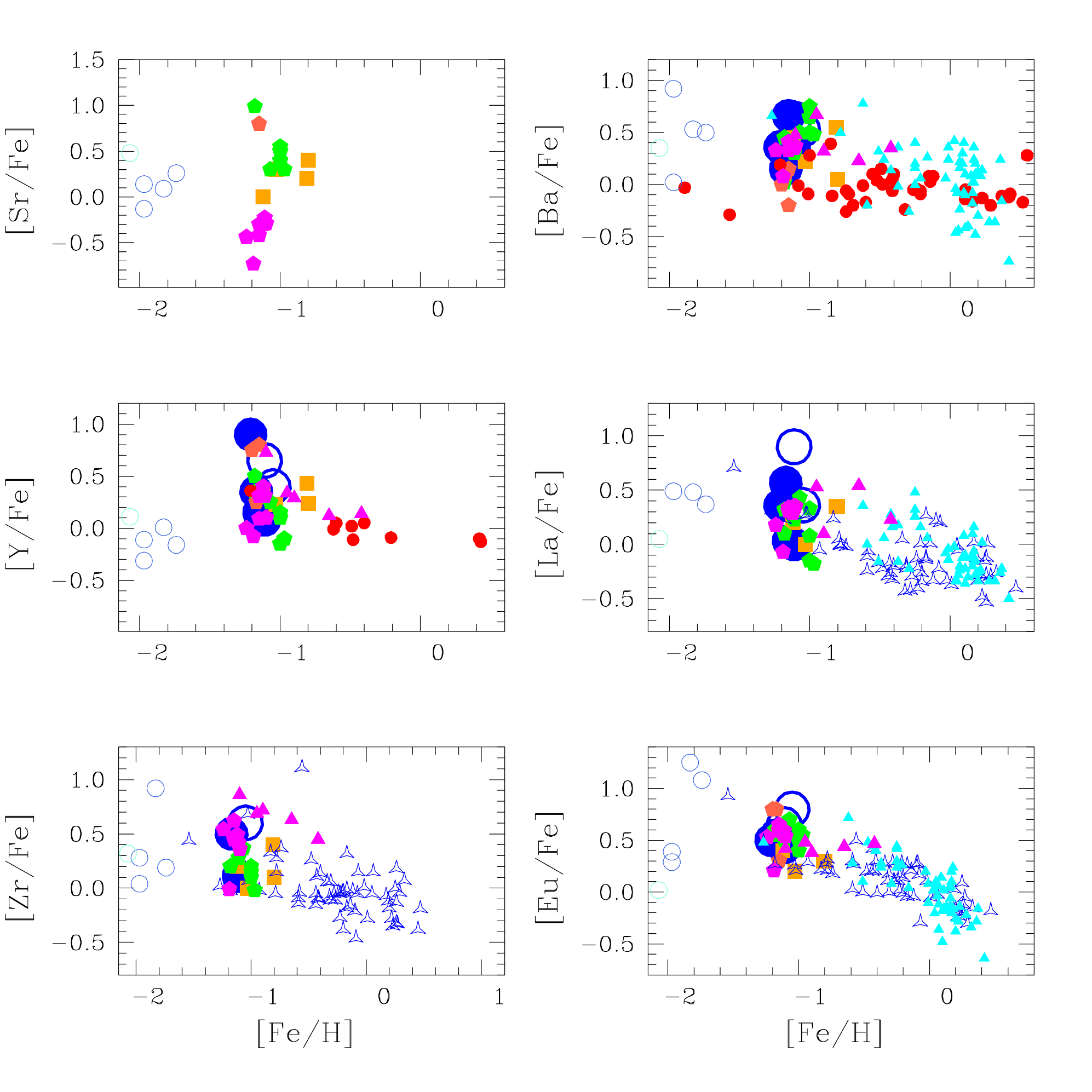}
  \label{newnewnew}
\caption{[Sr,Y,Zr,Ba,La,Eu/Fe] versus [Fe/H] in bulge stars. Symbols:
  Large,  blue-filled circles mark present results of NGC 6522 member stars;
  large,  blue open circles represent the present results of non-member stars.
  {\it Field stars:}
  Blue open triangles represent red giants by Johnson et al. (2012);
  magenta-filled triangles represent red giants by Siqueira-Mello et al. (2016);
  cyan-filled triangles denote red giants by van der Swaelmen et al. (2016); 
red-filled circles show dwarfs by Bensby et al. (2017);
light blue open circles mark metal-poor giants by Howes et al. (2016);
acquamarine open circles show metal-poor giants by Lamb et al. (2017).
{\it Globular cluster stars:}
 Orange-filled squares show NGC 6522 (Barbuy et al. 2014);
 green-filled pentagons mark HP~1 (Barbuy et al. 2016); 
 tomato-filled pentagons represent NGC 6558 (Barbuy et al. 2018);
magenta-filled pentagons show M62 (Yong et al. 2014).} 
\end{figure*}

\begin{figure}
  \includegraphics[width=9cm]{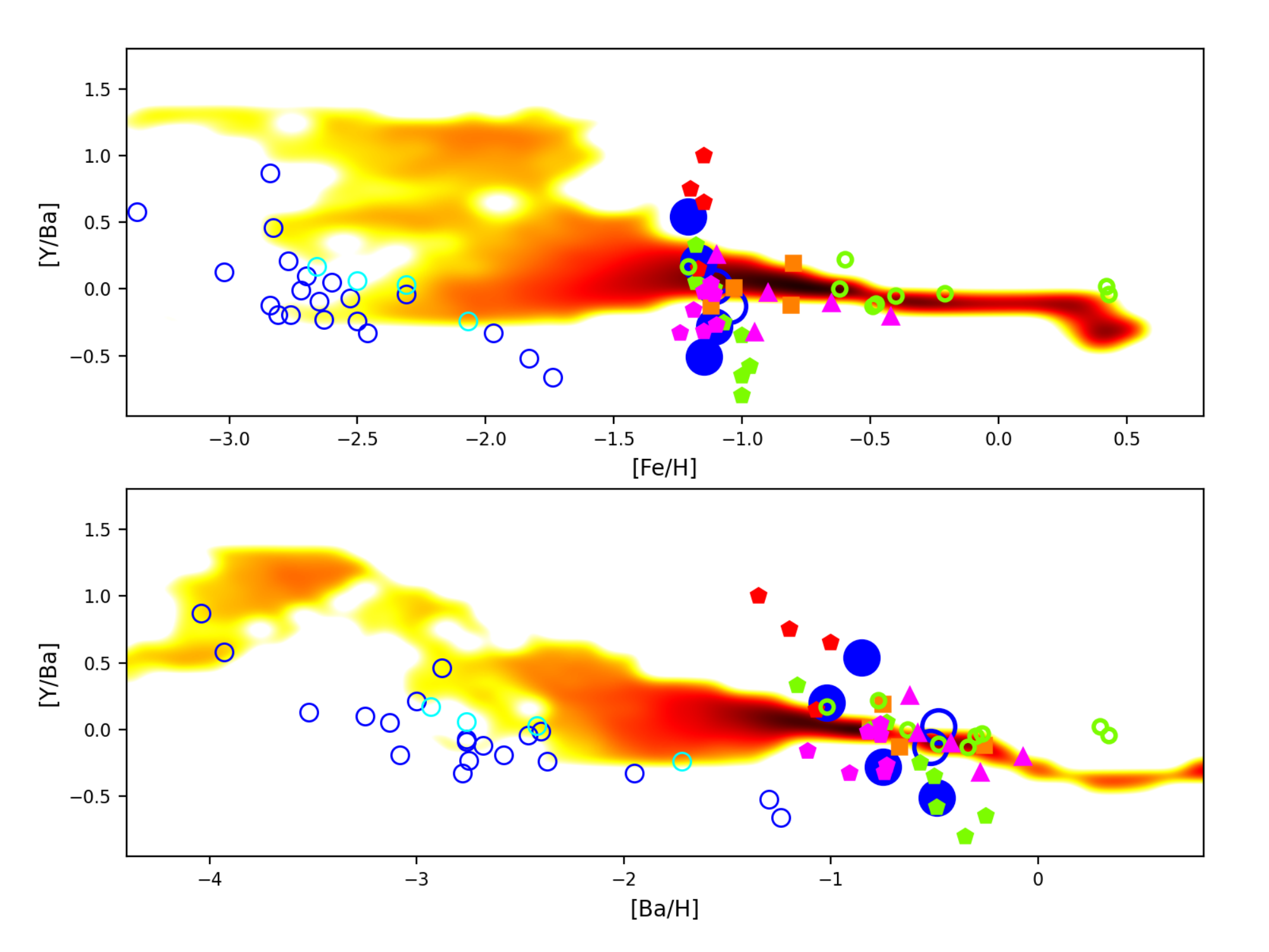}
  \label{cescutti}
  \caption{[Y/Ba] vs. [Fe/H] (upper panel) and [Y/Ba] versus [Ba/H] (lower panel).
  Symbols:
  Large blue-filled circles show present results on NGC 6522 member stars;
  large  blue open circles show present results on non-member stars.
  {\it Field stars:}
  Magenta-filled triangles denote red giants by Siqueira-Mello et al. (2016);
green open circles mark dwarfs by Bensby et al. (2017);
light blue open circles show metal-poor giants by Howes et al. (2016);
acquamarine open circles represent metal-poor giants by Lamb et al. (2017).
{\it Globular cluster stars:}
 Orange-filled squares show NGC 6522 (Barbuy et al. 2014);
 green-filled pentagons represent HP~1 (Barbuy et al. 2016); 
 tomato-filled pentagons show NGC 6558 (Barbuy et al. 2018);
 magenta-filled pentagons represent M62 (Yong et al. 2014).
 The underlying coloured model corresponds to calculations
   for nucleosynthesis taking place in  spinstars and in MRD supernovae
 (Cescutti et al. 2018 and references therein). } 
\end{figure}

\begin{figure}
  \includegraphics[width=9cm]{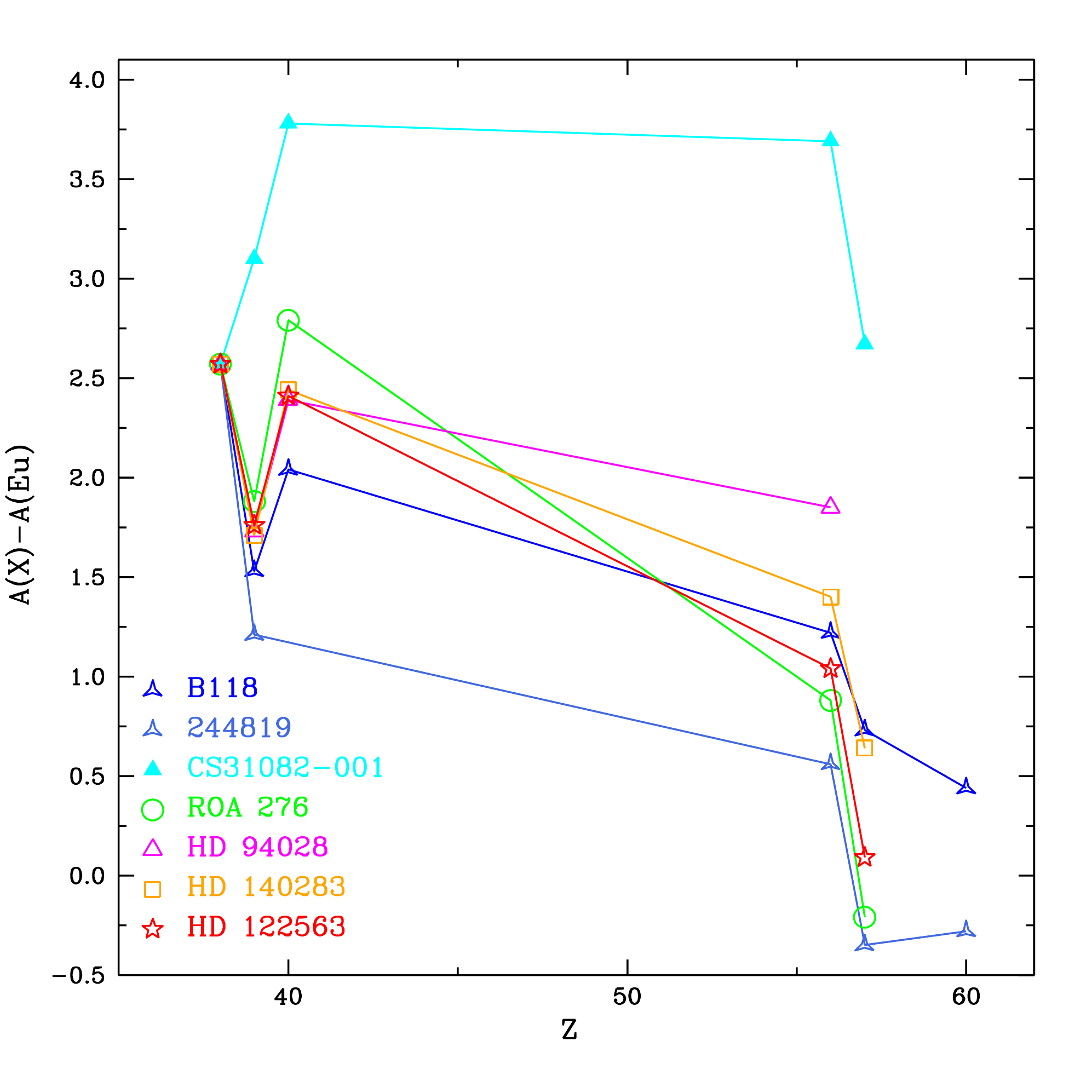}
  \label{pattern}
  \caption{A(X)-A(Eu) versus atomic number Z for stars B118 and
    244819 compared with
  stars CS 31082-001, ROA 276, HD 94028, HD 140283, and HD 122563.
  All abundances are normalised to the Sr abundance of B118.
    Symbols are identified in the figure panel.} 
\end{figure}

\begin{figure}
  \includegraphics[width=9cm]{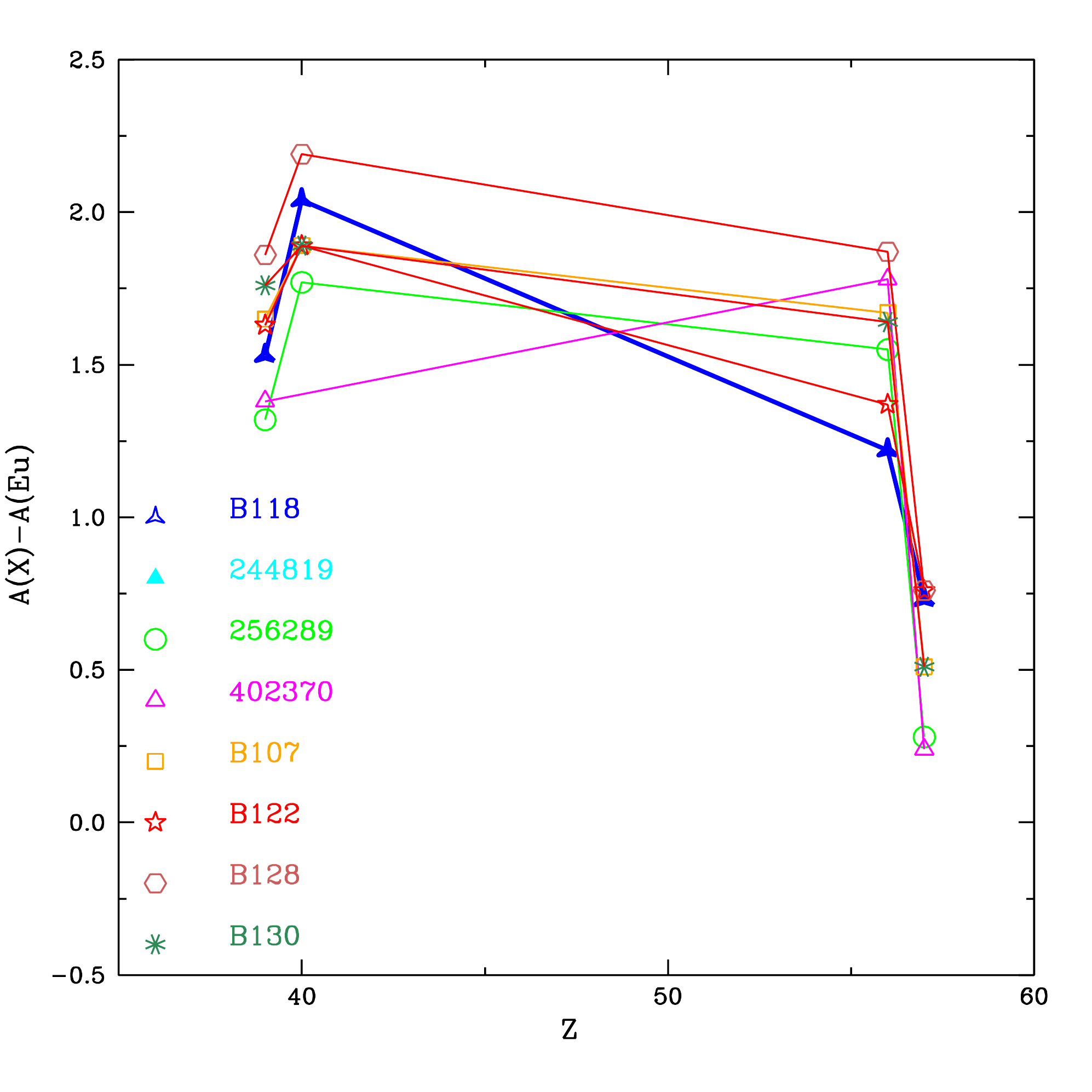}
  \label{patternstars}
  \caption{A(X)-A(Eu) versus atomic number Z for star B118 compared with
  the other seven sample stars. The abundances are not normalised. 
    Symbols are identified in the figure panel.} 
\end{figure}

\subsection{The two non-member stars 234816 and 244523}

The stars 234816 and 244523 have the correct magnitudes, radial velocities and
metallicities to be members of NGC~6522. However, the Gaia proper motions
reported in Table \ref{gaia} rule out their membership.

Could these two stars be former members that are evaporating
from the cluster? Madrid et al. (2017) studied evaporation rates
as a function of galactocentric distance R$_{\rm GC}$ and time
and predicted a very high evaporation at low  R$_{\rm GC}$ due to
the strong tidal field in the central parts of the Galaxy.
NGC~6522 is estimated to have a mass of 5.93$\times$10$^{4}$ M$_{\odot}$
(Gnedin \& Ostriker 1997), which is not high for a globular cluster.
It is located at R$_{\rm GC}$ $\approx$ 1 kpc and has an age above
12.1 - 12.4 Gyr using Dotter et al. (2008) isochrones and even
older (with 14.1 - 14.2 Gyr) using BaSTI isochrones (Kerber et al. 2018).
It can be seen from Fig. 6 of Madrid et al. (2017) that the
evaporation rate in a bulge cluster like NGC~6522 should be extremely
high. It is therefore acceptable to suggest that the two stars
could be evaporating from the cluster.
 However, a more detailed orbital calculation would be needed to check
  this possibility, such as the one carried out by Hanke et al. (2020). In particular,
  star 234816 has different alpha-element abundances and should not be
  a member.

\section{Conclusions}

We  derived  a mean metallicity of [Fe/H]=$-$1.16$\pm$0.05 from
the four sample stars. Combined with the other four stars from Barbuy et al. (2014),
the result is a mean metallicity of [Fe/H]=$-$1.05$\pm$0.20.

Among the six stars analysed, two of them are indicated to be non-members
from Gaia proper motions; still, they have the correct magnitude, radial
velocity, and metallicity to be members. Only a fraction of about
0.5\% of stars in the Galactic bulge have metallicities below
[Fe/H]$\simless$-1.0 (Barbuy et al. 2018a).  Therefore, we suggest that these
stars could be evaporating from the cluster; but even so,
we do not include their abundances in the discussion below.
Star 244523 has abundances compatible with the member stars. Star
234816 shows different high alpha-element abundances,
and it could be an intruder; hence,
  it could have been a bulge star with the correct magnitude and metallicity to be
  considered a member before we had Gaia measurements, but it was eventually
  revealed as a non-member star.

For the present results on the four confirmed member stars,
together with those by Barbuy et al. (2014),
the alpha-elements show  enhancements of 
[O/Fe]=+0.38, [Mg/Fe]=$\approx$+0.28, [Si/Fe]$\approx$+0.19, and
[Ca/Fe]$\approx$+0.13,  [Ti/Fe]$\approx$+0.13.
A higher enhancement in O and Mg, and a lower one in Si, Ca, and Fe can be explained by their
formation in hydrostatic conditions for the former, and in explosive nucleosynthesis for the latter (e.g. Woosley \& Weaver 1995; McWilliam 2016). 

 The r-process element Eu is enhanced by [Eu/Fe]=+0.40.
The $\alpha$-element enhancements in  O and Mg, together with that of 
the r-process element Eu, are indicative of a fast early enrichment
by type II supernovae.
With regard to the indicators of multiple stellar populations, we
suggest that Na shows an anti-correlation with O, and
more clearly a correlation
with N and Al, whereas Mg and Al are also correlated.

A main objective of this study is the verification of the enhancement of
s-element abundances, and the possibility of an early enrichment by spinstars. The neutron-capture elements typically indicated as s-process elements are enhanced with [Y/Fe]=+0.33, [Zr/Fe]=+0.23,
[Ba/Fe]=+0.35, and [La/Fe]=+0.23.  In addition to this observation
we find 
the following:

a) There are significant relative abundance variations between neutron-capture elements, where
[Y/Ba] is particularly enhanced in two stars.

b) [Ba/Eu] = $-$0.14, $-$0.07, $-$0.40, +0.16, in the four member stars,
as a measure of the s- to r- process nucleosynthesis, tends to be
slightly below solar. This result is still compatible with
 the interpretation given in B09 and C11 that 
their production cannot be attributed to the r-process only, 
as first suggested by Truran (1981) for very old stars.

As discussed in C11, the presence of s-process element enhancements in very old stars 
could be due to an s-process enrichment of the primordial matter from which the cluster formed,
processed in spinstars (e.g. Frischknecht et al. 2016).
Alternatively, 
the production of heavy elements could be due 
to a combination of different nucleosynthesis processes, in particular for the atomic mass region of Sr. Another possibility
would be to have spinstars producing the s-process elements during its
hydrostatic phase and producing the r-process elements at the
supernova explosion, and
therefore to be the source of both. This is possible if the spinstars
rotate fast enough to produce an MHD explosion with the right conditions
to produce an r process (Nishimura et al. 2017 and references therein).
However, within this scenario it is extremely uncertain to predict the observed relative contribution of the s-process and r-process elements, since the r-process-rich material could be ejected asymmetrically and/or could carry a large range of efficiency in r-process production. Therefore, a possible outcome could be 
that the final enrichment produced by such a spinstar
and magnetic-rotationally driven (MRD) SN
is dominated by the r-process signature,
because of higher 
yields of the MRD SNe  
compared to those of the s process (Spite et al. 2018).
On the other hand, the enrichment of the local interstellar medium could also be s-process rich, depending on the spatial distribution of different nucleosynthesis products in the SN ejecta. Finally, nucleosynthesis taking place in AGB 
stars and the i process might be alternative possibilities that should be further
inspected.

Taking into account the different uncertainties at play, we confirm the conclusions from Barbuy et al. (2014)
that the observed abundances are compatible with the s-process production in spinstars. However, we cannot rule out
that the same enrichment signature could be produced by a combination of nucleosynthesis processes active in
the early generations of stars.

\begin{acknowledgements}
  BB, EC, LM, SO acknowledge partial financial support
  from CAPES- Finance code 001, CNPq and Fapesp.
  SOS acknowledges the
  FAPESP PhD fellowship no. 2018/22044-3 and the DGAPA-PAPIIT grant IG100319.
 RH and MP acknowledges support from the IReNA AccelNet
Network of Networks, supported by the National Science Foundation under Grant
No. OISE-1927130 and from the World Premier International Research Centre
Initiative (WPI Initiative), MEXT, Japan. CC, 
  RH, MP and GC acknowledge support from the ChETEC COST Action (CA16117), 
supported by COST (European Cooperation in Science and Technology).
  MP thanks the support to NuGrid from STFC (through the University of
  Hull Consolidat ed Grant ST/R000840/1), and access to {\sc viper},
  the University of Hull High Performance Computing Facility. MP acknowledges
  the support from the Ledulet-2014 Program of the Hungarian Academy of Sciences (Hungary).
SO acknowledges the Italian Ministero dell'Universit\`a e della Ricerca
Scientifica e Tecnologica (MURST), Italy.
This work has made use of data from the European Space Agency (ESA) missionGaia (http://www.cosmos.esa.int/gaia), processed by the Gaia Data Processing and Analysis Consortium (DPAC,http://www.cosmos.esa.int/web/gaia/dpac/consortium). Funding for the DPAC has been provided by national institutions, in particular the institutions participating in the Gaia Multilateral Agreement.
\end{acknowledgements}


\begin{appendix}
\section{Candidate members observed with GIRAFFE}

The radial velocities of GIRAFFE observations were
measured by fitting the cross-correlation peak with the solar spectra
available at
the ESO portal\footnote{http://www.eso.org/observing/dfo/quality/GIRAFFE/pipeline/solar.html},
and the heliocentric radial velocities were derived.
The final heliocentric radial velocities of GIRAFFE spectra were determined by taking
the average of the mean heliocentric velocity of each setup. This method showed a better
agreement with their counterparts in the more reliable UVES spectra velocities.
Signal to noise was measured in points of continua in each of the
setups, resulting in a mean S/N$\sim$ 70 in the HR11 setup and S/N$\sim$ 93
in the HR12 setup.

We applied a selection of stars with radial velocities within $\pm$12 Km/s of that
of NGC 6522 to both the 2012 and the 2016 observations.
On the selected sample, we combined the radial velocities with
Gaia proper motions to derive membership probabilities.
Table \ref{girmag}
reports
the selected stars, their identification, coordinates,
magnitudes, and respective heliocentric radial velocities.

\begin{table*}
\label{girmag} 
\caption[1]{OGLE and 2MASS numbers, coordinates, magnitudes including
  $JHK_{s}$ from both 2MASS and VVV surveys, and final radial velocities
  for stars observed with GIRAFFE in 2012 and 2016. }
\scalefont{0.75}
\begin{flushleft}
\tabcolsep 0.15cm
\begin{tabular}{lcccccccccccccccccccc}
\noalign{\smallskip}
\hline
\noalign{\smallskip}
\hline
\noalign{\smallskip}
{\rm  OGLE}& \phantom{-}\phantom{-}\phantom{-}2MASS ID & $\alpha_{2000}$ & \phantom{-}\phantom{-}\phantom{-}$\delta_{2000}$ 
& \phantom{-}\phantom{-}\phantom{-}$V$ & \phantom{-}\phantom{-}\phantom{-}$I$ & \phantom{-}\phantom{-}\phantom{-}$J$ 
& \phantom{-}\phantom{-}\phantom{-}$H$ & \phantom{-}\phantom{-}\phantom{-}$K_{\rm s}$ &   \phantom{-}\phantom{-}\phantom{-}$J_{\rm VVV}$ 
& \phantom{-}\phantom{-}\phantom{-}\hbox{H$_{\rm VVV}$} & {\rm K$_{\rm VVV}$} &
\hbox{v$_{\rm r}^{hel}$}& B$_{Johnson}$ & member \\
\noalign{\smallskip}
\noalign{\hrule}
\noalign{\smallskip}
\noalign{\centerline{GIRAFFE SAMPLE}}
\noalign{\smallskip}
\noalign{\hrule}
\noalign{\smallskip}
& & & & \multicolumn{4}{c}{\hbox{GIRAFFE sample (2012)}} &&&&&\cr
\noalign{\smallskip}
244853 & 18033424-3002109 &  18:03:34.13& $-$30:02:11.1 &  16.085 &  14.457 & 11.304 & 10.994&  11.374 &   ---  &   ---  &   ---    &$-$11.79&17.363&99.5 \\
402384 & 18034256-3001409 &  18:03:42.55& $-$30:01:40.5 &  16.055 &  14.388 & 13.259 & 12.506&  12.336 & 13.0877& 12.4530& 12.2813&-20.04&17.402&99.0\\
\noalign{\smallskip}
& & & & \multicolumn{4}{c}{\hbox{GIRAFFE sample (2016)}} &&&&&\cr
\noalign{\smallskip}
244551 & 18033361-3002389 &  18:03:33.61& $-$30:02:38.9 &  16.134 &  14.455 & 13.321 & 12.551&  12.523 & 13.1628& 12.5364& 12.3582&-8.63&17.486&97.8   \\
244555 & 18033467-3002305 &  18:03:34.67& $-$30:02:32.2 &  16.536 &  14.480 & 11.436 & 10.595&  10.478 &   ---  &    --- &   ---    &$-$2.33&18.231&99.3    \\
244813 & ---              &  18:03:29.00& $-$30:02:28.3 &  16.147 &  14.578 &  ---   &   --- &   ---   & 13.1904& 12.5908& 12.4422&$-$10.80&17.365&43.7   \\
256298 & 18033214-3000350 &  18:03:32.13& $-$30:00:34.9 &  16.038 &  14.430 & 13.133 & 12.306&  12.173 & 13.1264& 12.4863& 12.2987&$-$22.39&17.329&99.8   \\ 
402371 & 18033854-3002075 &  18:03:38.57& $-$30:02:07.3 &  16.065 &  14.417 & 12.932 & 12.218&  12.254 & 13.1894& 12.6119& 12.4397&$-$17.52&17.394&99.6   \\
402508 & 18034025-3003178 &  18:03:40.16& $-$30:03:18.1 &  16.278 &  14.590 & 11.112 & 11.373&  11.216 & 12.8682& 12.0634& 12.1385&$-$4.32&17.593&91.9    \\
\noalign{\smallskip} \hline \end{tabular}           
\end{flushleft}                                         
\end{table*}

\section{ Hyperfine structure of \ion{Ba}{II} lines}

\begin{table*}
\begin{flushleft}
\caption{Atomic constants for BaII used to compute hyperfine structure.
A constants are from Rutten (1978), B constants from Biehl (1976), and 
B constants not available in the literature are assumed as null.
}
\small
\label{balines1}      
\centering          
\begin{tabular}{lccccc@{}c@{}c@{}c@{}c@{}c@{}c@{}c@{}c}     
\noalign{\smallskip}
\hline\hline    
\noalign{\smallskip}
\noalign{\vskip 0.1cm} 
species & $\lambda$ ({\rm \AA}) & Lower level & J &A(mK)& A(MHz) &B(mK)& B(MHz) & Upper level & J &A(mK)& A(MHz) &B(mK)& B(MHz)  \\
\noalign{\vskip 0.1cm}
\noalign{\hrule\vskip 0.1cm}
\noalign{\vskip 0.1cm}
$^{135}$BaII & 5853.668 & 5d $^{2}$D$_{3/2}$ &  3/2 &3.56& 106.7261 &0& 0 & 6p $^2$P$^{\circ}$$_{3/2}$ & 3/2 &+3.47& 104.028 &+2.2& 65.9544  \\
$^{137}$BaII & 5853.668 & 5d $^{2}$D$_{3/2}$ &  3/2 &3.97& 119.0176 &0& 0 & 6p $^2$P$^{\circ}$$_{3/2}$ & 3/2 &+3.88& 116.3195 &+3.25& 97.4326  \\
$^{135}$BaII & 6141.713 & 5d $^2$D$_{5/2}$ &  5/2 &1.49& 44.6691 &0& 0 & 6p $^2$P$^{\circ}$$_{3/2}$ & 3/2 &+3.47& 104.028 &+2.2& 65.9544  \\
$^{137}$BaII & 6141.713 & 5d $^2$D$_{5/2}$ &  5/2 &1.66& 49.7655 &0& 0 & 6p $^2$P$^{\circ}$$_{3/2}$ & 3/2 &+3.88& 116.3195 &+3.25& 97.4326  \\
\noalign{\vskip 0.1cm}
\noalign{\hrule\vskip 0.1cm}
\noalign{\vskip 0.1cm}  
\hline                  
\end{tabular}
\end{flushleft}
\end{table*} 

\begin{table}
\caption{Hyperfine structure for \ion{Ba}{II} 5853.675 {\rm \AA} line. }
\label{hfsBa}
\scalefont{0.9}
\centering
\begin{tabular}{ccccccccccccc}
\hline
\noalign{\smallskip}
\multicolumn{3}{c}{5853.675 $\rm \AA$;  $\chi$=0.604321 eV}  & \\
 \multicolumn{3}{c}{log gf(total) = $-$1.10} & \\
\noalign{\smallskip}
$\lambda$ ($\rm \AA$) & log gf & iso && \\
\noalign{\smallskip}
 5853.673  &   $-$2.3441& 135  & \\
 5853.674  &   $-$2.3441& 135  & \\
 5853.674  &   $-$2.7421& 135  & \\
 5853.673  &   $-$2.1400& 135  & \\
 5853.677  &   $-$2.1400& 135  & \\
 5853.676  &   $-$2.0431& 135  & \\
 5853.671  &   $-$2.1980& 135  & \\
 5853.679  &   $-$2.1980& 135  & \\
 5853.675  &   $-$1.5959& 135  & \\
 5853.673  &   $-$2.3441& 137  & \\
 5853.674  &   $-$2.3441& 137  & \\
 5853.674  &   $-$2.7421& 137  & \\
 5853.673  &   $-$2.1400& 137  & \\
 5853.677  &   $-$2.1400& 137  & \\
 5853.676  &   $-$2.0431& 137  & \\
 5853.670  &   $-$2.1980& 137  & \\
 5853.680  &   $-$2.1980& 137  & \\
 5853.674  &   $-$1.5959& 137  & \\
 5853.675  &   $-$2.7562& 134  & \\.
 5853.675  &   $-$2.2451& 136  & \\
 5853.675  &   $-$1.2845& 138  & \\
\noalign{\smallskip}
\hline
\end{tabular}
\end{table}

In Table \ref{balines1}, we give the hyperfine structure constants
for the BaII 5853.675 {\rm \AA} line, and in Table \ref{hfsBa}
we report the list of lines sub-divided due to the hyperfine
structure, thus completing the
similar results for the 6141.713 and 6496.897 {\rm AA} lines
reported in Barbuy et al. (2014). 

\section{Equivalent widths and atomic data}

\begin{longtable}{l@{}rrrrl@{}r@{}rrrrrrrrr}
  \caption{\label{EW} Equivalent widths of
\ion{Fe}{I} and \ion{Fe}{II} lines employed, their wavelengths,
excitation potential (eV), oscillator strengths,and van der Waals 
broadening constant C$_{6}$ adopted. }\\
\hline\hline
\endfirsthead
\caption{continued.}\\
\hline\hline
\hbox{species} & \hbox{$\lambda$} & \hbox{$\chi_{ex}$} &
 \multispan4 {loggf}
& \phantom{-}\phantom{-}C$_{6}$ & 234816 & 244523 & 244819 & 256289 & 402322 & 402370 & \\
\hline
\endhead
\hline
\endfoot
\noalign {}
\hline
\noalign {}
\hbox{species} & \hbox{$\lambda$} & \hbox{$\chi_{ex}$} &
\multispan4 {loggf}
& \phantom{-}\phantom{-}C$_{6}$ & 234816 & 244523 & 244819 & 256289 & 402322 & 402370 &\\
& (\AA) &  (eV) &
KUR & VALD & NIST & ADOPT &  &  &  & & &  &  &\\
\noalign {}
\hline
\hline
\noalign {}

 \ion{Fe}{I}  & 5853.15   & 1.48 &---&$-$5.280&---  & \phantom{-}\phantom{-}$-$5.13  &  \phantom{-}\phantom{-}0.174E-31 & 43.4 & 13.9 & ---  & ---  &  9.6 & ---  &    \\
 \ion{Fe}{I}  & 5856.08   & 4.29 &$-$1.640&$-$1.328&$-$1.328  & $-$1.64  &  \phantom{-}0.423E-30 & 38.3 & ---  & ---  & ---  & 17.4 & ---  &     \\
 \ion{Fe}{I}  & 5858.78   & 4.22 &$-$2.260&$-$2.260&---  & $-$2.26  &  0.376E-30 &  5.0 & ---  & ---  & ---  & ---  & ---  &     \\
 \ion{Fe}{I}  & 5859.60   & 4.55 &$-$0.386&$-$0.418&---  & $-$0.60  &  0.674E-30 & 66.8 & 66.8 & ---  & ---  & 46.7 & ---  &     \\
 \ion{Fe}{I}  & 5861.11   & 4.28 &$-$2.450&$-$2.450&---  & $-$2.45  &  0.415E-30 &  7.0 & ---  & ---  & ---  & ---  & ---  &     \\
 \ion{Fe}{I}  & 5881.28   & 4.61 &$-$1.840&$-$1.840&---  & $-$1.84  &  0.415E-30 & 99.  & ---  & ---  & ---  & 10.4 & ---  &     \\
 \ion{Fe}{I}  & 5902.48   & 4.59 &$-$1.810&$-$1.810&---  & $-$1.81  &  0.708E-30 &  6.6 & ---  & ---  & ---  & 10.2 & ---  &     \\
 \ion{Fe}{I}  & 5905.67   & 4.65 &$-$0.730&$-$0.730&$-$0.77  &  $-$0.73 &  0.623E-30 & 65.9 & 38.7 & 32.9 & 30.3 & 31.2 & ---  &      \\
 \ion{Fe}{I}  & 5916.25   & 2.45 &$-$2.994&$-$2.994&$-$2.99  & $-$2.99  &  0.428E-31 & 98.4 & 43.2 & 65.9 & 61.5 & 46.3 & 56.5 &      \\ 
 \ion{Fe}{I}  & 5927.79   & 4.65 &$-$1.090&$-$1.090&$-$1.07  & $-$1.09  &  0.786E-30 & 27.6 & 42.0 &  9.7 &  --- & 16.1 & 12.4 &      \\
 \ion{Fe}{I}  & 5929.67   & 4.55 &$-$1.410&$-$1.410&$-$1.38  & $-$1.38  &  0.644E-30 & 45.4 & 35.7 & 26.1 &  --- & 13.0 & 10.6 &      \\
 \ion{Fe}{I}  & 5930.18   & 4.65 &$-$0.230&$-$0.230&---  & $-$0.23  &  0.784E-30 &110.5 & ---  & ---  & 60.8 & 54.9 & 46.7 &     \\
 \ion{Fe}{I}  & 5934.65   & 3.93 &$-$1.170&$-$1.170&$-$1.12  & $-$1.12  &  0.234E-30 & 64.0 & 71.7 & 57.1 & ---  & 73.0 & 41.7 &       \\
 \ion{Fe}{I}  & 5952.72   & 3.98 &$-$1.440&$-$1.440&$-$1.39  & $-$1.39  &  0.249E-30 & 50.9 & 46.7 & 47.2 & 39.6 & 30.5 & 43.7 &       \\
 \ion{Fe}{I}  & 5956.69   & 0.86 &$-$4.605&$-$4.605&$-$4.61  & $-$4.60  &  0.948E-32 &133.2 & 97.5 & 70.8 & ---  & 72.2 & 70.3 &       \\
 \ion{Fe}{I}  & 5975.35   & 4.84 &$-$0.822&$-$1.292&---  & $-$0.82  &  0.300E-31 & 37.1 & 42.0 & 26.3 & 39.5 & 44.3 & 36.3 &       \\
 \ion{Fe}{I}  & 5983.68   & 4.55 &$-$1.521&$-$0.485&---  & $-$0.78  &  0.622E-30 & 99.  &119.2 & 42.0 & 45.3 & 28.0 & 23.1 &       \\
 \ion{Fe}{I}  & 5987.06   & 4.79 &$-$0.654&$-$0.432&---  & $-$0.42  &  0.102E-29 & 99.  & ---  & 40.0 & ---  & 33.9 & 37.2 &       \\
 \ion{Fe}{II} & 5991.38   & 3.15 &$-$3.557&$-$3.540&$-$3.60  & $-$3.54  &  0.775E-32 & 18.4 & 18.4 & 19.6 & 29.0 & 27.2 &  8.9 &        \\
 \ion{Fe}{I}  & 6003.01   & 3.88 &$-$1.120&$-$1.120&---  & $-$1.12  &  0.211E-30 & 91.2 & 81.1 & 62.5 & 63.8 & 75.3 & 44.2 &       \\
 \ion{Fe}{I}  & 6005.54   & 2.59 &$-$2.922&$-$3.602&---  & $-$3.61  &  0.300E-31 & 18.4 & 29.7 & 18.4 & 24.3 & 20.8 & 14.9 &       \\
 \ion{Fe}{I}  & 6008.56   & 3.88 &$-$1.291&$-$0.982&---  & $-$0.99  &  0.300E-31 & 89.3 & 81.7 & ---  & 61.3 & ---  & 55.1 &       \\
 \ion{Fe}{I}  & 6012.21   & 2.22 &$-$4.200&$-$4.038&$-$4.04  & $-$3.94  &  0.362E-31 & ---  &  --- & ---  & 17.8 & ---  & ---  &       \\
 \ion{Fe}{I}  & 6020.17   & 4.61 &$-$0.270&$-$0.270&---  & $-$0.27  &  0.300E-31 & 62.8 & 92.0 & 64.3 & ---  & 90.2 & ---  &       \\
 \ion{Fe}{I}  & 6024.05   & 4.55 &$-$0.120&$-$0.120&---  & $-$0.11  &  0.606E-30 & 99.9 & ---  & 81.9 & ---  & 68.7 & 77.8 &       \\
 \ion{Fe}{I}  & 6027.06   & 4.07 &$-$1.210&$-$1.089&$-$1.09  & $-$1.09  &  0.273E-30 & 65.6 & 54.1 & ---  & 52.3 & 42.8 & 39.0 &       \\
 \ion{Fe}{I}  & 6056.01   & 4.73 &$-$0.460&$-$0.460&---  & $-$0.46  &  0.849E-30 & 51.2 & 39.2 & 51.0 & 42.7 & 42.5 & 25.1 &      \\
 \ion{Fe}{I}  & 6065.49   & 2.61 &$-$1.530&$-$1.530&$-$1.530  & $-$1.53  &  0.477E-31 &145.0 &116.9 &116.0 & ---  &131.3 &112.4 &       \\
 \ion{Fe}{I}  & 6078.50   & 4.79 &$-$0.481&$-$0.323&---  & $-$0.40  &  0.951E-30 & 60.1 & 47.5 & 50.5 & ---  & 60.2 & 53.0 &       \\
 \ion{Fe}{I}  & 6079.00   & 4.65 &$-$1.120&$-$1.120&$-$1.10  & $-$1.10  &  0.710E-30 & 47.5 & 18.1 & 20.0 & ---  & 28.2 & 12.7 &       \\
 \ion{Fe}{I}  & 6082.71   & 2.22 &$-$3.573&$-$3.573&$-$3.57  & $-$3.57  &  0.300E-31 & 59.9 & 59.1 & 42.5 & ---  & 36.9 & 26.3 &       \\
 \ion{Fe}{II} & 6084.11   & 3.20 &$-$3.808&$-$3.780 &$-$3.90  & $-$3.79  &  0.787E-32 &  7.9 & 19.1 &  9.1 & 15.1 & 12.0 &  5.2 &  \\
 \ion{Fe}{I}  & 6093.64   & 4.61 &$-$1.500&$-$1.500&$-$1.47  & $-$1.47  &  0.638E-30 & 17.0 & 10.8 &  3.7 & 14.8 & 14.2 & 10.2 &  \\
 \ion{Fe}{I}  & 6094.37   & 4.65 &$-$1.940&$-$1.940&$-$1.92  & $-$1.92  &  0.703E-30 & 11.7 & 99.0 & ---  & ---  & ---  &  2.3 &  \\
 \ion{Fe}{I}  & 6136.99   & 2.20 &$-$2.950&$-$2.950&$-$2.950  & $-$2.95  &  0.282E-31 & ---  &  --- & ---  & 75.8 & ---  & ---  &       \\
 \ion{Fe}{I}  & 6137.70   & 2.59 &$-$1.403&$-$1.403&$-$1.403  & $-$1.40  &  0.457E-31 &110.4 &158.0 &131.0 &122.2 & ---  &120.7 &  \\ 
 \ion{Fe}{II} & 6149.25   & 3.89 &$-$2.724&$-$2.720&$-$2.80  & $-$2.69  &  0.943E-32 & 15.6 & 34.3 & 17.6 & 20.0 & 18.9 & 12.1 &     \\
 \ion{Fe}{I}  & 6151.62   & 2.18 &$-$3.299&$-$3.299&$-$3.30  & $-$3.30  &  0.305E-31 & 84.0 &108.9 & 60.0 & 57.8 & 68.4 & 50.7 &  \\
 \ion{Fe}{I}  & 6157.73   & 4.08 &$-$1.260&$-$1.260&$-$1.22  & $-$1.25  &  0.261E-30 & 71.1 & 24.5 & 59.8 & 43.4 & 59.8 & 43.2 &  \\
 \ion{Fe}{I}  & 6159.38   & 4.61 &$-$1.970&$-$1.970&---  & $-$1.85  &  0.625E-30 & 12.1 & 99.0 & ---  & ---  & 11.5 &  5.0 &  \\
 \ion{Fe}{I}  & 6165.36   & 4.14 &$-$1.550&$-$1.474&$-$1.47  & $-$1.55  &  0.284E-30 & 40.7 & 32.4 & 33.7 & 24.1 & 26.3 & 23.8 &    \\
 \ion{Fe}{I}  & 6173.33   & 2.22 &$-$2.880&$-$2.880&$-$2.88  & $-$2.88  &  0.882E-30 & ---  &  --- & ---  & 75.8 & 80.1 & ---  &       \\
 \ion{Fe}{I}  & 6180.21   & 2.73 &$-$2.780&$-$2.586&$-$2.65  & $-$2.65  &  0.519E-31 & 82.6 & 94.8 & 60.0 & 52.1 & 81.3 & 49.0 &  \\
 \ion{Fe}{I}  & 6187.99   & 3.94 &$-$2.204&$-$1.720&$-$4.16  & $-$1.72  &  0.211E-30 & 63.6 & 42.9 & 30.0 & ---  & 20.2 & 29.4 &      \\
 \ion{Fe}{I}  & 6200.31   & 2.61 &$-$2.437&$-$2.437&$-$1.67  & $-$2.44  &  0.507E-31 & ---  &  --- & ---  & 75.5 &134.1 & ---  &       \\
 \ion{Fe}{I}  & 6213.43   & 2.22 &$-$2.660&$-$2.482&$-$2.48  & $-$2.49  &  0.311E-31 &111.6 & 89.2 & ---  & 92.0 & 99.4 & 87.7 &      \\
 \ion{Fe}{I}  & 6219.28   & 2.20 &$-$2.433&$-$2.433&$-$2.43  & $-$2.43  &  0.305E-31 & 83.0 &143.6 & 98.4 & 98.3 & ---  & 97.3 &      \\
 \ion{Fe}{I}  & 6220.78   & 3.88 &$-$2.460&$-$2.460&---  & $-$2.46  &  0.192E-30 & 19.7 & ---  & ---  & ---  &  9.8 &  6.7 &  \\
 \ion{Fe}{I}  & 6226.73   & 3.88 &$-$2.220&$-$2.220&---  & $-$2.20  &  0.191E-30 & 36.6 & 13.9 & 17.3 & 22.1 & 21.1 &  9.2 &  \\
 \ion{Fe}{I}  & 6229.23   & 2.84 &$-$2.970&$-$2.805&$-$2.805  & $-$2.97  &  0.571E-31 & 54.5 & 49.4 & 28.4 & 37.3 & 24.0 & 28.8 &  \\
 \ion{Fe}{I}  & 6240.65   & 2.22 &$-$3.380&$-$3.233&$-$3.17  & $-$3.21  &  0.309E-31 & 75.8 & 66.0 & 56.0 & 51.1 & 50.2 & 49.7 &  \\
 \ion{Fe}{I}  & 6246.32   & 3.60 &$-$0.960&$-$0.733&$-$0.88  & $-$0.88  &  0.133E-30 &119.2 &117.3 & 92.4 & 93.0 & ---  & 91.6 &     \\
 \ion{Fe}{II} & 6247.56   & 3.89 &$-$2.329&$-$2.310&$-$2.40  & $-$1.98  &  0.881E-32 & 22.4 & 51.1 & 31.8 & 47.8 & 50.1 & 22.3 &    \\
 \ion{Fe}{I}  & 6252.56   & 2.40 &$-$1.687&$-$1.687&$-$1.687  & $-$1.69  &  0.366E-31 & 51.3 &126.6 &130.0 &118.7 &145.6 &119.7 &  \\
 \ion{Fe}{I}  & 6254.25   & 2.28 &$-$2.480&$-$2.443&$-$2.426  & $-$2.43  &  0.326E-31 &166.0 &129.7 & ---  &113.1 &113.5 &107.9 &    \\
 \ion{Fe}{I}  & 6265.14   & 2.18 &$-$2.550&$-$2.550&$-$2.55  & $-$2.53  &  0.295E-31 &170.3 &118.1 &104.2 & ---  & 93.5 & 85.4 &  \\
 \ion{Fe}{I}  & 6270.23   & 2.86 &$-$2.710&$-$2.464&$-$2.61  & $-$2.61  &  0.575E-31 & 76.1 & 70.1 & 57.5 & 57.5 & 61.4 & 54.8 &  \\
 \ion{Fe}{I}  & 6271.28   & 3.33 &$-$2.950&$-$2.703&$-$2.70  & $-$2.81  &  0.945E-31 & 30.1 & 34.8 & 11.4 & ---  &  8.2 & 17.3 &  \\
 \ion{Fe}{I}  & 6297.80   & 2.22 &$-$2.740&$-$2.740&$-$2.74  & $-$2.74  &  0.304E-31 &106.6 &100.4 & 98.6 & 81.8 & 90.4 & 74.6 &    \\
 \ion{Fe}{I}  & 6301.50   & 3.65 &$-$0.672&$-$0.718&$-$0.72  & $-$0.60  &  0.138E-30 &109.0 &108.3 &101.6 & 95.3 &182.0 & 95.1 &  \\
 \ion{Fe}{I}  & 6302.50   & 3.69 &$-$1.131&$-$0.968&---  & $-$0.91  &  0.145E-30 & 90.2 & 82.9 & ---  & 72.4 & ---  & ---  &  \\
 \ion{Fe}{I}  & 6311.50   & 2.83 &$-$3.230&$-$3.141&$-$3.14  & $-$3.22  &  0.551E-31 & 71.9 & 31.1 & 20.0 & 26.0 & 22.0 & 13.2 &  \\
 \ion{Fe}{I}  & 6315.31   & 4.14 &$-$1.232&$-$1.232&$-$1.232  & $-$1.23  &  0.265E-30 & 44.4 & 27.0 & 39.2 & 42.2 & 44.6 & 35.5 &  \\
 \ion{Fe}{I}  & 6315.81   & 4.08 &$-$1.710&$-$1.710&$-$1.66  & $-$1.66  &  0.243E-30 & 38.8 & 40.8 & 25.0 & ---  & 31.6 & 15.4 &  \\
 \ion{Fe}{I}  & 6322.69   & 2.59 &$-$2.426&$-$2.426&$-$2.43  & $-$2.43  &  0.485E-31 & ---  &  --- & ---  & 74.6 & 96.5 & ---  &       \\
 \ion{Fe}{I}  & 6335.33   & 2.20 &$-$2.230&$-$2.177&$-$2.18  & $-$2.18  &  0.295E-31 &135.3 & 96.1 &120.0 & ---  & ---  &105.0 &    \\
 \ion{Fe}{I}  & 6336.83   & 3.69 &$-$1.050&$-$0.856&$-$0.86  & $-$1.05  &  0.143E-30 & 94.9 &107.9 & 98.3 & 83.9 & ---  & 79.7 &  \\
 \ion{Fe}{I}  & 6344.15   & 2.43 &$-$2.923&$-$2.923&$-$2.923  & $-$2.92  &  0.366E-31 & 81.1 & 81.8 & 62.0 & 61.9 & 68.5 & 61.8 &  \\
 \ion{Fe}{I}  & 6355.03   & 2.85 &$-$2.420&$-$2.350&$-$2.29  & $-$2.29  &  0.549E-31 & 90.9 & 93.2 & 90.0 & 68.7 & 83.0 & 48.2 &   \\
 \ion{Fe}{II} & 6369.46   & 2.89 &$-$4.253&$-$4.160&$-$4.29  & $-$4.11  &  0.742E-32 & .... & ---  & ---  & 15.5 & 23.1 & ---  &  \\
 \ion{Fe}{I}  & 6380.74   & 4.19 &$-$1.400&$-$1.376&$-$1.38  & $-$1.38  &  0.277E-30 & 74.3 &141.1 & ---  & 35.8 & 27.2 & 14.7 &  \\ 
 \ion{Fe}{I}  & 6392.54   & 2.28 &$-$4.030&$-$4.030&---  & $-$4.03  &  0.313E-31 & 42.2 & 26.6 & ---  & ---  & 28.6 & 11.5 &  \\
 \ion{Fe}{I}  & 6393.60   & 2.43 &$-$1.620&$-$1.432&$-$1.576  & $-$1.58  &  0.361E-31 &156.4 &134.6 & ---  &120.9 &135.0 &123.1 &    \\
 \ion{Fe}{I}  & 6400.00   & 3.60 &$-$0.520&$-$0.290&$-$0.290  & $-$0.29  &  0.402E-30 & ---  &  --- & ---  &111.7 & ---  & ---  &       \\
 \ion{Fe}{I}  & 6408.02   & 3.69 &$-$0.970&$-$1.018&$-$1.02  & $-$1.00  &  0.139E-30 & 88.8 & 75.0 & ---  & 82.4 & 88.7 & ---  &       \\
 \ion{Fe}{I}  & 6411.11   & 4.73 &$-$2.026&$-$1.935&---  & $-$2.21  &  0.679E-30 & ---  & ---  & ---  & ---  & ---  & ---  &  \\
 \ion{Fe}{I}  & 6411.65   & 3.65 &$-$0.820&$-$0.595&$-$0.72  & $-$0.72  &  0.132E-30 &128.7 &111.8 &102.0 & 95.6 &109.0 & 96.2 &   \\
 \ion{Fe}{II} & 6416.92   & 3.89 &$-$2.740&$-$2.650&$-$2.90  & $-$2.64  &  0.930E-32 & 28.3 & 36.4 & 20.5 & 26.1 & 44.0 & 19.8 &  \\
 \ion{Fe}{I}  & 6419.94   & 4.73 &$-$0.240&$-$0.240&$-$0.27  & $-$0.25  &  0.675E-30 & 54.0 & 60.2 & 54.2 & 52.9 & 66.6 & 45.7 &    \\
 \ion{Fe}{I}  & 6421.35   & 2.28 &$-$2.027&$-$2.027&$-$2.027  & $-$2.03  &  0.310E-31 &122.5 &121.9 &120.0 & ---  & ---  &108.1 &    \\
 \ion{Fe}{I}  & 6430.85   & 2.18 &$-$2.006&$-$2.006&$-$2.006  & $-$2.01  &  0.281E-31 &137.9 &129.9 &131.0 & ---  &130.0 &113.3 &  \\
 \ion{Fe}{II} & 6432.68   & 2.89 &$-$3.708&$-$3.520&$-$3.50  & $-$3.57  &  0.742E-32 & 35.3 & 28.6 & 32.3 & 32.8 & 36.0 & 29.0 &    \\
 \ion{Fe}{II} & 6456.38   & 3.90 &$-$2.075&$-$2.100&$-$2.20  & $-$2.05  &  0.930E-32 & 41.4 & 33.9 & 27.2 & 51.9 & 54.7 & 24.7 &  \\
 \ion{Fe}{I}  & 6469.20   & 4.83 &$-$0.770&$-$0.770&$-$0.81  & $-$0.81  &  0.802E-30 & 63.1 & 18.1 & 30.3 & ---  & 33.4 & 27.3 &  \\
 \ion{Fe}{I}  & 6475.62   & 2.56 &$-$2.940&$-$2.942&$-$2.94  & $-$2.94  &  0.400E-31 & 95.2 & 75.1 & 76.2 & 57.8 & 88.0 & 49.8 &  \\
 \ion{Fe}{I}  & 6481.87   & 2.28 &$-$2.984&$-$2.984&$-$2.98  & $-$2.98  &  0.305E-31 & 98.9 & 50.2 & 76.2 & 67.3 & 78.8 & ---  &  \\
 \ion{Fe}{I}  & 6494.98   & 2.40 &$-$1.273&$-$1.273&$-$1.273  & $-$1.27  &  0.340E-31 &155.2 &138.2 &157.4 &138.7 &140.0 &140.1 &    \\
 \ion{Fe}{II} & 6516.08   & 2.89 &$-$3.450&$-$3.320&$-$3.37  & $-$3.31  &  0.742E-32 & 52.2 & 55.3 & 29.3 & 55.2 & 48.9 & 35.1 &       \\
 \ion{Fe}{I}  & 6518.37   & 2.83 &$-$2.750&$-$2.460&$-$2.30  & $-$2.30  &  0.516E-31 & 58.1 &102.4 & ---  & 48.2 & 56.4 & ---  &   \\
 \ion{Fe}{I}  & 6533.93   & 4.56 &$-$1.460&$-$1.460&$-$1.430  & $-$1.43  &  0.497E-30 & 34.7 &  6.1 & 13.7 & ---  & ---  & 18.6 &    \\
 \ion{Fe}{I}  & 6546.24   & 2.76 &$-$1.650&$-$1.536&$-$1.54  & $-$1.54  &  0.472E-31 &131.8 &106.9 &112.2 &106.3 & ---  &111.2 &  \\
 \ion{Fe}{I}  & 6569.21   & 4.73 &$-$0.420&$-$0.127&$-$0.45  & $-$0.45  &  0.622E-30 & 73.6 & 82.9 & 60.6 & 59.3 & ---  & 38.1 &  \\
 \ion{Fe}{I}  & 6574.23   & 0.99 & --- &$-$5.023&$-$5.004  & $-$5.00  &  0.129E-31 & ---  &  --- & ---  & 48.2 & ---  & ---  &       \\
 \ion{Fe}{I}  & 6575.02   & 2.59 &$-$2.820&$-$2.710&$-$2.710  & $-$2.71  &  0.468E-31 & ---  &  --- & ---  & 66.9 & 67.3 & ---  &       \\
 \ion{Fe}{I}  & 6581.21   & 1.48 &$-$4.860&$-$4.679&$-$4.68  & $-$4.85  &  0.142E-31 & 75.2 & 43.8 & 33.9 & ---  & 41.2 & 16.2 &  \\
 \ion{Fe}{I}  & 6591.31   & 4.59 &$-$2.070&$-$2.070&---  & $-$2.00  &  0.476E-30 &  4.1 & ---  & ---  & ---  & ---  &  2.9 &  \\
 \ion{Fe}{I}  & 6593.87   & 2.43 &$-$2.422&$-$2.422&$-$2.42  & $-$2.42  &  0.341E-31 &115.0 & 99.0 & ---  & 86.9 & 87.2 & 82.4 &  \\
 \ion{Fe}{I}  & 6597.56   & 4.79 &$-$1.070&$-$1.070&$-$1.05  & $-$1.05  &  0.701E-30 & 46.0 & 31.5 & ---  & 19.8 & ---  &  8.4 &  \\
 \ion{Fe}{I}  & 6608.04   & 2.28 &$-$4.030&$-$4.030&---  & $-$4.03  &  0.294E-31 & 16.7 & 33.8 &  7.4 & 16.3 & 18.2 & 10.5 &  \\
 \ion{Fe}{I}  & 6609.11   & 2.56 &$-$2.692&$-$2.692&$-$2.69  & $-$2.69  &  0.385E-31 & 94.6 & 29.4 & ---  & 69.3 & 54.9 & 67.1 &  \\
 \ion{Fe}{I}  & 6627.54   & 4.55 &$-$1.680&$-$1.680&---  & $-$1.68  &  0.437E-30 & 13.3 & ---  & 10.7 & ---  & ---  &  9.4 &  \\
 \ion{Fe}{I}  & 6678.00   & 2.69 &$-$1.470&$-$1.418&$-$1.418  & $-$1.42  &  0.428E-31 &160.0 &200.0 &124.0 &120.9 &130.0 &117.4 &  \\
 \ion{Fe}{I}  & 6699.14   & 4.59 &$-$2.190&$-$2.101&$-$2.101  & $-$2.10  &  0.452E-30 &  5.2 & ---  & ---  & ---  & ---  & ---  &  \\
 \ion{Fe}{I}  & 6705.10   & 4.61 &$-$1.470&$-$1.382&---  & $-$1.06  &  0.467E-30 & 38.1 & 30.1 & ---  & 26.4 & ---  &  3.3 &  \\
 \ion{Fe}{I}  & 6726.67   & 4.61 &$-$0.952&$-$1.094&---  & $-$1.09  &  0.447E-30 & 32.5 & 36.4 & ---  & ---  & ---  & 18.7 &  \\
 \ion{Fe}{I}  & 6739.52   & 1.56 &$-$4.950&$-$4.794&$-$4.79  & $-$4.80  &  0.147E-31 & 42.7 & 18.5 & ---  & ---  & ---  &  8.9 &  \\
\noalign{\smallskip} \hline 
\end{longtable}

\end{appendix}
\end{document}